\documentclass[acmtog]{./ref/acmart}
\AtBeginDocument{%
  }

\usepackage{tikz}
\usepackage[export]{adjustbox}

\usepackage{colortbl}
\usepackage{siunitx}
\usepackage[dvipsnames]{xcolor}
\usepackage{array}
\usepackage{stfloats}
\usepackage{multirow}
\usepackage{url}
\usepackage{wrapfig}
\usepackage{algorithm}
\usepackage{natbib}
\setcitestyle{authoryear,open={[},close={]}} %Citation-related commands
\usepackage[noend]{algpseudocode}
\usepackage{sidecap}
\usepackage{tabularx}
\usepackage{gensymb}
\usepackage{mathtools}
\usepackage{ccicons}
\usepackage{tikz}
\usepackage{placeins}
\usepackage{tcolorbox}

\algdef{SE}[DOWHILE]{Do}{doWhile}{\algorithmicdo}[1]{\algorithmicwhile\ #1}
\algnewcommand\algorithmicforeach{\textbf{for each}}
\algdef{S}[FOR]{ForEach}[1]{\algorithmicforeach\ #1\ \algorithmicdo}

\usepackage{hhline}
\usepackage{prettyref}
\newrefformat{fig}{Fig.~\ref{#1}}
\newrefformat{par}{Section~\ref{#1}}
\newrefformat{appen}{Appendix~\ref{#1}}
\newrefformat{sec}{Section~\ref{#1}}
\newrefformat{sub}{Section~\ref{#1}}
\newrefformat{table}{Table~\ref{#1}}
\newrefformat{tab}{Table~\ref{#1}}
\newrefformat{ass}{Assumption~\ref{#1}}
\newrefformat{alg}{Algorithm~\ref{#1}}
\newrefformat{def}{Definition~\ref{#1}}
\newrefformat{thm}{Theorem~\ref{#1}}
\newrefformat{cor}{Corollary~\ref{#1}}
\newrefformat{lem}{Lemma~\ref{#1}}
\newrefformat{step}{Step~\ref{#1}}
\newrefformat{ln}{Line~\ref{#1}}
\newrefformat{rem}{Remark~\ref{#1}}
\newrefformat{eq}{Equation~\ref{#1}}
\newrefformat{pb}{Problem~\ref{#1}}
\newrefformat{it}{Item~\ref{#1}}
\newrefformat{te}{Term~\ref{#1}}
\def\Eqref Eq:#1:{\eqref{eq:#1}}
\newrefformat{Eq}{Equation~\Eqref#1:}
\newrefformat{App}{Appendix~\ref#1:}

\newcommand{\vect}[1]{\textbf{#1}}

\begin{document}
\title{Differential Locally Injective Grid Deformation and Optimization}

\author{Julian Knodt}
\email{julianknodt@gmail.com}
\orcid{0000-0003-4461-2036}
\affiliation{%
  \institution{POSTECH}
  \city{Pohang}
  \country{South Korea}
  \postcode{37673}
}
\thanks{Both authors are corresponding authors for this work.}

\author{Seung-Hwan Baek}
\email{shwbaek@postech.ac.kr}
\orcid{0000-0002-2784-4241}
\affiliation{%
  \institution{POSTECH}
  \city{Pohang}
  \country{South Korea}
  \postcode{37673}
}

\begin{abstract}
Grids are a general representation for capturing regularly-spaced information, but since they are uniform in space, they cannot dynamically allocate resolution to regions with varying levels of detail. There has been some exploration of indirect grid adaptivity by replacing uniform grids with tetrahedral meshes or locally subdivided grids, as inversion-free deformation of grids is difficult.
This work develops an inversion-free grid deformation method that optimizes differential weight to adaptively compress space. The method is the first to optimize grid vertices as differential elements using vertex-colorings, decomposing a dense input linear system into many independent sets of vertices which can be optimized concurrently. This method is then also extended to optimize UV meshes with convex boundaries. Experimentally, this differential representation leads to a smoother optimization manifold than updating extrinsic vertex coordinates.
By optimizing each sets of vertices in a coloring separately, local injectivity checks are straightforward since the valid region for each vertex is fixed. This enables the use of optimizers such as Adam, as each vertex can be optimized independently of other vertices. We demonstrate the generality and efficacy of this approach through applications in isosurface extraction for inverse rendering, image compaction, and mesh parameterization.
\end{abstract}

\setcopyright{acmlicensed}
\acmJournal{TOG}
\acmYear{2025} \acmVolume{0} \acmNumber{0} \acmArticle{0} \acmMonth{0} \acmDOI{None}

    


\begin{teaserfigure}
    \centering
    \setlength{\tabcolsep}{0.5pt}
    \includegraphics[width=0.95\linewidth]{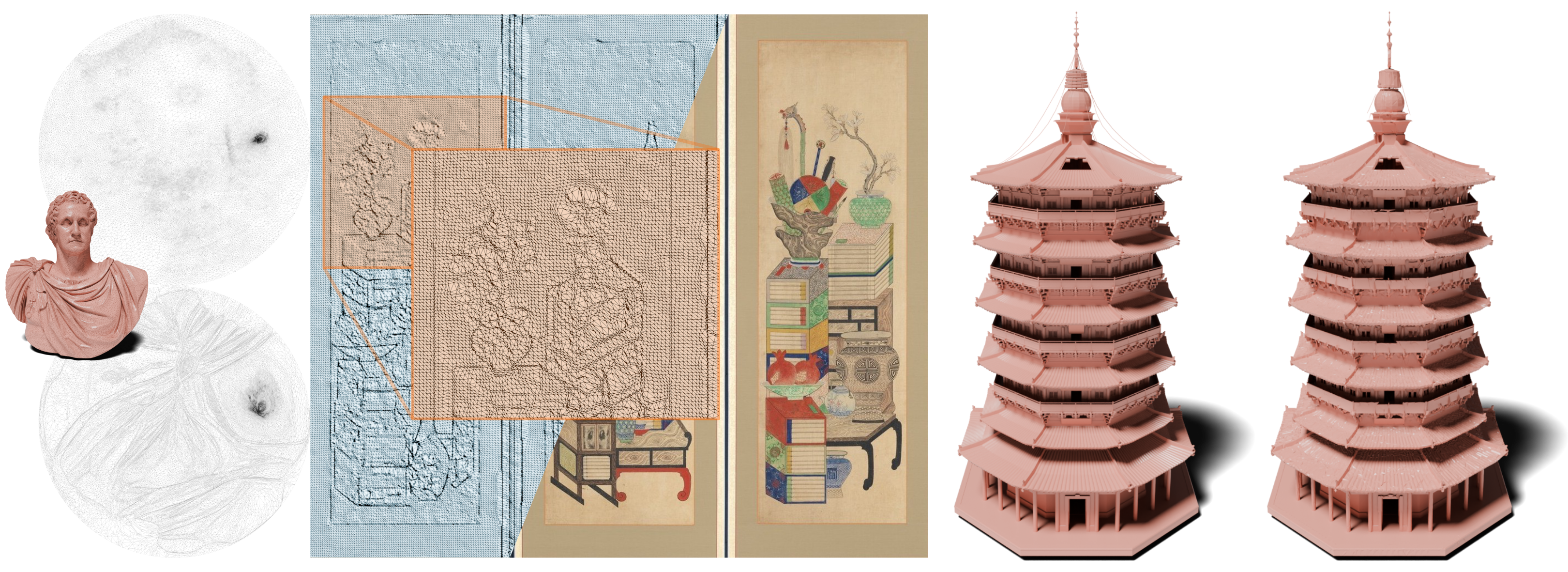}
    \put(-0.925\linewidth,-0.02\linewidth){\textit{UV Optimization}}
    \put(-0.64\linewidth,-0.02\linewidth){\textit{Image Compaction}}
    \put(-0.32\linewidth,-0.02\linewidth){\textit{Differentiable Isosurface Extraction}}
    \put(-0.95 \linewidth, 0.325\linewidth){\footnotesize Init.}
    \put(-0.95 \linewidth, 0.015\linewidth){\footnotesize Optim.}
    \put(-0.755\linewidth, 0.325\linewidth){\footnotesize Pixel Deformation}
    \put(-0.4975\linewidth, 0.015\linewidth){\footnotesize Compacted Image}
    \put(-0.37 \linewidth, 0.325\linewidth){\footnotesize Target}
    \put(-0.20 \linewidth, 0.325\linewidth){\footnotesize Optim.}
    \caption{
    This work develops a locally injective grid deformation framework that adaptively warps grids to capture spatial data by representing sets of vertices as differential elements. The method is simple yet effective, yielding improved quality across applications such as UV optimization, image-compaction and isosurface extraction, while preserving injectivity of the input using barrier. \cczero George Washington Bust (Smithsonian), \cczero Scholar's Possessions (The Met) triangulated in rendering, \ccby Chinese Pagoda (Aaron Huo).}
    \label{fig:teaser}
    \Description{Reconstruction of a Chinese Pagoda, with reconstruction of a buddha and the inside of the pagoda (which is visible through windows) zoomed in.}
\end{teaserfigure}
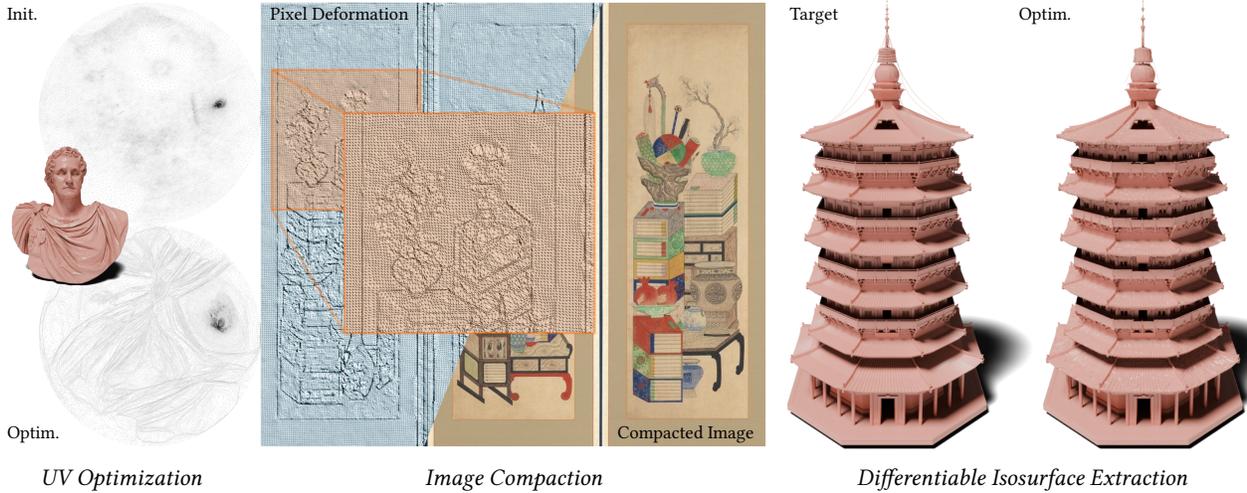

\maketitle

\section{Introduction}
Grids are versatile data structures that can represent diverse data such as images, signed distance functions and volumes, and thus are widely used in applications including fluid and physics simulations, image and volume processing, and rendering.
However, conventional grids are typically defined with uniform resolution, making them inefficient for representing data with spatially varying levels of detail.
To address this limitation, recent works have explored grid deformation~\citep{deformablegrid, dmtet, flexicubes, neumanifold}, enabling grids to adaptively capture regions with higher-frequency details.
A major challenge in grid deformation, however, is inversion, which occurs when neighboring grid cells overlap during deformation. Inversion makes the grid invalid for downstream tasks, since each position is associated with multiple values.
Previous approaches~\citep{flexicubes} mitigated this issue by restricting vertex displacement to within half a grid cell, thereby avoiding inversion but significantly limiting the expressive power and adaptability of grid deformation.
Another inversion-free optimization approach would be to rely on computationally expensive line-search evaluations, which can easily get stuck in near-degenerate configurations as mentioned in~\cite{scalable_locally_injective_mappings}.

This work addresses these challenges by introducing an efficient, inversion-free grid deformation method built upon two key ideas, first by using a differential representation~\citep{tutte, differential_representations}, where each vertex is constrained to a convex combination of its neighbor, and second is the use of graph colorings allowing sets of vertices to be optimized independently. The use of a differential representation contrasts alternative approaches which use a differential prior for the optimization gradient~\citep{large_steps}.

Representing vertices as a convex combination of neighbors is commonly expressed using the Laplacian formulation $M\vect{v}' =  L\vect{v}$, where $M\in|V|$ is a per-vertex mass matrix, and $L\in|V|\times|V|$ is the sparse Laplacian operator that relates each vertex to its neighbors. This representation has been well-studied, and has numerous applications in geometry processing, including mesh smoothing, stylization, and rigid deformation~\citep{normal_driven_spherical_shape_analogies, laplacian_mesh_smoothing, a_signal_processing_approach_to_fair_surface_design, arap_modeling}. This formulation works well for the forward case, as it straightforward to compute $\vect{v}'$ by solving for the sparse inverse of $L$. On the other hand, it is difficult to optimize the weights of the Laplacian operator $L$, since it has a dense gradient, requiring $O(|V|^2)$ memory, and thus no previous work has directly done so. Instead, optimization is made possible by decomposing the single fully-connected system into multiple independent subsystems via vertex coloring. This vertex coloring allows all vertices of one color to be optimized concurrently, while other colors remain fixed. Since all neighbors of one color are fixed, it is straightforward to prevent their inversion without getting stuck. Through this, graphs and grid vertices can be optimized as differential elements, using a vertex coloring to create many independent subsystems. This differential representation is then experimentally shown to lead to better results as compared to direct optimization on a few tasks.

In summary, this work describes methods that enables optimization of grid and graph structures with vertices represented as convex combinations of their neighbors. This is useful in arbitrary domains which require manipulating graphs without inversion, including deformation due to  physical simulation, optimization as we show with UV optimization, or can be useful for more compactly storing information as we show with image compaction. The method also enables more general inversion-free optimization, as we show with grid deformation. As compared to previous approaches which use line search and directly modify vertex positions, this work enables more efficient optimization of arbitrary energies.

\section{Related Work}

The approach outlined is based on a large body of research into adaptive grids, UV optimization, and differential representations. We review some key works, and relate them to the approach described herein.

\paragraph{Adaptive Grids}

There are many works related to subdividing grid representations, primarily focused on subdivision of grid structures around higher frequency details~\citep{acorn, flexicubes, alliez2009aabb, adaptive_multigrid_solver, adaptive_view_dependent_tessellation, adaptive_subdivision_for_surface_fitting, adaptive_subdivision_for_realistic_image_synthesis, adaptive_grid_generation_for_discretizing_implicit_complexes, poisson_surface_reconstruction} for efficient collision detection~\citep{quadtree, octree} and ray tracing~\citep{compact_fast_and_robust_grids}. There are also a number of irregular approaches based on the construction of Delauney triangulations and voronoi diagrams~\citep{radfoam, tetweave, simplicial_complex_augmentation_framework, triangle}. We make the distinction between  ``regular'' topology where each vertex has uniform degree (except boundaries) and ``irregular'' where the number of neighbors is arbitrary. Often, regular topology also implies that the embedding is uniform such that vertices are evenly spaced in $\mathbb{R}^N$. Adaptive subdivision of both regular and irregular approaches may introduce additional elements and lead to irregular output, making them less general than a regular grid. We also take ``fully-adaptive'' to mean that the topology of the grid remains unchanged, but its embedding changes. This is unlike prior work which modifies the topology.

This work maintains a regular grid, and specifically vertex deformations can be encoded as a regular grid, similar to encoding geometry on a regular grid~\citep{geometry_images, multichart_geo_images}. Maintaining topology should permit easier use in downstream applications.

\paragraph{Local Injectivity}
In mesh optimization and deformation (including animation and physics simulation), maintaining local injectivity of elements is a fundamental constraint. Local injectivity specifically refers to each position in space mapping to a single mesh element and having non-negative signed area, for mesh faces in 2D or polyhedral elements in 3D. A number of works have revolved around maintaining local injectivity during optimization, either by using barrier energies with line-search~\citep{locally_injective_mappings, ipc, c-ipc}, explicit backtracking with inversion~\citep{angle_based_flattening, air_meshes_for_robust_collision_handling, simplicial_complex_augmentation_framework, strict_minimizers_for_geometric, inter_surface_mapping}, or optimization with energies that explicitly prevent inversion~\citep{scalable_locally_injective_mappings, bijective, finite_element_image_warping}. Similar to our work, \shortcite{computing_locally_injective_mappings_by_advanced_mips} also uses colorings for optimizing MIPS within blocks of a mesh, but does not directly optimize convex weights. Other approaches focus on global bijectivity~\citep{bijective}, optimizing boundaries while preventing overlap. This work focuses on local injectivity, while maintaining global bijectivity by fixing boundaries.

\paragraph{Differential Representations}

A key component of this work is representing vertices as the convex sum of their neighbors. This representation has been well-studied in the forward manner~\citep{arap_modeling, tutte, cubic_stylization, normal_driven_spherical_shape_analogies, laplacian_mesh_smoothing, polygon_laplacian_made_simple, spectral_mesh_processing, nonmanifold_mesh_laplacian} where a known Laplacian is used to derive the position of vertices, but has not been as closely studied in the inverse setting where a differential representation is updated to minimize a target energy as no prior work has optimized differential weights. The most similar work on inverse differential representations is ~\citep{large_steps}, which preconditions the optimization gradient with the mesh Laplacian.

Another similar approach to this work is ``How to Morph Tiles Injectively''~\citep{how_to_morph_tilings_injectively}, which morphs between two convex, bounded triangulations with fixed correspondences injectively. Under identical boundary conditions, injective transformations between tilings are guaranteed by averaging convex combinations. Many optimization approaches under the surface are quite similar, requiring inversion-free optimization, possibly by augmenting ambient space with a triangulation or tetrahedralization~\citep{scalable_locally_injective_mappings, simplicial_complex_augmentation_framework, bijective, shape_blending_using_the_star_skeleton, air_meshes_for_robust_collision_handling}.

\paragraph{Differentiable Rendering Mesh Representations}

One specific application shown in this work for grid deformation is differentiable rendering, which has recently been of interest for recovering surfaces from photogrammetry~\citep{neural_radiance_fields, instant_ngp, gaussian_splatting}. There are a large variety of representations built to represent object structure of 3D shapes~\citep{sugar_gaussian_splat, radfoam, deftet, dmtet, deformablegrid, flexicubes, neumanifold, deep_marching_cubes, dmesh, dmesh++, neural_radiance_fields}, and some use standard mesh representations~\citep{large_steps, nvdiffmod, nvdiffrast}. There are also many isosurface extraction methods~\citep{marching_cubes, marching_primitives, dual_contouring, surface_nets, dual_marching_cubes, manifold_dual_contouring, marching_cubes_33, dmtet, marching_tetrahedra}. This work uses an existing isosurface extraction algorithm of \shortcite{dmtet} on a regular grid while preventing self-intersections. 

\section{Method}

\begin{figure}[h]
    \centering
    \begin{tabular}{c c}
        \multicolumn{2}{c}{Single Vertex Deformation} \\
        Half-Cell ~\citep{flexicubes} & Kernel (This Work) \\
        \includegraphics[width=0.36\linewidth]{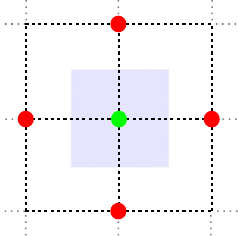} &
        \includegraphics[width=0.36\linewidth]{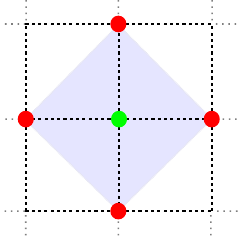} \\
    \end{tabular}
    \caption{Previous work~\citep{flexicubes} deforms each \textcolor{green}{vertex} along half of each grid cell. This work optimizes vertices within the kernel of its \textcolor{red}{neighbors}. Valid positions in the current configuration are shown in \textcolor{blue}{blue}.}
    \label{fig:single-vertex}
    \Description{Visualization of optimization of prior work versus this work. Left column is one grid vertex surrounded by a grid cell with a box around it half the size of the surrounding cell. Right column is one vertex with a diamond shape around it, the diamond's corners lay on the midpoints of the outer cell.}
\end{figure}

\begin{figure}[!hbt]
    \centering
    \setlength{\tabcolsep}{2pt}
    \begin{tabular}{c c c}
        \multicolumn{3}{c}{Non-Convex Star-Shaped Polygon Regions} \\
        Convex Sum & Interior & Kernel  \\
        \includegraphics[width=0.25\linewidth]{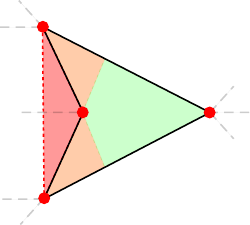} &
        \includegraphics[width=0.25\linewidth]{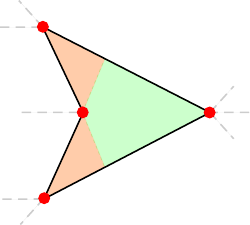} &
        \includegraphics[width=0.25\linewidth]{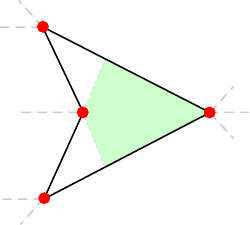} \\
    \end{tabular}
    \caption{The unconstrained convex combination of a polygon may lead to vertices exiting the polygon's interior, shown in \textcolor{red}{red}. When vertices are inside the polygon, some regions lead to non-injectivity, shown in \textcolor{orange}{orange}. The region which has no inversion is the kernel of the polygon, shown in \textcolor{green}{green}. To ensure vertices stay in the kernel, explicit checks for inversion are required and optimization is reverted if elements flipped.}
    \label{fig:convex_sum_counterexample}
    \Description{The same polygon 3 times which is shaped like an arrow-head, with different possible output regions highlighted, including the region for the convex sum, the interior, and the kernel.}
\end{figure}

Recent optimization of grids has limited deformation to half of each grid cell where injectivity is guaranteed, as in~\citep{flexicubes}:
\begin{align}
    v' = v + \delta && \delta\in[-0.5, 0.5]
\end{align}
Instead, this work treats grid deformation similarly to UV optimization, by guaranteeing each vertex lies within the kernel of its neighbors, guaranteeing injectivity, as shown in Fig.~\ref{fig:single-vertex}. 
To more closely capture the valid deformation region, deformations are represented using a convex sum:
\begin{align}
    v_i' = \sum_{x\in \text{nbr}(v_i)} w_i x_i && w_i \geq 0 && \sum w_i  = 1
\end{align}
where $w_i$ is a positive weight for each adjacent vertex in $\text{nbr}(v_i)$, which sum to unity, and $x_i$ is an adjacent vertex's neighbor. When the neighboring $x_i$ form a convex polygon, and thus are identical to the kernel, this formulation prevents inversion. When the neighboring $x_i$ otherwise form a concave polygon, this formulation does not guarantee injectivity as shown in Fig.~\ref{fig:convex_sum_counterexample}. To handle both cases, optimization is augmented with a barrier energy and explicit checks that prevent inversion.

\paragraph{Why a Differential Representation?}
Since a convex combination does not prevent inversion, it is unclear whether there are benefits as compared to an extrinsic deformation $x+\delta,\delta\in\mathbb{R}^3$. While not technically more expressive, there are a few reasons why an implicit representation is better than an extrinsic representation. First, the set of valid configurations for convex combinations is larger compared to extrinsic deformations. Extrinsic deformations are locked to a small convex region, and thus the valid set of values is infinitesimal compared to all possible values of $\delta$. In contrast, while a convex combination has some invalid configurations, a finite portion of configurations are valid, making optimization smoother which we show experimentally. Second, by representing vertices differentially, when an adjacent vertex deforms the neighbors will adaptively update as a kind of regularization, similar to ~\citep{large_steps} but baked into the representation itself.

\paragraph{Convex Formulation}
Optimizing each vertex as a convex sum of its neighbors is conceptually similar to optimizing a Laplacian formulation of the grid, which can be characterized by the following equation: 
\begin{align}\label{eq:laplacian}
    M\vect{v}' &= L\vect{v} \\
    M_{ii} &= \text{BarycentricCellArea(v)} \nonumber\\
    L_{ij} &= \begin{cases}
        w \geq 0 \text{ if } j \in\text{nbr}(i) \\
        -\sum_{k\in\text{nbr}(v_i)} L_{ik} \text{ if } i =j\\
        0 \text{ otherwise}
    \end{cases} \nonumber
\end{align}
Where $\text{nbr}(i)$ are the neighbors of vertex $i$, the mass matrix $M$ represents per vertex weights, and the sparse symmetric Laplacian matrix $L\in\mathbb{R}^{|V|\times|V|}$ represents the influence of each vertex on its neighbors. In Eq.~\ref{eq:laplacian}, per edge weights $w$ are usually chosen to be either uniform ($w = 1$) or, for triangle meshes, the cotangent Laplacian weights $w = \frac{1}{2}(\cot\alpha + \cot\beta)$. 
For the purpose of grid deformation, it is necessary to optimize $w$ to deform the underlying grid structure. Unfortunately, while it is possible to compute $\vect{v}'$ efficiently using sparse inverse solvers, the gradient of $w$ with respect to each input vertex is dense. Specifically, $\forall i,j,\frac{dv'_i}{dw_j} > 0$. This makes it computationally intractable to compute the gradient, since the size of the gradient and thus the required computation increases quadratically with the number of vertices.

Thus, two components of this formulation make direct optimization difficult: it is not possible to optimize the Laplacian matrix directly, since the memory scales quadratically with the number of vertices, $O(|V|^2)$. Second, when optimizing one vertex as a convex combination of its neighbors, it is assumed that a vertex's neighbors are fixed, which is not true for global optimization approaches which updates all vertices concurrently. This leads to incorrect optimization and possible inversion when neighboring vertices changes. To get around both problems, it is possible to optimize each vertex one-by-one (as in ~\citep{angle_based_flattening}), but this is too inefficient. Instead, this work proposes to optimize \textit{independent sets} of vertices concurrently, which minimizes the efficiency loss of updating vertices while maintaining their neighbors.

\paragraph{Alternating Optimization for 2D Grids}

To use a differential representation per vertex on a regular 2-dimensional grid, a useful observation is that when a vertex's neighbors are fixed, inversion-free optimization is straightforward. Requiring each vertex's neighbors to be fixed leads to the simple alternating optimization where each vertex $i,j\in\mathbb{Z}$ in a 2D uniform grid is labeled ``even'' or ``odd'', as in the following: \begin{equation}\label{eq:grid-alt}
    \text{label(i,j)} = \textcolor{green}{\textbf{even}} \text{ if } (i+j)\text{ mod } 2 = 0 \text{ else } \textcolor{red}{\textbf{odd}}
\end{equation}

\begin{figure}[!htb]
    \centering
    \begin{tabular}{c c}
        \textcolor{green}{Even} Step & \textcolor{red}{Odd} Step \\
        \includegraphics[width=0.4\linewidth]{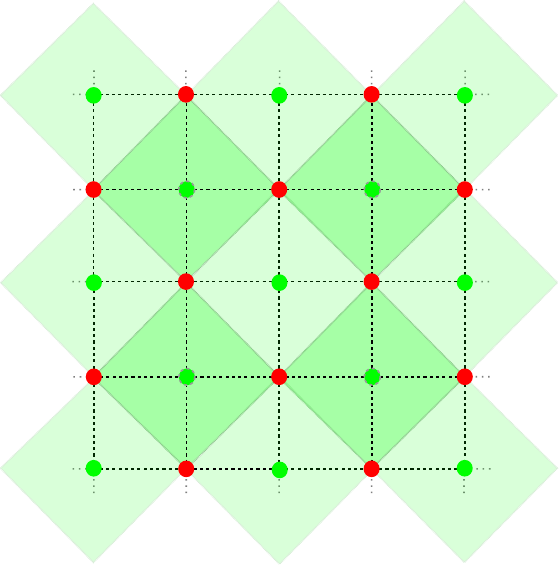} &
        \includegraphics[width=0.4\linewidth]{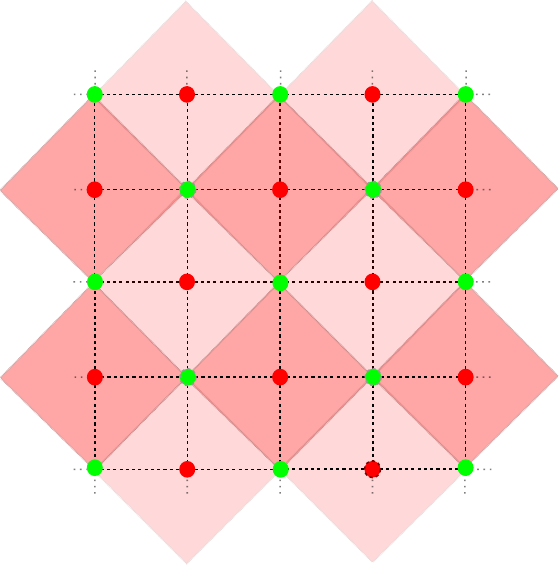} \\
    \end{tabular}
    \caption{To optimize vertices in a uniform grid, alternating steps are taken (Eq.~\ref{eq:grid-alt}), fixing all \textcolor{red}{odd vertices}, and optimizing \textcolor{green}{even vertices}. Then, all \textcolor{green}{even vertices} are fixed, and \textcolor{red}{odd vertices} are optimized. The checkerboard region indicates regions vertices can be optimized within initially.}
    \label{fig:grid_alt_opt}
    \Description{Two grids, with alternating vertex colors of green and red. On the left, around each green vertex is a green diamond, and on the right around each red vertex is a red diamond, to indicate different optimization steps.}
\end{figure}

Optimization is performed alternating between fixing odd vertices, then fixing even vertices, as shown in Fig.~\ref{fig:grid_alt_opt}. If a vertex is moved out of the kernel of the polygon formed by its fixed neighbors, it is reset to its previous position ensuring injectivity without affecting any other vertices. This allows the use of arbitrary optimization approaches, and does not require global search. Since vertices are updated independently, optimizers such as Adam~\citep{adam} can be used, whereas previous approaches required line search with a global step-size~\citep{scalable_locally_injective_mappings}.

\paragraph{Grid Boundary Conditions}

For boundary vertices, additional Dirichlet boundary constraints must be imposed to ensure that the grid does not shrink. Specifically, vertices along each edge of the grid are pinned along their corresponding dimension. For example the vertices on the left side of a 2-dimensional grid are pinned with $x\stackrel{\text{fixed}}{=}-1$. These constraints reduce the degree of freedom of each vertex by one for each boundary it lays along, while letting it move in other dimensions.

\paragraph{Alternating Optimization for Meshes}

\begin{figure}[!ht]
    \centering
    \setlength{\tabcolsep}{0pt}
    \begin{tabular}{c c c c}
        \textcolor{red}{Red} Opt & \textcolor{blue}{Blue} Opt & \textcolor{green}{Green} Opt & \textcolor{Goldenrod}{Yellow} Opt \\
        \includegraphics[width=0.24\linewidth]{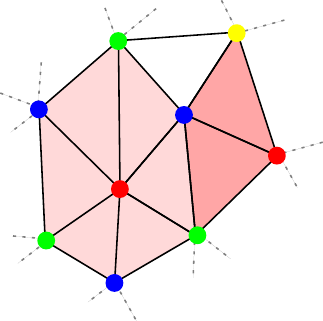} &
        \includegraphics[width=0.24\linewidth]{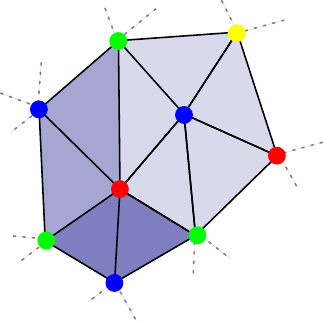} &
        \includegraphics[width=0.24\linewidth]{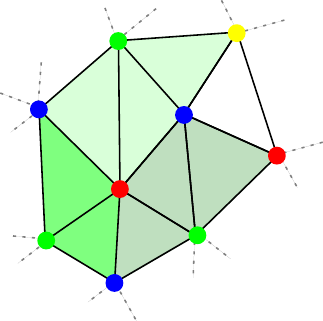} &
        \includegraphics[width=0.24\linewidth]{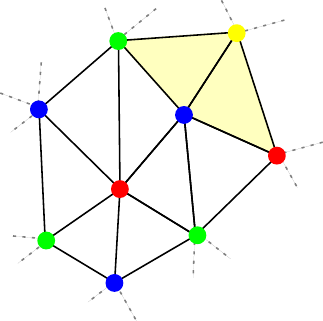} \\
    \end{tabular}
    \caption{To optimize a differential mesh representation, first vertices are colored, then all vertices of one color are optimized concurrently while other colors are fixed. From this it is easy to verify if a vertex is not in its one-ring neighborhood. Different shading indicates the one-ring for different vertices, showing the optimizable regions for each.}
    \label{fig:mesh-opt}
    \Description{Similar to above, on a small subset of a mesh there is a 4-vertex coloring shown. In each column, the area around vertices of each color is shown to indicate the regions of optimization.}
\end{figure}

Extending this alternating optimization to general already-injective meshes is possible by generalizing the choice of ``even'' and ``odd'' labeling as a vertex coloring on the input graph. Inspiration for this is drawn from ~\citep{computing_locally_injective_mappings_by_advanced_mips}, which uses a vertex coloring to optimize the MIPS energy.
Constructing a vertex coloring on a graph and optimizing each color works similarly to grids, except with more optimization steps as shown in Fig.~\ref{fig:mesh-opt}. Since each color requires one optimization step, the fewer colors used, the more efficient the optimization. For planar mesh optimization, such as UV optimization, it is possible to color planar graphs with at most four colors~\citep{four_color_theorem}, but in practice a simple greedy coloring has more than four colors.

\paragraph{Generalizing to N-Dimensional Grids}

The alternating optimization for 2D grids is easily extended to N-dimensional grids:
\begin{equation}
    \text{label(c $\in\mathbb{Z}^\text{N}$)} = \textcolor{green}{\textbf{even}} \text{ if } \sum_{i=0}^N c[i]\text{ mod } 2 = 0 \text{ else } \textcolor{red}{\textbf{odd}}
\end{equation}

This makes it clear that the number of dimensions of the grids does not affect the number of independent sets allowing efficient optimization in arbitrary dimensions, unlike arbitrary meshes. Pseudocode for optimization of differential grids is shown in Alg.~\ref{alg:differential-opt}.

\begin{algorithm}[!h]
\caption{Differential Grid Optimization\label{alg:differential-opt}}
\begin{algorithmic}[1]
    \Statex \textbf{Input: } $V_0\in\mathbb{R}^{N\times K}, F\subseteq V$, injective w/ fixed boundaries 
    \Statex \textbf{Output: } $V'\in\mathbb{R}^{N\times K}$
    \State $W[i,j]\in\mathbb{R}^{|V|\times|V|\times d}\stackrel{\text{init}}{=}\begin{cases}1 \text{ if } v_i \text{ adj } v_j\\0 \text{ otherwise}
    \end{cases}$
    \For{$k \in 1 :N$}
        \State $\text{Mask}= \text{\textcolor{green}{Even} if } k\bmod2 = 0 \text{ else \textcolor{red}{Odd}}$
        \State $W'[i,j] = \frac{\text{Softplus}(W[i,j])}{\sum_{j' \in\text{nbr($V_i$)}} \text{Softplus}(W[i,j'])}$ \\
        \State $V_k[i] = \begin{cases}\sum_{v_j\in \text{nbr($v_i$)}} V_{k-1}[j] W'[i,j] \text{ where Mask} \\
        V_{k-1} \text{ Otherwise}\end{cases}$
        \State Fix Boundaries of $V_k$
        \State $\text{L} = \ell(V_k, F) + 
        \text{Barrier}(V_k, F)$ \Comment{Per Problem Loss Function}
        \State $W[i,j] \leftarrow W[i,j]-\eta \frac{\partial W[i,j]}{\partial L}$
        \While{any face $F^*\subseteq V_k$ inverted}
            \State $V_k[F] \leftarrow V_{k-1}[F]$ \Comment{Reset vertices of inverted face}
        \EndWhile
    \EndFor
    \Return $V_N$
\end{algorithmic}
\end{algorithm}

\paragraph{Benefits Compared to Global Line Search}
While it is still necessary to evaluate barrier energies and backtrack optimization if inversion is found, there are multiple benefits of optimizing each vertex in a coloring. First, previous approaches optimize all vertices concurrently, requiring global optimization with one step size. By optimizing independent vertex color sets, each vertex has no affect on others. This prevents one nearly-degenerate element from blocking optimization of all other elements. Furthermore, and key to this work, it enables representing vertices as differential elements, which is not computationally feasible if all vertices are updated concurrently.

% \paragraph{Generality of Mesh Optimization} With this approach, all possible meshes where vertices are a convex sum of its neighbors is representable. Furthermore, it is always possible to represent a vertex as a convex combination of its neighbors, by constructing the neighbor's convex hull, and finding the barycentric coordinates of any simplex containing the nested vertex, or for certain dimensions alternative formulations have been shown~\citep{mean_value_coordinates, wachspress_coordinates}. Thus, it should be possible to represent all possible configurations with our approach.

\paragraph{Per Dimension Differential} While prior approaches to differential representations operate per vertex, experimentally we observe that for \textit{a few} cases representing vertices as differential \textit{per dimension} leads to better results. To be specific, rather than representing vertices as $v_i = \sum_{j\in nbr(v_i)} w_{ij}v_j$, vertices are represented differentially per dimension as $v_{i,d} = \sum_{j\in nbr(v_i)} w_{ij,d}v_{j,d}, d\in[x,y,z,\ldots]$. This allows for more degrees of freedom for optimization, at the cost of requiring $\times d$ more parameters, and to the authors' knowledge has not been explored in prior work.

For experiments in this work, per-dimension weights are used only for the toy example on the $\ell_\text{spin}$ loss (Section. ~\ref{sec:per-dim-diff}), where vertices must ``wrap'' around their neighbors. Other experiments do not require this kind of ``wrapping'', and have losses that are affected more by the density of any vertex in some regions. Thus, for all results in this text, aside from $\ell_\text{spin}$, results are shown with weights per-vertex, with additional results comparing weights per-dimension in the supplement.

\paragraph{Inversion Prevention with Barrier Energies}

Since it is possible for elements to invert with a differential representation, to prevent inversion there are three kinds of approaches. First, is \textit{computational} post-conditional checks for inversion with barrier energies after each optimization step. Second, is \textit{explicit} computation of the kernel, where rather than using convex weights of the one-ring, convex weights are instead applied to the vertices of the kernel. Third is \textit{implicitly} parameterizing points in the kernel directly, and optimization of this parameterization. Unfortunately, it has been shown that it is impossible to create a generalized barycentric coordinate for polygons with more than three sides that maps to the kernel~\citep{bijective_mappings_with_generalized_barycentric_coordinates_counterexample}, and we are left with choosing between either the explicit or computational optimizations. We chose to use the computational version, and we chose not to explicitly compute the kernel to avoid is that it is non-trivial to compute with consistent ordering and number of vertices, and hard to extend to 3D. A more detailed explanation of reasoning is given in the supplement.

To incorporate barrier energies, we compute the signed area (resp. volume) for 2D and 3D triangle (tetrahedral) subdivisions, and apply the IPC~\citep{ipc} energy to each subdivided simplex. For 2D, each square cell is subdivided into 4 triangles, and for 3D each cubic cell is subdivided into 10 tetrahedron. Without subdivision, these subdivided elements may flip leading to invalid output for further processing such as bilinear sampling of pixels, and isosurface extraction. Furthermore, by subdividing along multiple axes, symmetry is maintained, not biasing towards certain flip-free solutions.
\section{Experiments}

To isolate the effects of the differential representation with alternating optimization, we show this approach on four problems. First, we demonstrate each effect in isolation on a simple 2D grid, and then expand these ideas to UV optimization, image compaction, and differentiable rendering with isosurface extraction.

All experiments were run on an 8 Core AMD Ryzen 7800X3D CPU, with an NVIDIA RTX 3060 GPU. \textcolor{blue}{Code will be made public.}

\subsection{2D Grid Examples}

\begin{figure}[htb]
    \centering
    \setlength\fboxsep{0pt}
    \setlength{\tabcolsep}{\arraystretch pt}
    \begin{tabular}{c c c}
        Initialization & X Optimization & X,Y Optimization \\
        \colorbox{teal!5}{\includegraphics[width=0.25\linewidth]{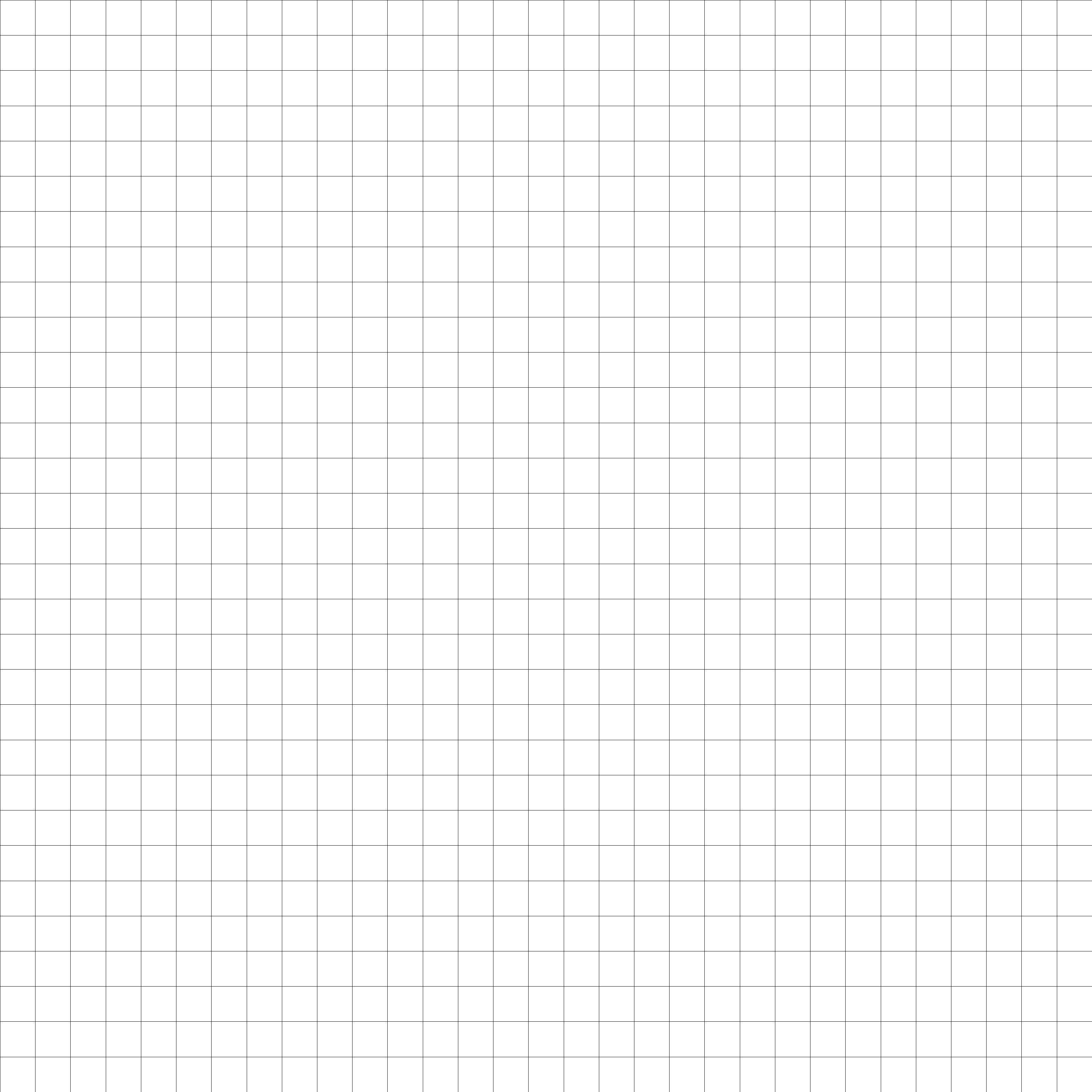}}
        \put(-0.14025\linewidth, 0.119\linewidth){
        \tikz\node[opacity=0.5,fill=red,inner sep=\fboxsep,anchor=base]{
        \phantom{\includegraphics[width=0.012\linewidth, height=0.012\linewidth]{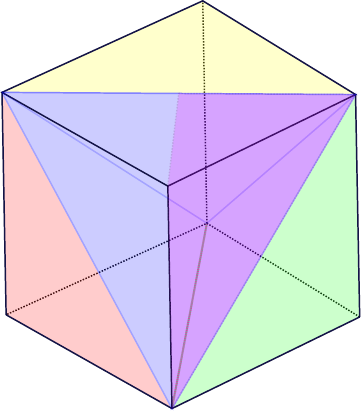}}
        };
        }
        &
        \colorbox{teal!5}{\includegraphics[width=0.25\linewidth]{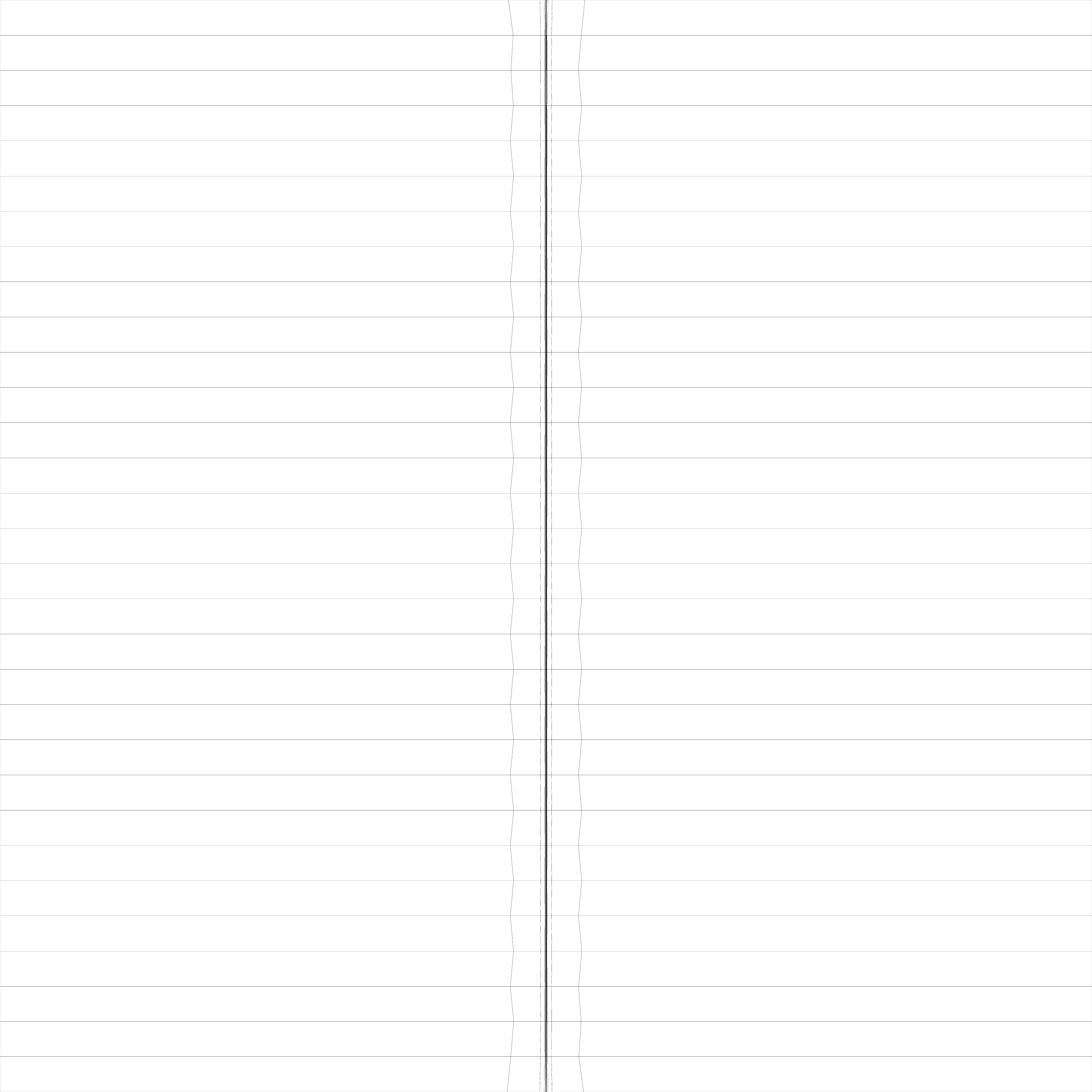}} &
        \colorbox{teal!5}{\includegraphics[width=0.25\linewidth]{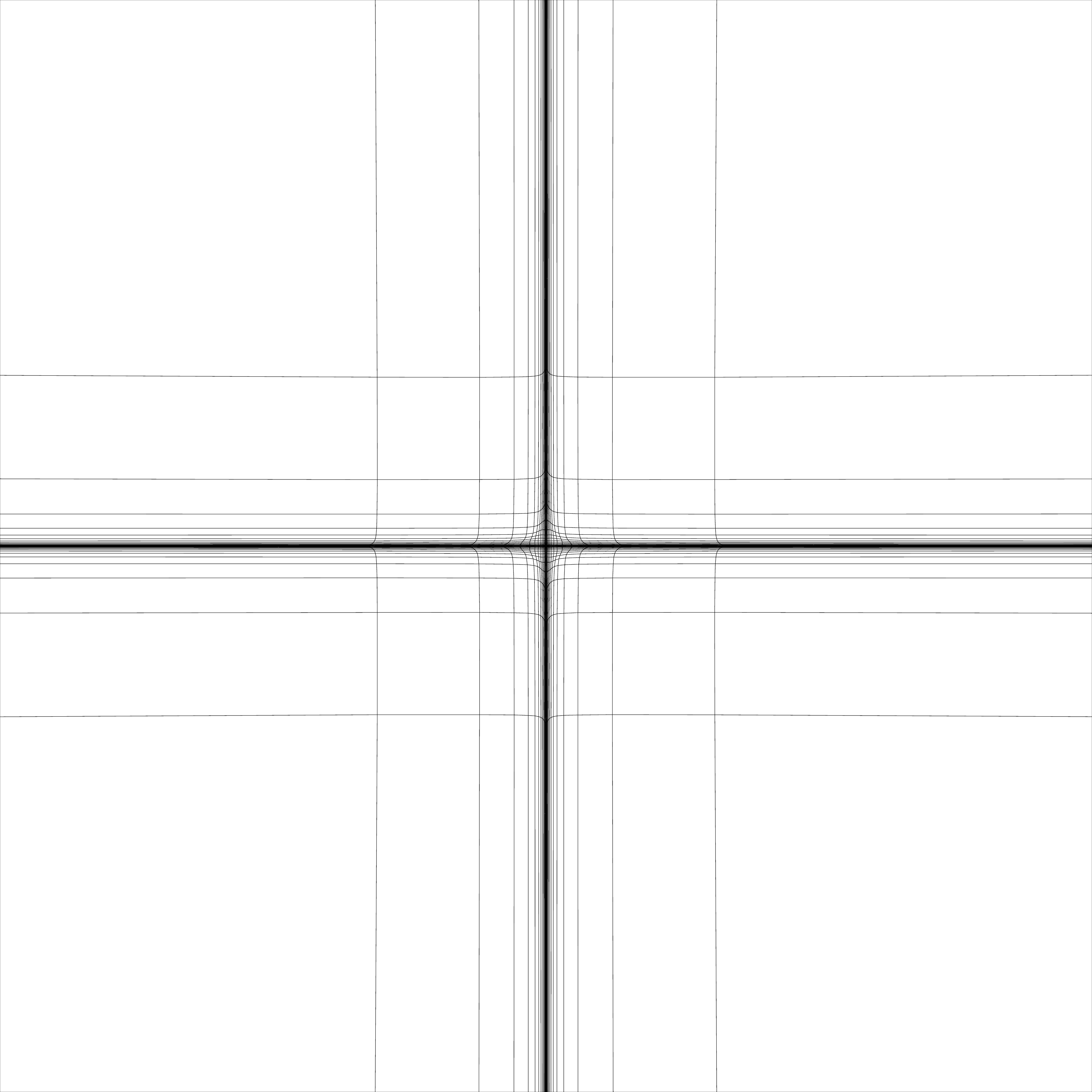}} \\
        \put(-10pt,0.012\linewidth){\rotatebox{90}{30$\times$ Ctr. \textcolor{red}{Zoom}}}
        \colorbox{teal!5}{\includegraphics[width=0.25\linewidth]{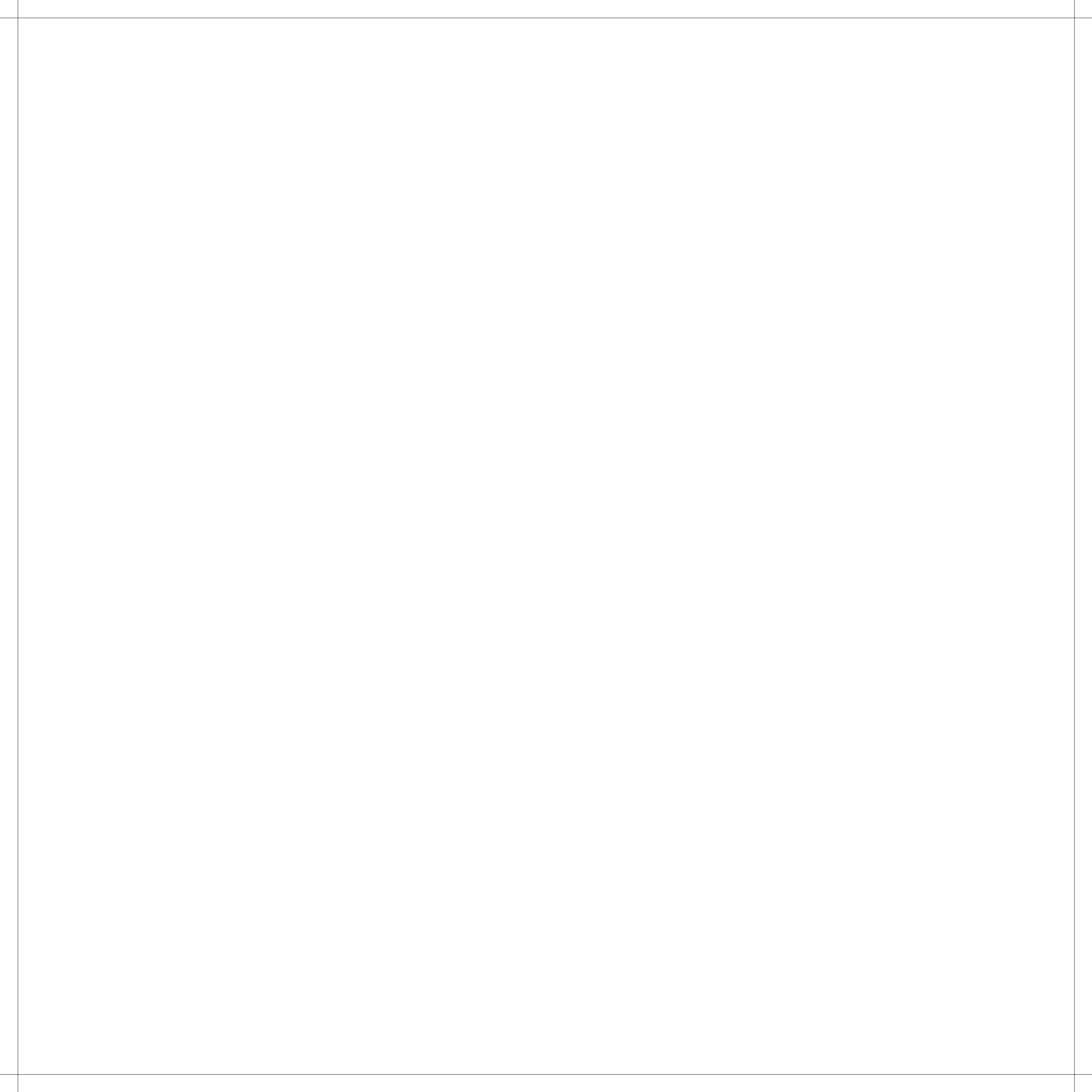}} &
        \colorbox{teal!5}{\includegraphics[width=0.25\linewidth]{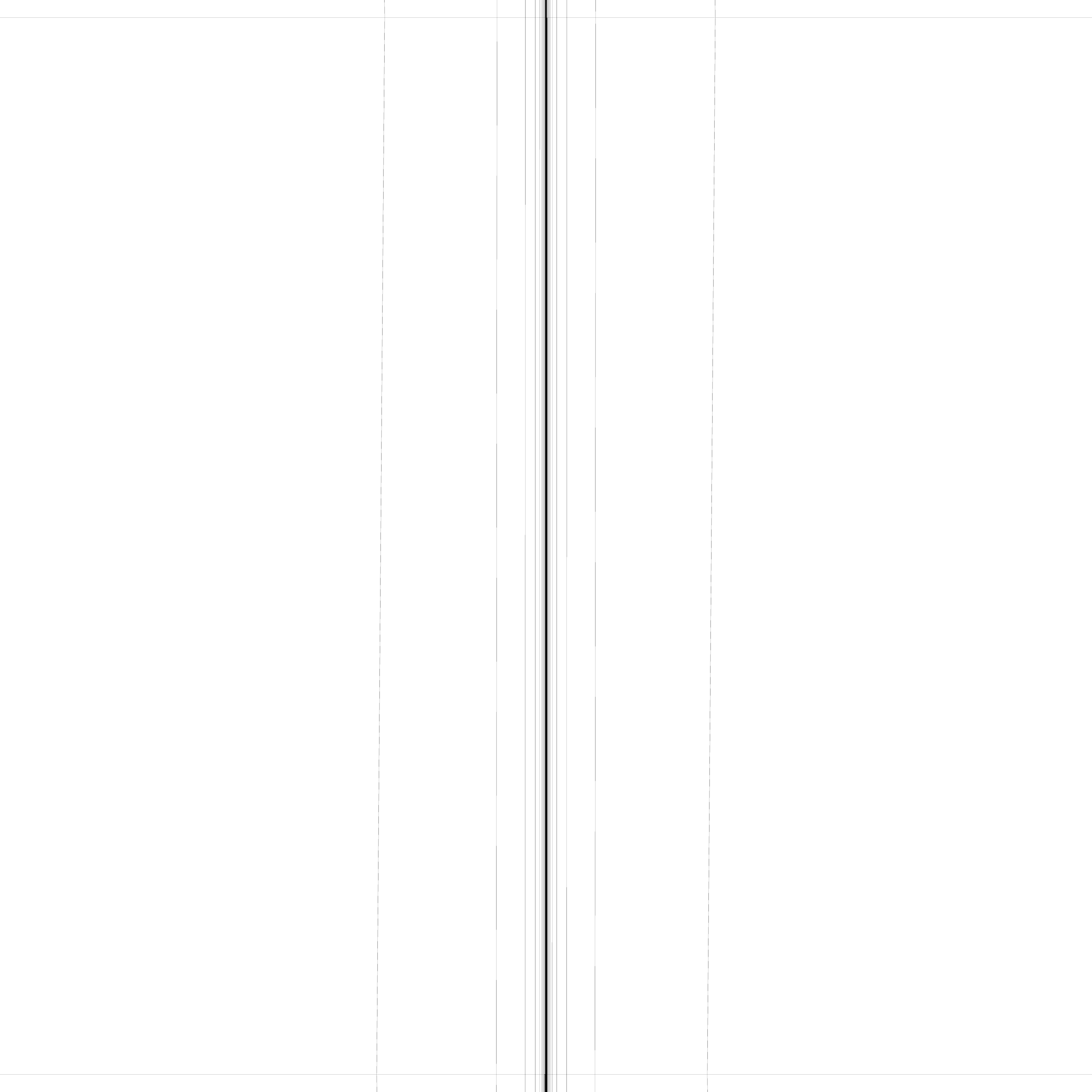}} &
        \colorbox{teal!5}{\includegraphics[width=0.25\linewidth]{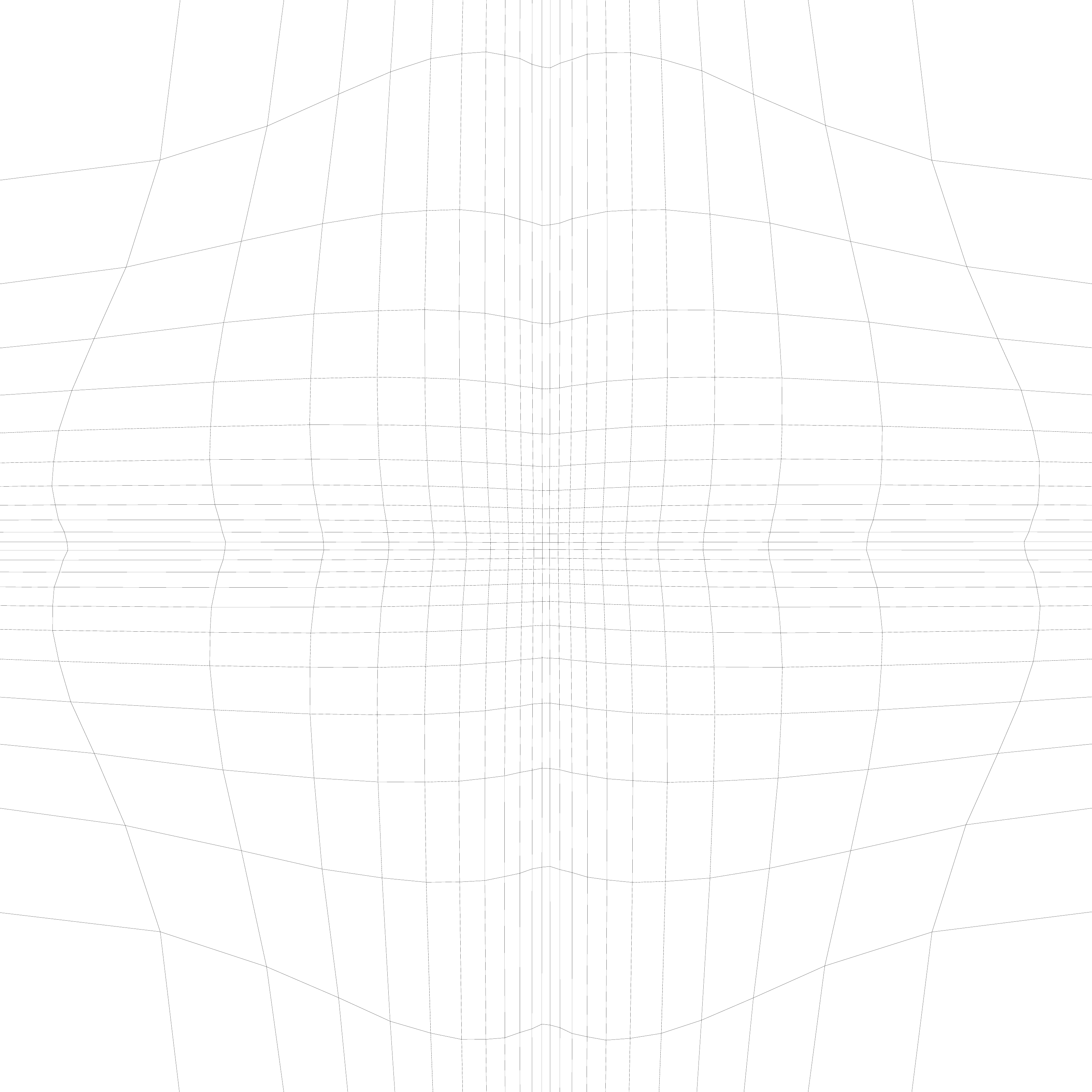}} \\
    \end{tabular}
    \caption{Example of direct optimization of position values per grid vertex, with alternating optimization on even and odd vertices, and representing vertices as convex combinations of their neighbors. There is no visible bias introduced, and vertices optimize smoothly. The zoomed region is indicated by the red box.}
    \label{fig:simple_grid_opt}
    \Description{Left column: A grid, and a zoom in of a grid-cell.
    Center column: All vertices in the grid are squished onto the X-axis. A zoom in of the squished vertices. Right Column: All vertices are squished towards the center. A zoom in of the squish, showing no overlaps.}
\end{figure}

To start, we show both the convex representation and vertex coloring while deforming a grid and minimizing the two following loss functions per vertex: \begin{align}
    \ell_x &= \lvert v_x \rvert & \text{X Opt} \nonumber\\
    \ell_{x,y} &= |v_x| + |v_y| & \text{XY Opt}
\end{align}
$\ell_x$ effectively pulls all vertices towards the X-axis and $\ell_{x,y}$ towards the center of the grid, shown in Fig.~\ref{fig:simple_grid_opt}. Even without inversion checks, for these simple energies vertices deform smoothly using the Adam optimizer. Despite the mesh being heavily compressed, no inversions are introduced, although for this toy example it is difficult to introduce inversions. To prevent bias, we found it best to use a separate Adam optimizer for even and odd vertices (or more generally per color) due to momentum. We track multiple momenta for all experiments in this work.

\subsubsection*{Comparison of Per Dimension Weights}\label{sec:per-dim-diff}

To show more complex results with inversion checks and resets, we show another toy example, where vertices are spun around the center of the grid by $175\degree$: \begin{align}
    \ell_\text{spin} = (175\degree + \angle_{init} - \angle_\text{curr})^2 + (\sqrt{x_\text{curr}^2 + y_\text{curr}^2} - \sqrt{x_{init}^2 + y_{init}^2})^2 
\end{align}
Since boundary vertices are locked to the edges they lie on, this spinning introduces inversions, which are prevented through alternating optimizations with inversion checks.

\begin{figure}[!h]
    \centering
    \setlength\fboxsep{0pt}
    \setlength{\tabcolsep}{1.5pt}
    \begin{tabular}{c c c c}
        \multicolumn{4}{c}{Ablating Differential per Vertex or per Dimension} \\
        Per Dimension & 5$\times$ \textcolor{red}{Zoom} & Per Vertex & 5$\times$ \textcolor{red}{Zoom} \\
        \colorbox{teal!5}{\includegraphics[width=0.22\linewidth]{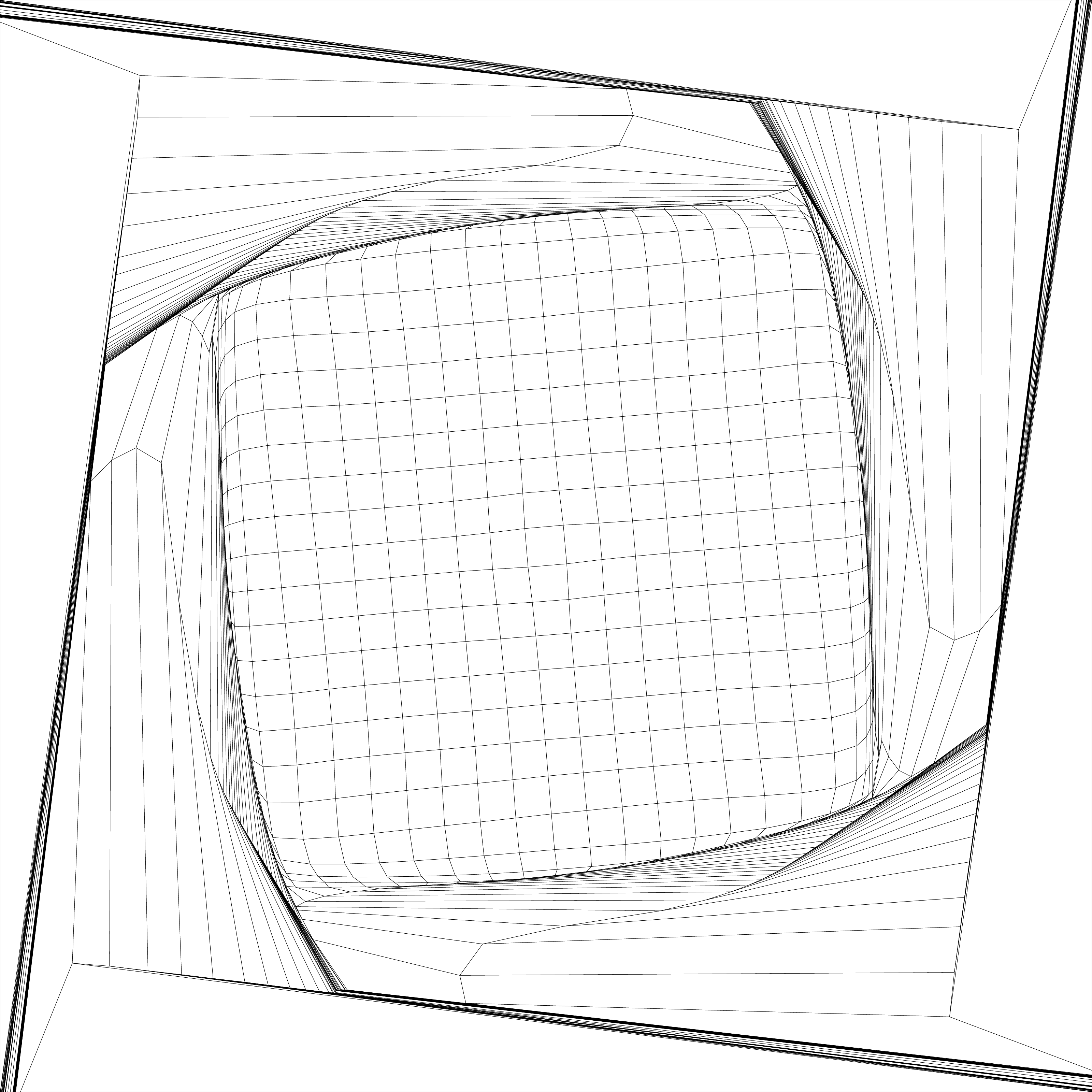}}
        \put(-0.139\linewidth,0.09\linewidth){
        \tikz\node[opacity=0.5,fill=red,inner sep=\fboxsep,anchor=base]{
        \phantom{\includegraphics[width=0.04\linewidth, height=0.04\linewidth]{assets/cube_subdiv.pdf}}
        };}
        &
        \colorbox{teal!5}{\includegraphics[width=0.22\linewidth]{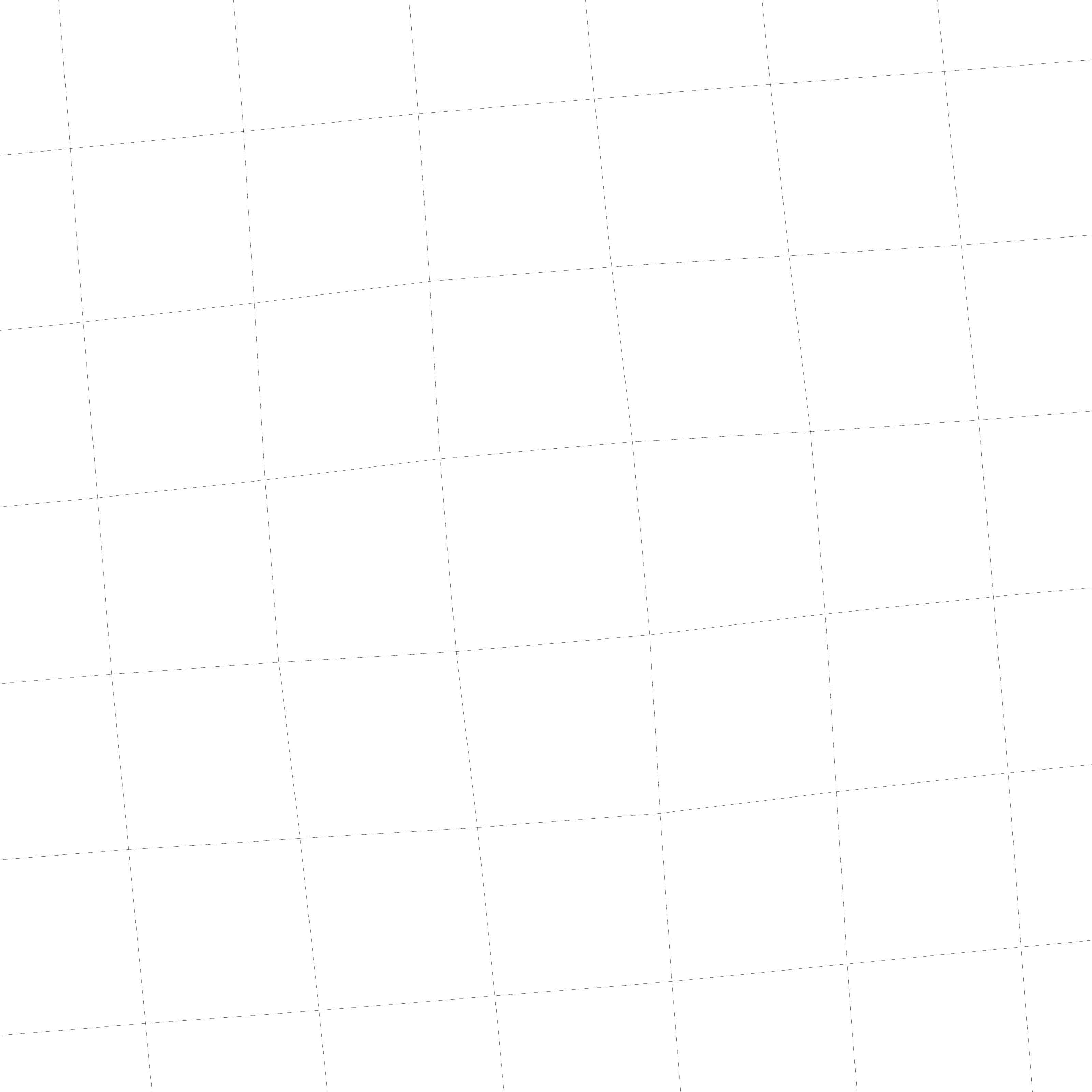}} &
        \colorbox{teal!5}{\includegraphics[width=0.22\linewidth]{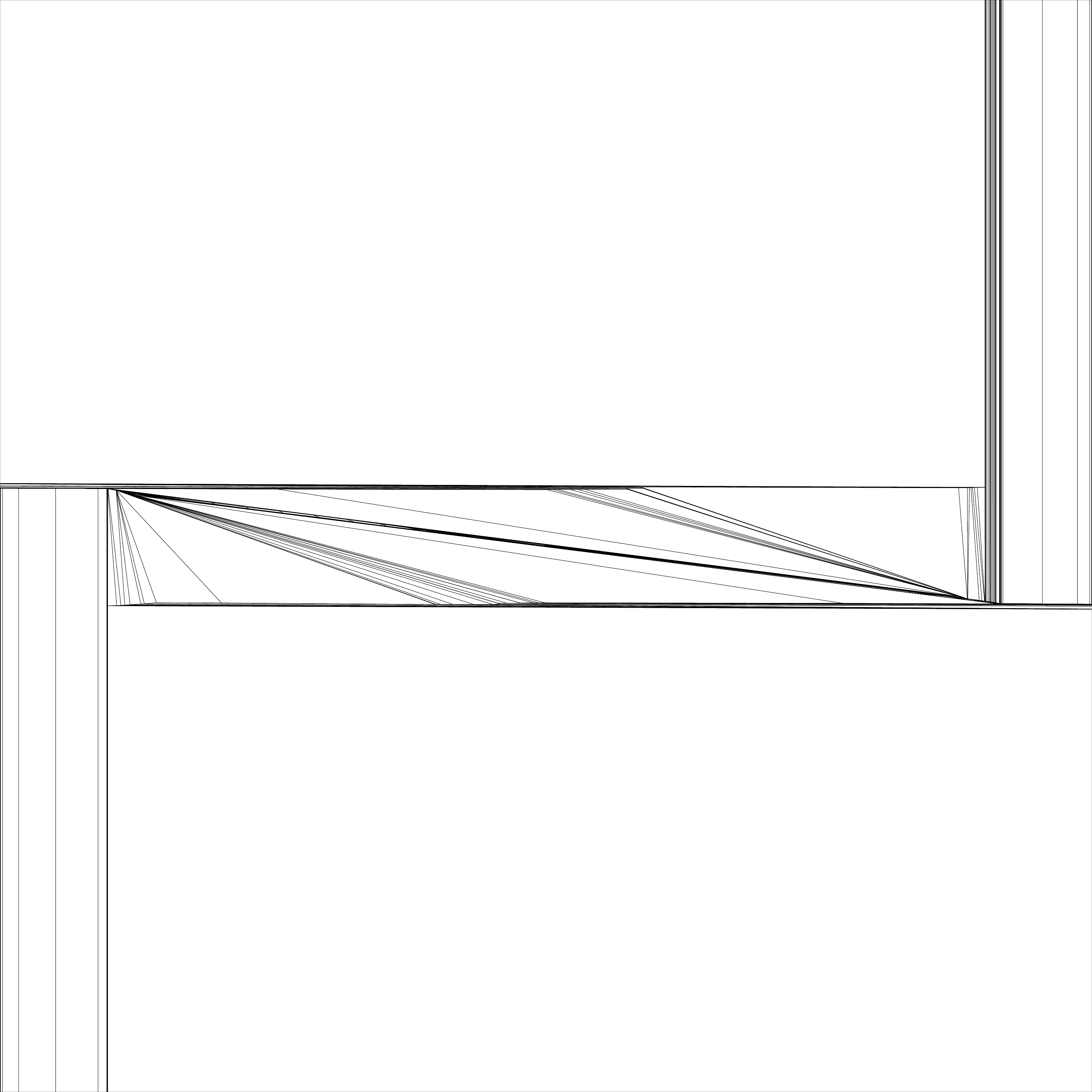}} &
        \colorbox{teal!5}{\includegraphics[width=0.22\linewidth]{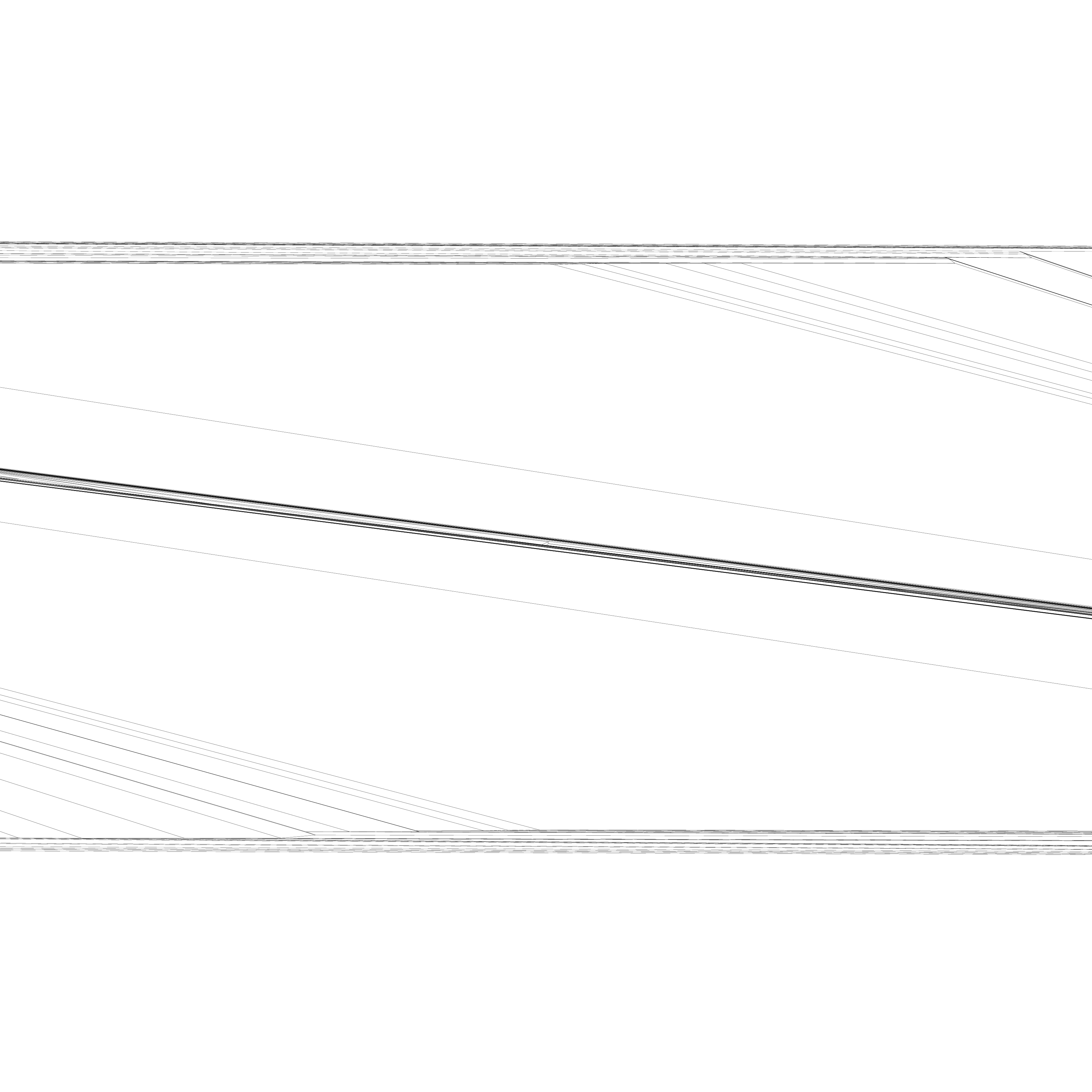}} \\
        Loss & 0.882 & & 1.930 \\
    \end{tabular}
    \caption{When using one weights for each dimension, i.e. for X and Y, results significantly improve compared to a single weight per neighboring vertex on $\ell_\text{spin}$. $\ell_x, \ell_{x,y}$ are not shown as they have exactly the same result.}
    \label{fig:per-dimension}
    \Description{Comparison of spinning loss on two uniform 2D grids. The right figure only captures a small region in the center of the grid.}
\end{figure}

For this specific loss function though, we find that per-dimensions weights lead to better results as compared to per-vertex and we show the effect of optimizing vertices as the differential of each neighboring vertex or optimizing each dimension independently in Fig.~\ref{fig:per-dimension}. For $\ell_x, \ell_{x,y}$, the outputs have precisely the same loss, and thus are not shown. For $\ell_\text{spin}$, using a differential per dimension leads to significantly better results. This is because having vertices wrap around each other is difficult when using one weight per vertex. While this shows that using one weight per dimension is more general, it is not always an improvement, as for example with $\ell_x$ and $\ell_{x,y}$ the results are identical.

For the rest of the comparisons on $\ell_\text{spin}$, we use weights per-dimensions.

\subsubsection*{Comparison of Alternating Optimization to Line Search}

\begin{figure}[th]
    \centering
    \setlength\fboxsep{0pt}
    \setlength{\tabcolsep}{1.5pt}
    \begin{tabular}{c c c}
        \multicolumn{3}{c}{Inversion Check Kind Ablation} \\
        No Checks & Line Search & Alternating Opt.  \\
        \colorbox{teal!5}{\includegraphics[width=0.25\linewidth]{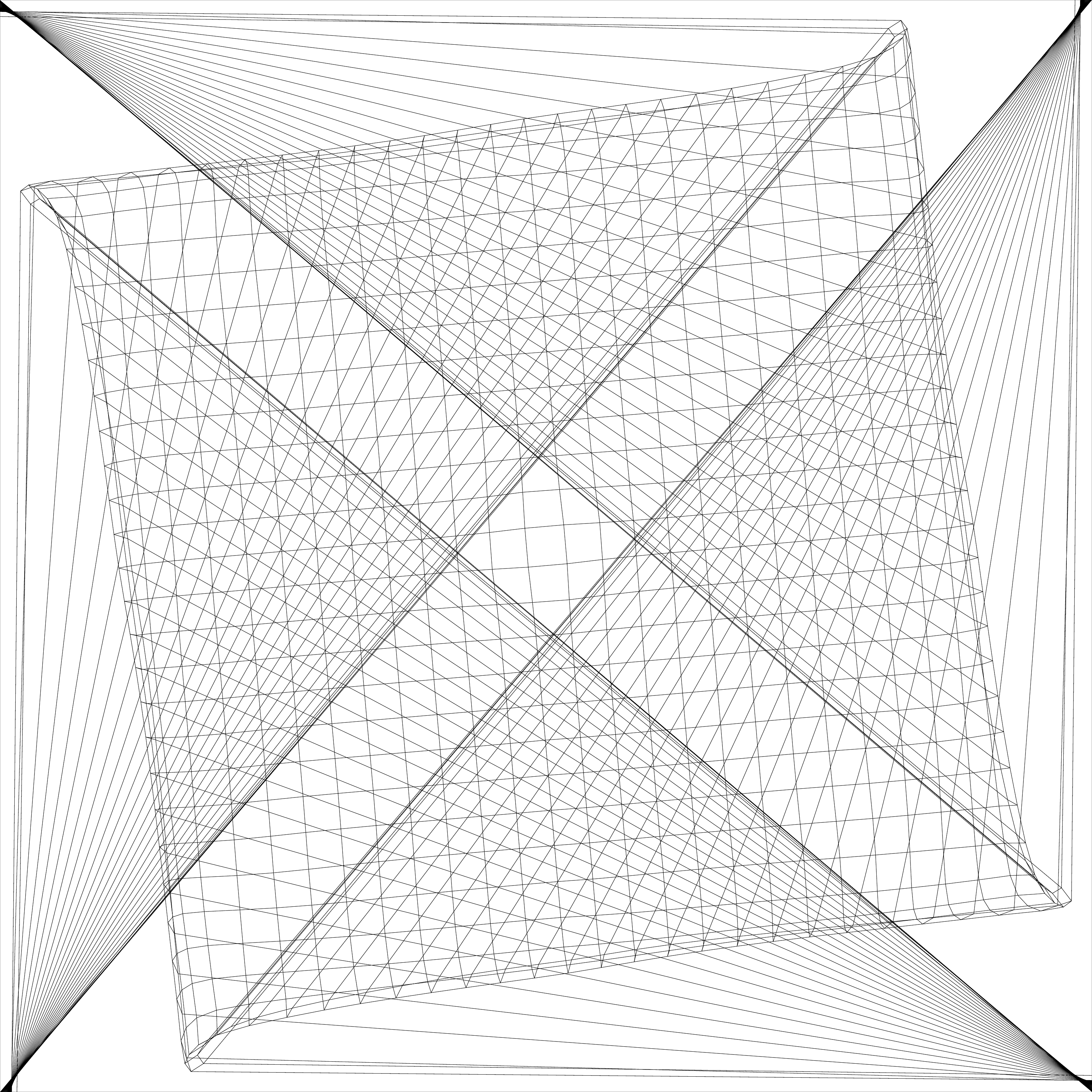}}
        \put(-0.161\linewidth,0.1\linewidth){
        \tikz\node[opacity=0.5,fill=red,inner sep=\fboxsep,anchor=base]{
        \phantom{\includegraphics[width=0.052\linewidth, height=0.052\linewidth]{assets/cube_subdiv.pdf}}
        };
        }
        &
        \colorbox{teal!5}{\includegraphics[width=0.25\linewidth]{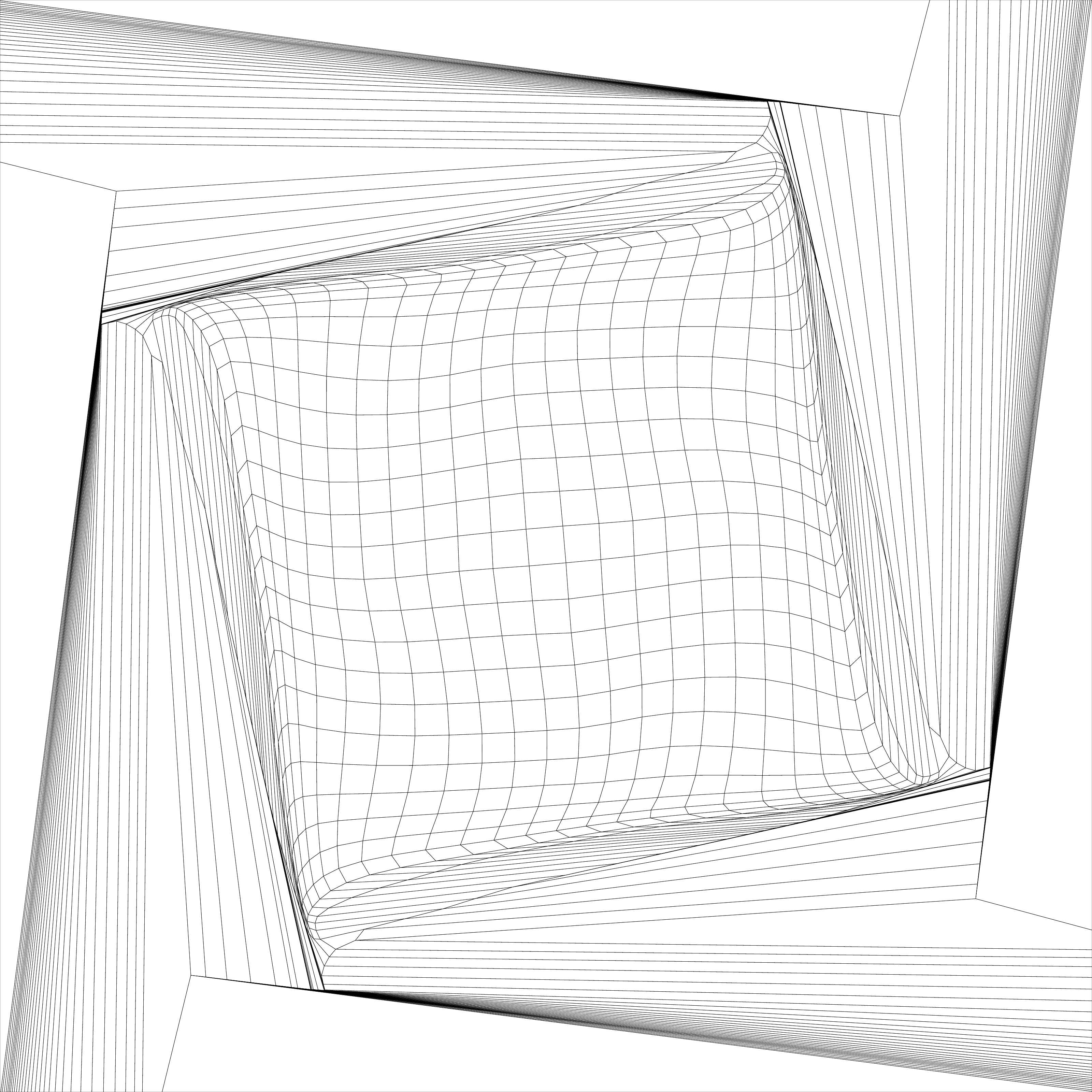}} &
        \colorbox{teal!5}{\includegraphics[width=0.25\linewidth]{assets/spin_test}} \\
        \put(-10pt,0.05\linewidth){\rotatebox{90}{5$\times$ \textcolor{red}{Zoom}}}
        \colorbox{teal!5}{\includegraphics[width=0.25\linewidth]{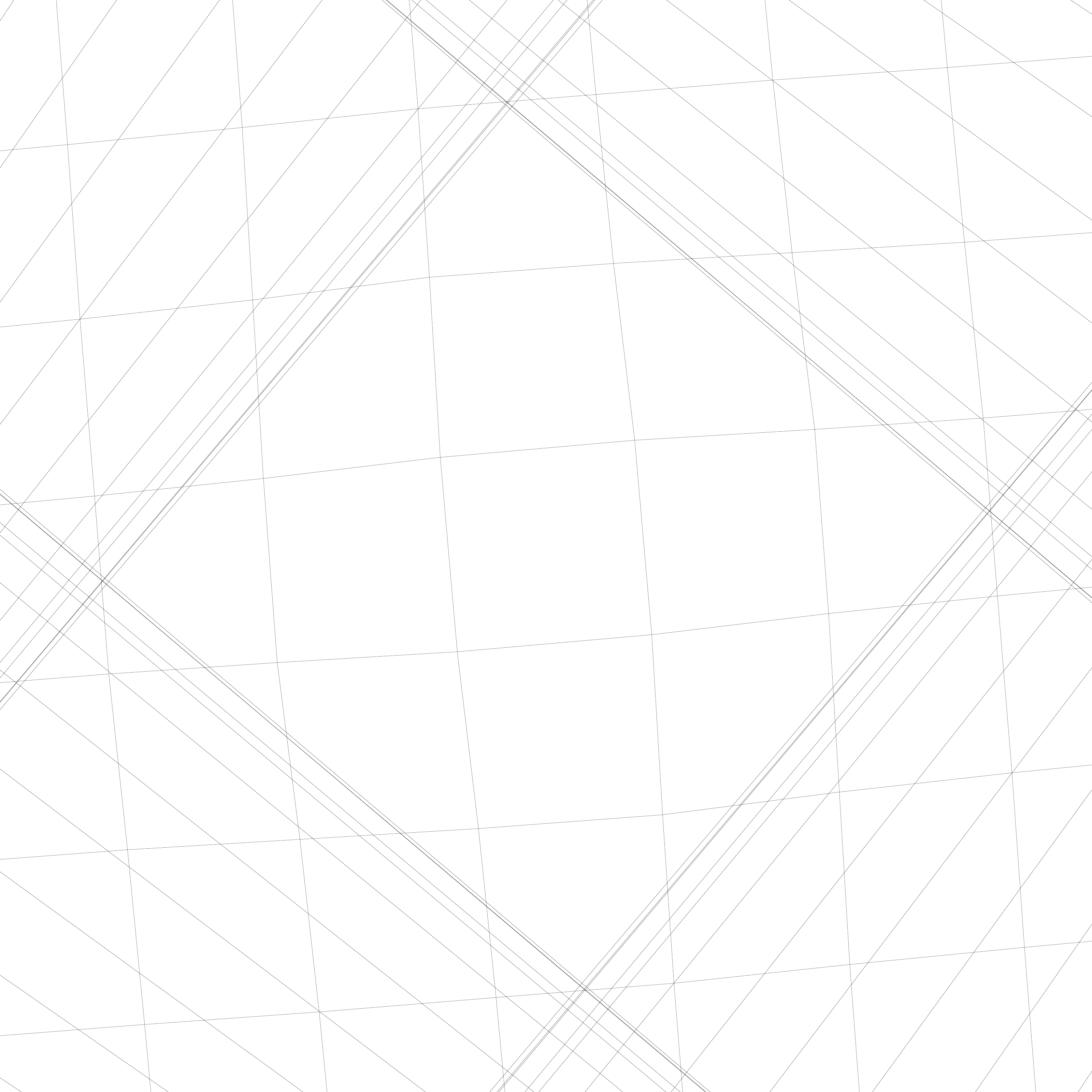}} &
        \colorbox{teal!5}{\includegraphics[width=0.25\linewidth]{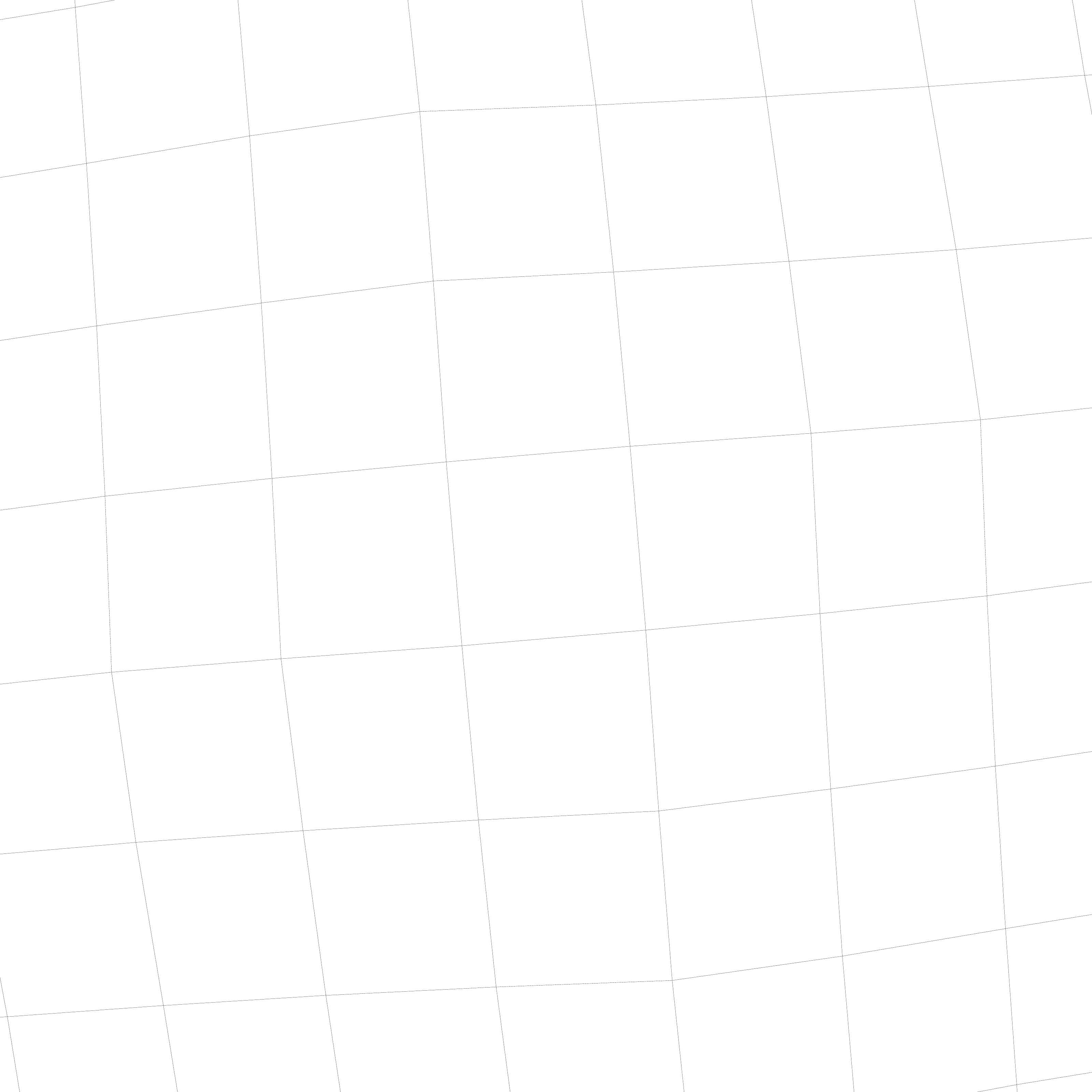}} &
        \colorbox{teal!5}{\includegraphics[width=0.25\linewidth]{assets/spin_test_zoom}} \\
        \put(-0.22\linewidth, 0){\scriptsize Runtime (sec):} 12.90 & 240.10 & 26.66 \\
        \put(-0.237\linewidth, 0){\scriptsize Injective:} \textcolor{red}{No} & \textcolor{ForestGreen}{Yes} & \textcolor{ForestGreen}{Yes} \\
        \put(-0.22\linewidth, 0){\scriptsize Loss:} $0.684$ & $0.897$ & $0.882$ \\
    \end{tabular}
    \caption{Comparison of inversion prevention approaches during optimization, all with equal iterations and vertices represented as the convex sum of their neighbors. Inversions are introduced when no checks are done, and line search slows optimization. Alternating optimization has less impact on efficiency, while performing similarly to line search.}
    \label{fig:inv-check}
    \Description{Same spinning grid as before. Underneath per-vertex, the grid is heavily compressed into the middle of the cell.}
\end{figure}

For $\ell_\text{spin}$ we then compare results using vertex-coloring based alternating optimization, which only resets inverted vertices, against line-search which must reduce a global step-size, greatly reducing efficiency. We compare to a simple version of line-search, which backtracks each step by $0.9\times$ until a valid solution is found. We show these results, along with optimization without inversion checks in Fig.~\ref{fig:inv-check}. We observed that simple back-tracking can get stuck due to numerical instability when using single-precision floating point, but did not see any issue with the alternating optimization.

\begin{figure}[!tbh]
    \centering
    \setlength\fboxsep{0pt}
    \setlength{\tabcolsep}{1.5pt}
    \begin{tabular}{c c c c}
        \multicolumn{4}{c}{Spinning Optimization Deformation Comparison} \\
        \multicolumn{3}{c}{Direct Deform} & Convex Sum \\
        \put(-0.155\linewidth, 0){\scriptsize Optimizer:}{\footnotesize Adam} & {\footnotesize Vector Adam} & {\footnotesize Uniform Adam} & {\footnotesize Adam} \\
        
        \colorbox{teal!5}{\includegraphics[width=0.22\linewidth]{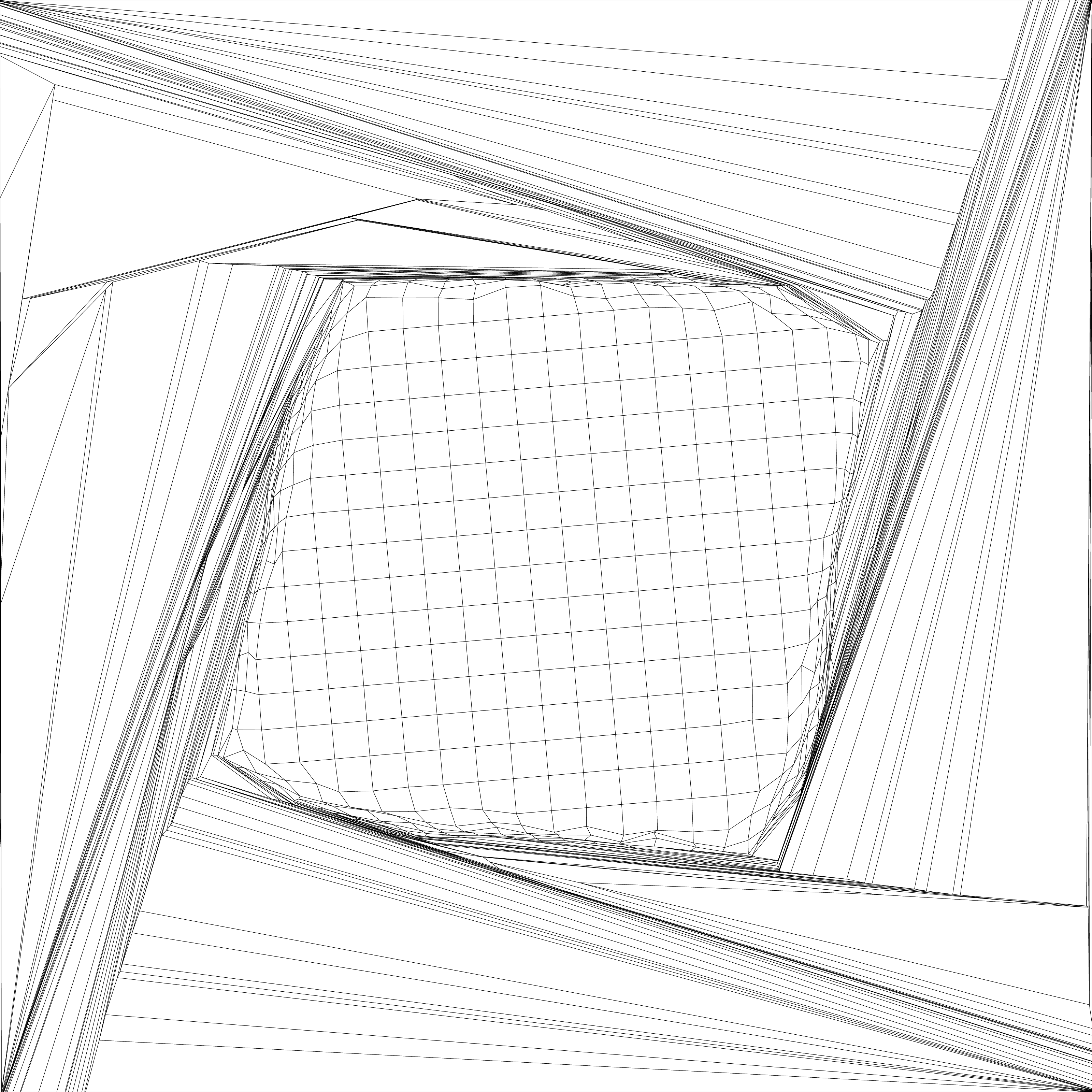}}
        \put(-0.139\linewidth,0.09\linewidth){
        \tikz\node[opacity=0.5,fill=red,inner sep=\fboxsep,anchor=base]{
        \phantom{\includegraphics[width=0.04\linewidth, height=0.04\linewidth]{assets/cube_subdiv.pdf}}
        };
        }
        &
        \colorbox{teal!5}{\includegraphics[width=0.22\linewidth]{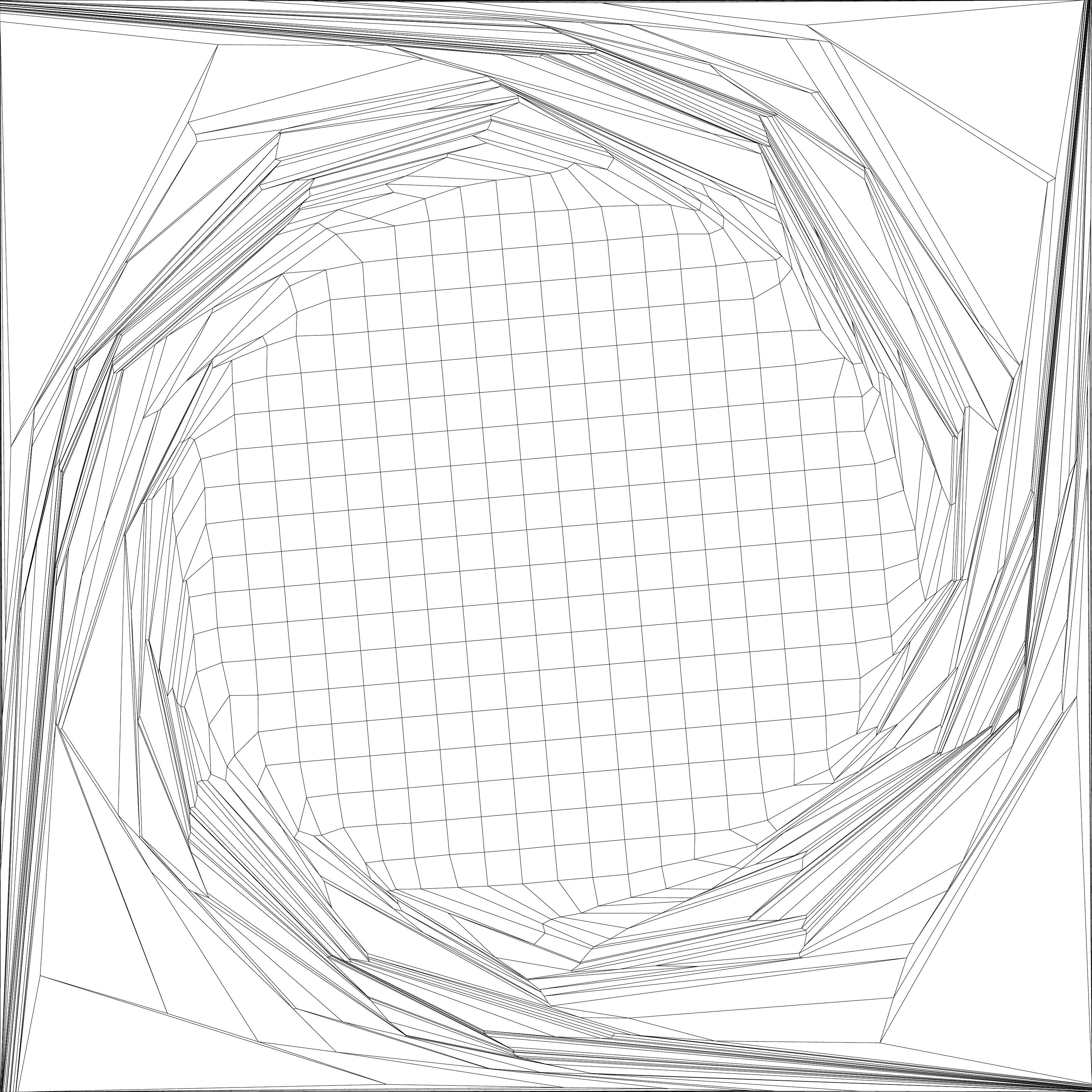}} &
        \colorbox{teal!5}{\includegraphics[width=0.22\linewidth]{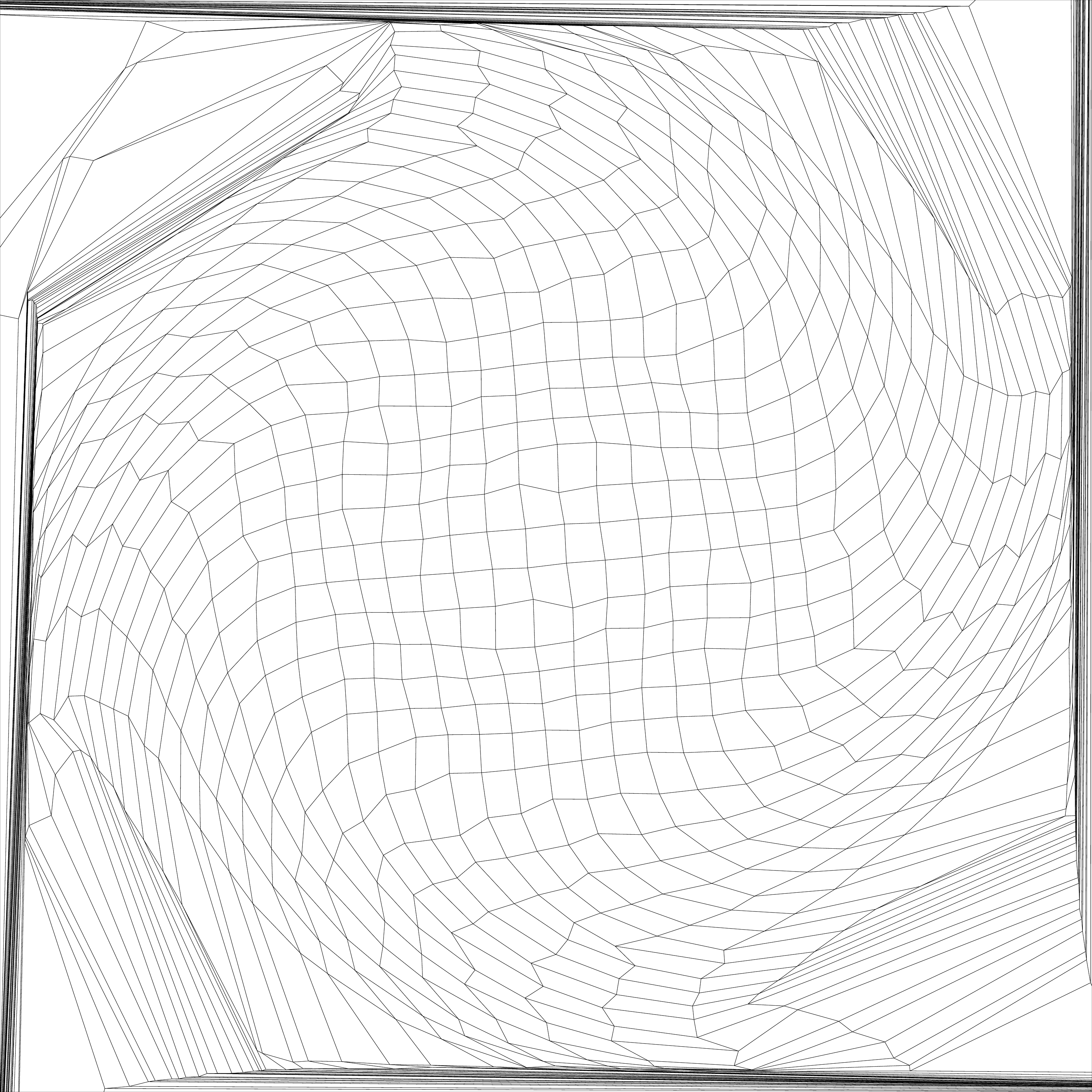}} &
        \colorbox{teal!5}{\includegraphics[width=0.22\linewidth]{assets/spin_test}} \\
        \put(-10pt,0.045\linewidth){\rotatebox{90}{5$\times$ \textcolor{red}{Zoom}}}
        \colorbox{teal!5}{\includegraphics[width=0.22\linewidth]{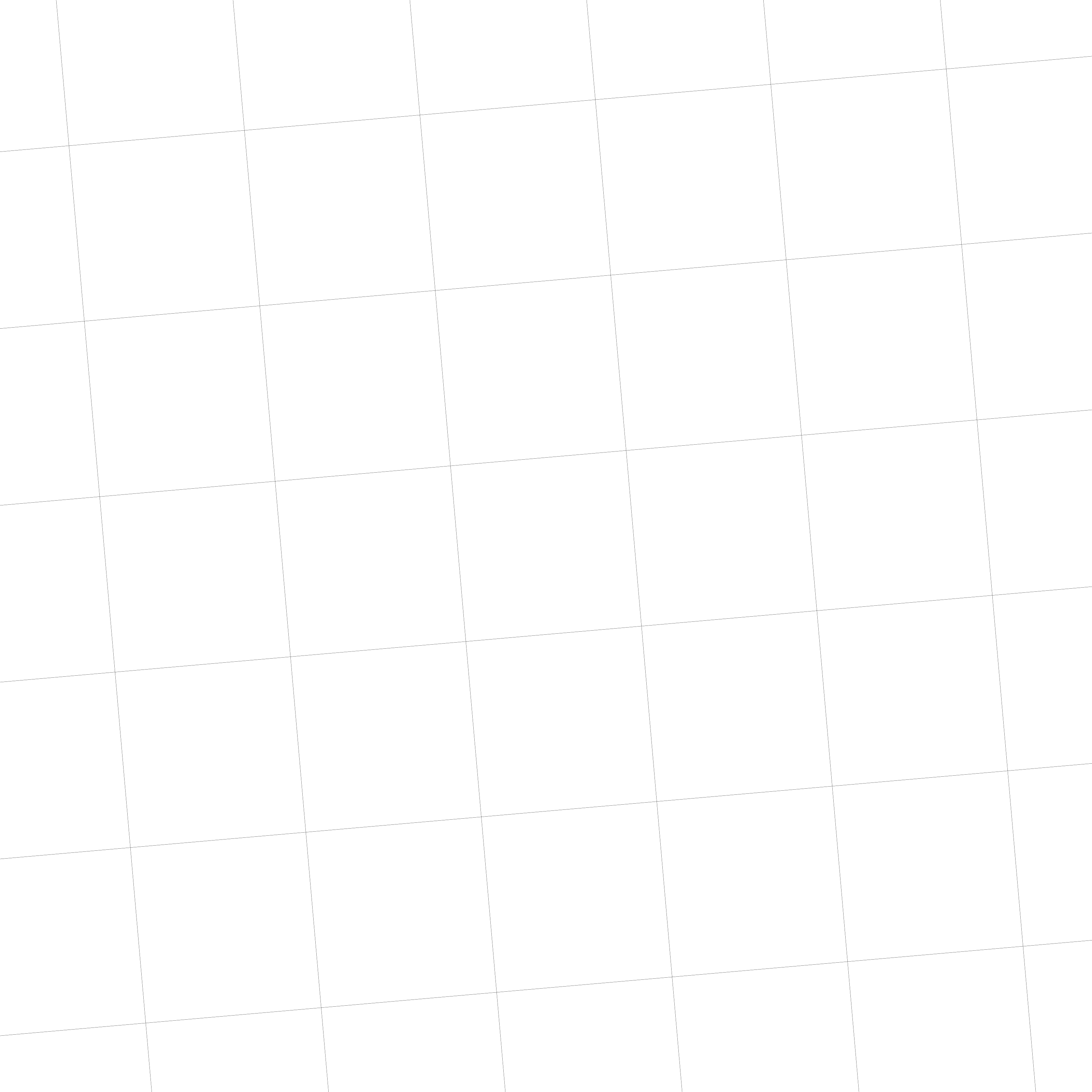}} &
        \colorbox{teal!5}{\includegraphics[width=0.22\linewidth]{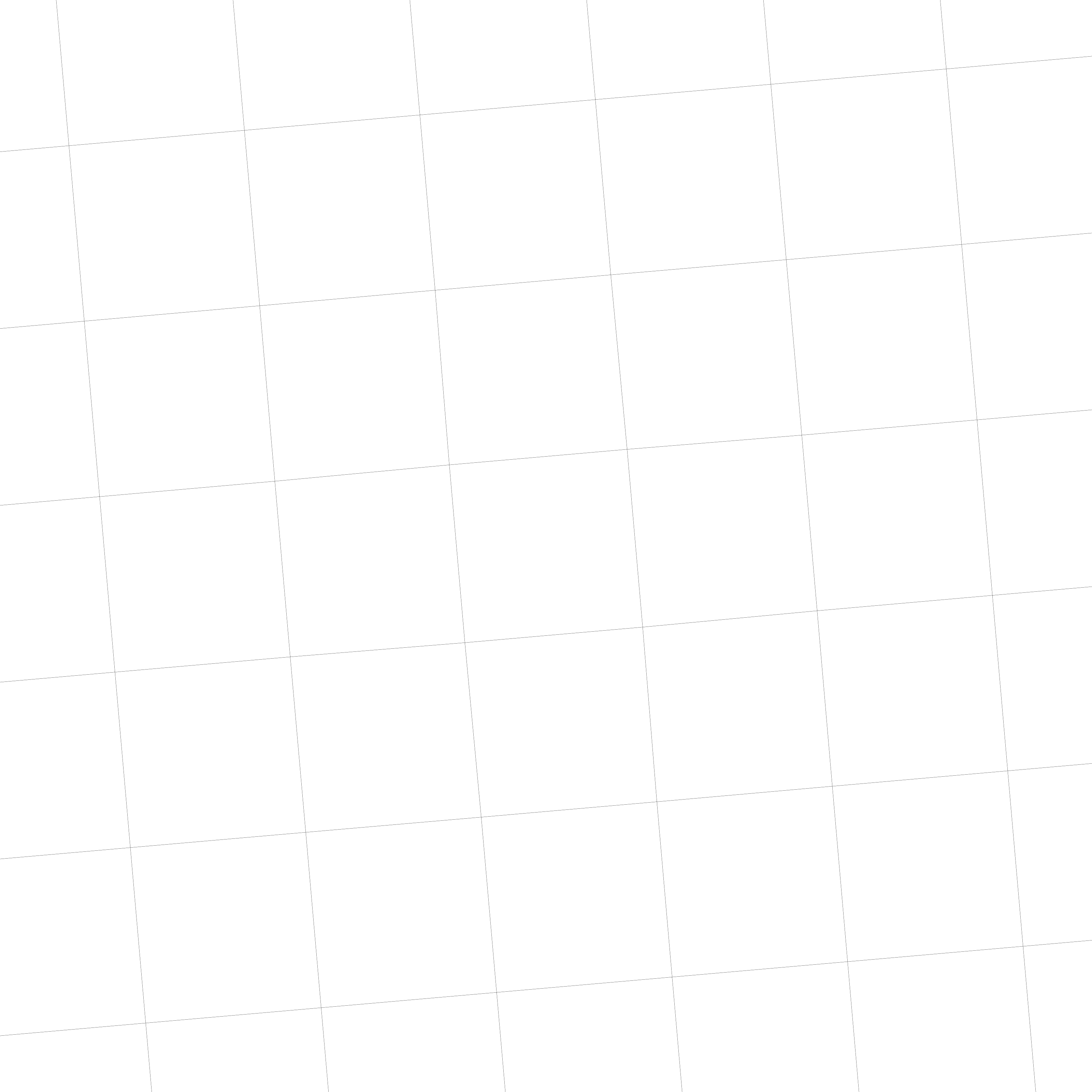}} &
        \colorbox{teal!5}{\includegraphics[width=0.22\linewidth]{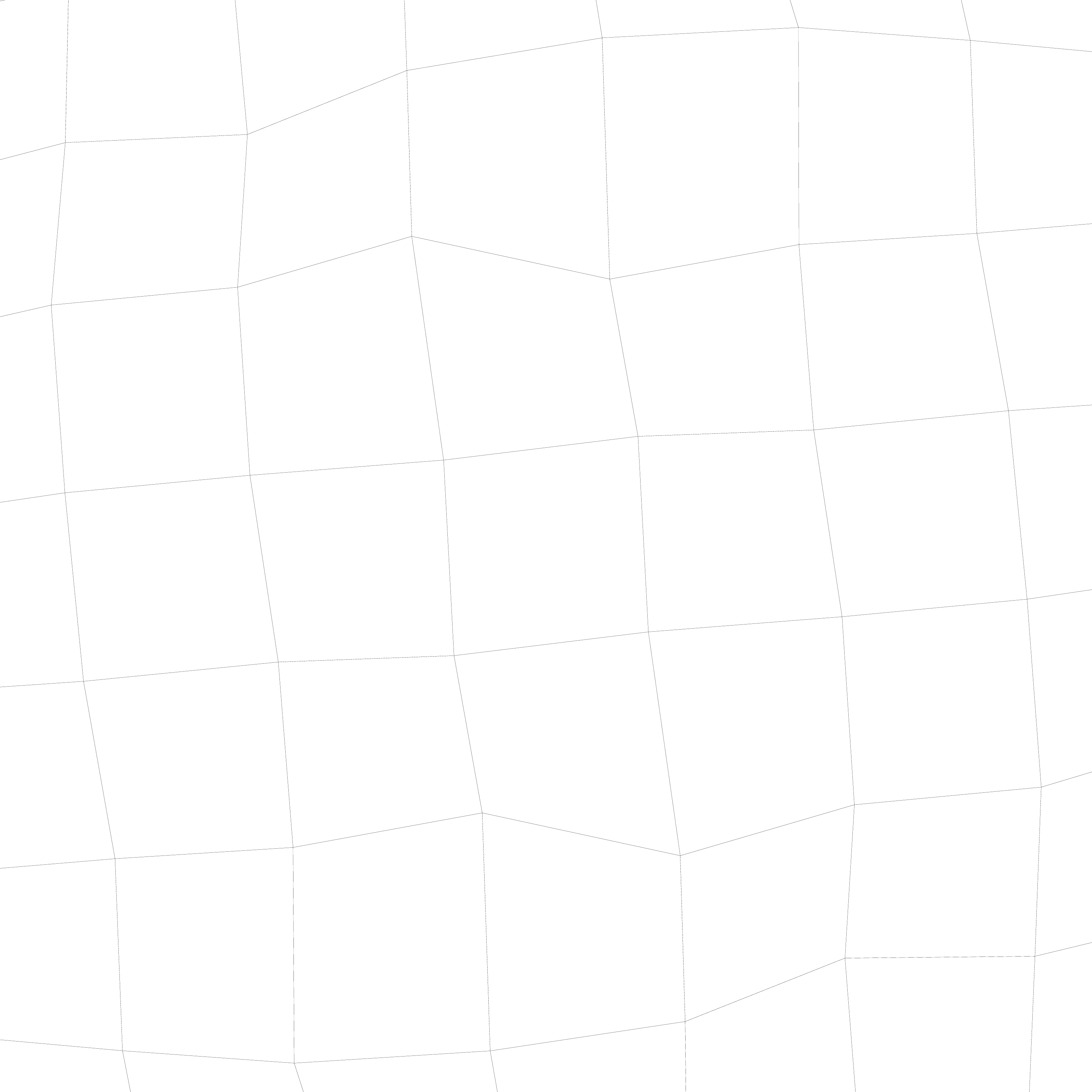}} &
        \colorbox{teal!5}{\includegraphics[width=0.22\linewidth]{assets/spin_test_zoom}} \\

        \put(-0.155\linewidth, 0){\scriptsize Loss:} $1.01$ & $1.56$ & $1.063$ & $0.882 $ \\
        \put(-0.12\linewidth, 0){\scriptsize Step Size:} $\num{1e-3}$ & $\num{1e-3}$ & $\num{5e-2}$ & $\num{1e-3}$ \\
        \put(-0.14\linewidth, 0){\scriptsize Iterations:} 10000 & 10000 & 20000 & 10000 \\
    \end{tabular}
    \caption{Comparison of optimizing vertices as the convex sum of their neighbors, leading to lower loss than optimizing direct deformation of vertices. Vector Adam~\citep{vector_adam} or Uniform Adam~\citep{large_steps} also appear less effective on this example.}
    \label{fig:direct-deform-cmp}
    \Description{Comparison of 4 different grids which are spun about their center with boundaries pinned. The center of each reigon varies in its straightness and how much of the grid is correctly placed.}
\end{figure}

\subsubsection*{Comparison of Convex Sum and $x+\delta$}

Finally, we compare the effect of optimizing vertices as the convex sum, versus optimizing them directly as $x+\delta$, both with the alternating optimization scheme, shown in Fig.~\ref{fig:direct-deform-cmp}. Note that we constrain the differential weights using $\text{softplus}(x, \beta \stackrel{\text{default}}{=} 1) = \frac{1}{\beta}\log(1 + \exp(\beta x))$, limiting the range of weights to $(0, \infty)$. We use softplus in all experiments, although there are other alternatives such as the exponential $e^x$.

When comparing to direct deformation, the convex sum is smoother than direct deformation, with a lower final loss. Prior work on optimizing vectors identified that normalization in Adam causes bias in the gradient, so we test two other optimizers~\citep{vector_adam, large_steps} but find they lead to higher loss than representing vertices directly as a differential elements.

\begin{figure*}[!tbh]
    \centering
    \setlength{\tabcolsep}{0pt}
    \begin{tabular}{c c c c c c}
        Mesh & Angle-Preserving & Area-Preserving & Sym. Dirichlet & Equilateral & Equiareal \\
        \put(-0.08\linewidth, 0){\includegraphics[height=0.14\linewidth]{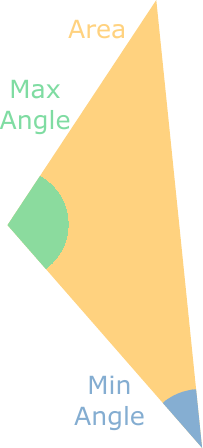}}
        \includegraphics[height=0.14\linewidth]{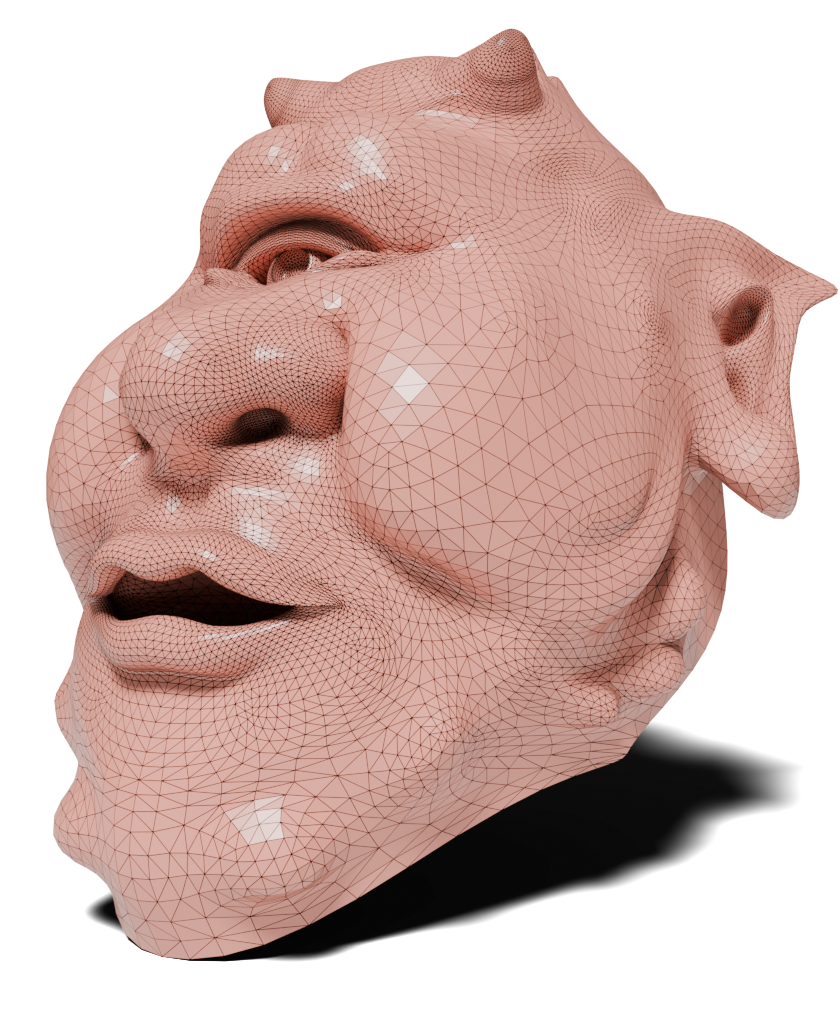} &
        \put(-6pt,0.035\linewidth){\rotatebox{90}{8$\times$ Zoom}}
        \colorbox{teal!8}{\includegraphics[width=0.14\linewidth]{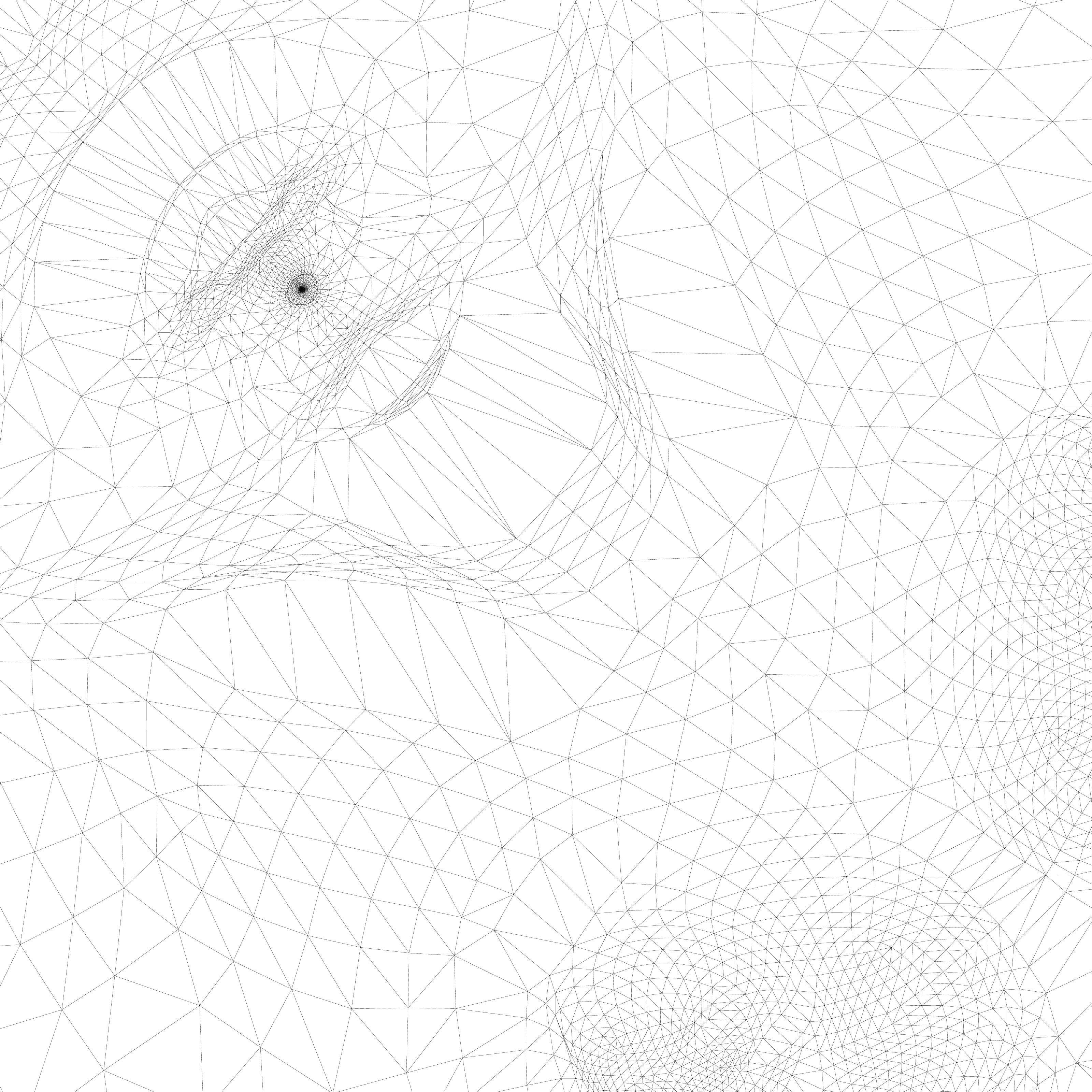}} &
        \colorbox{teal!8}{\includegraphics[width=0.14\linewidth]{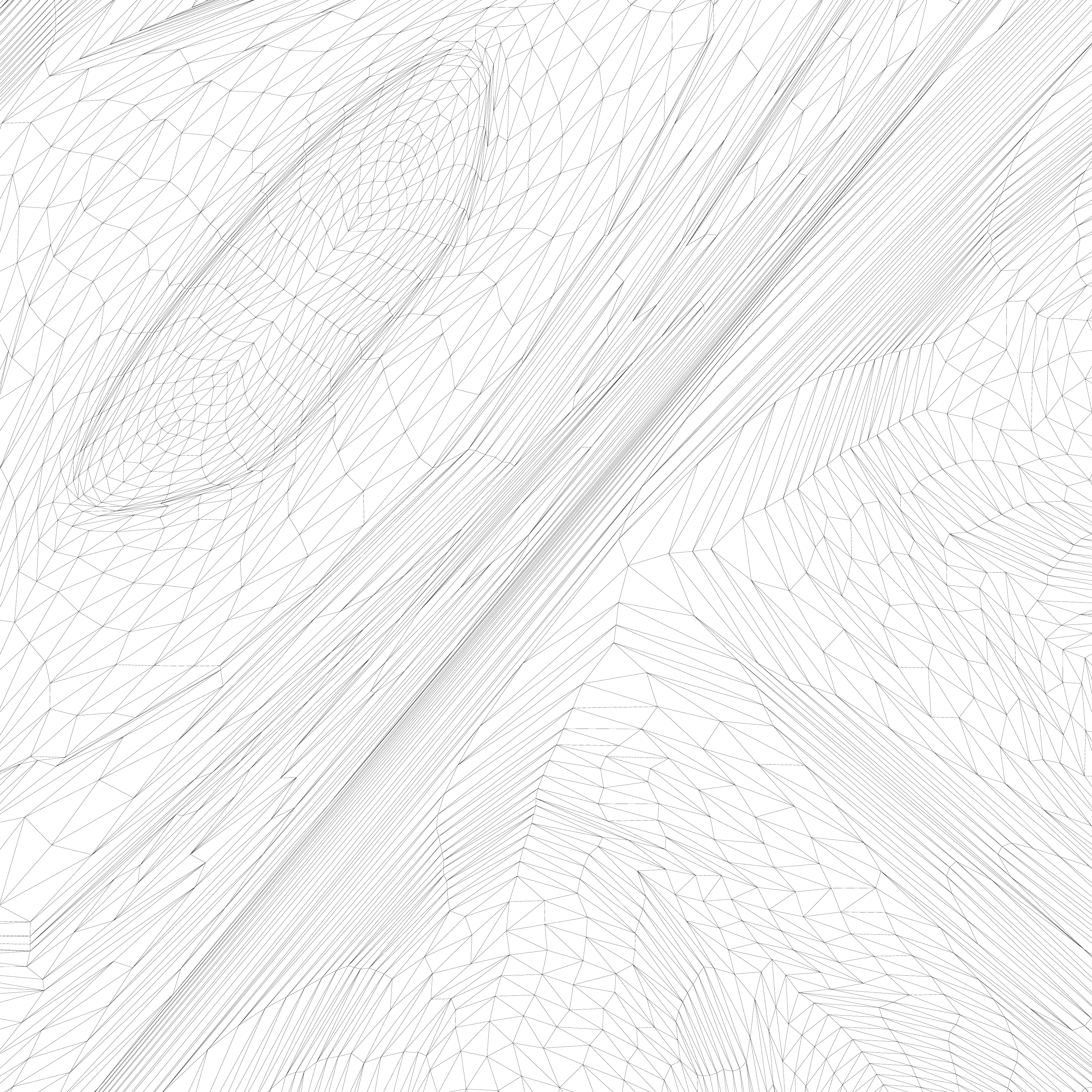}} &
        \colorbox{teal!8}{\includegraphics[width=0.14\linewidth]{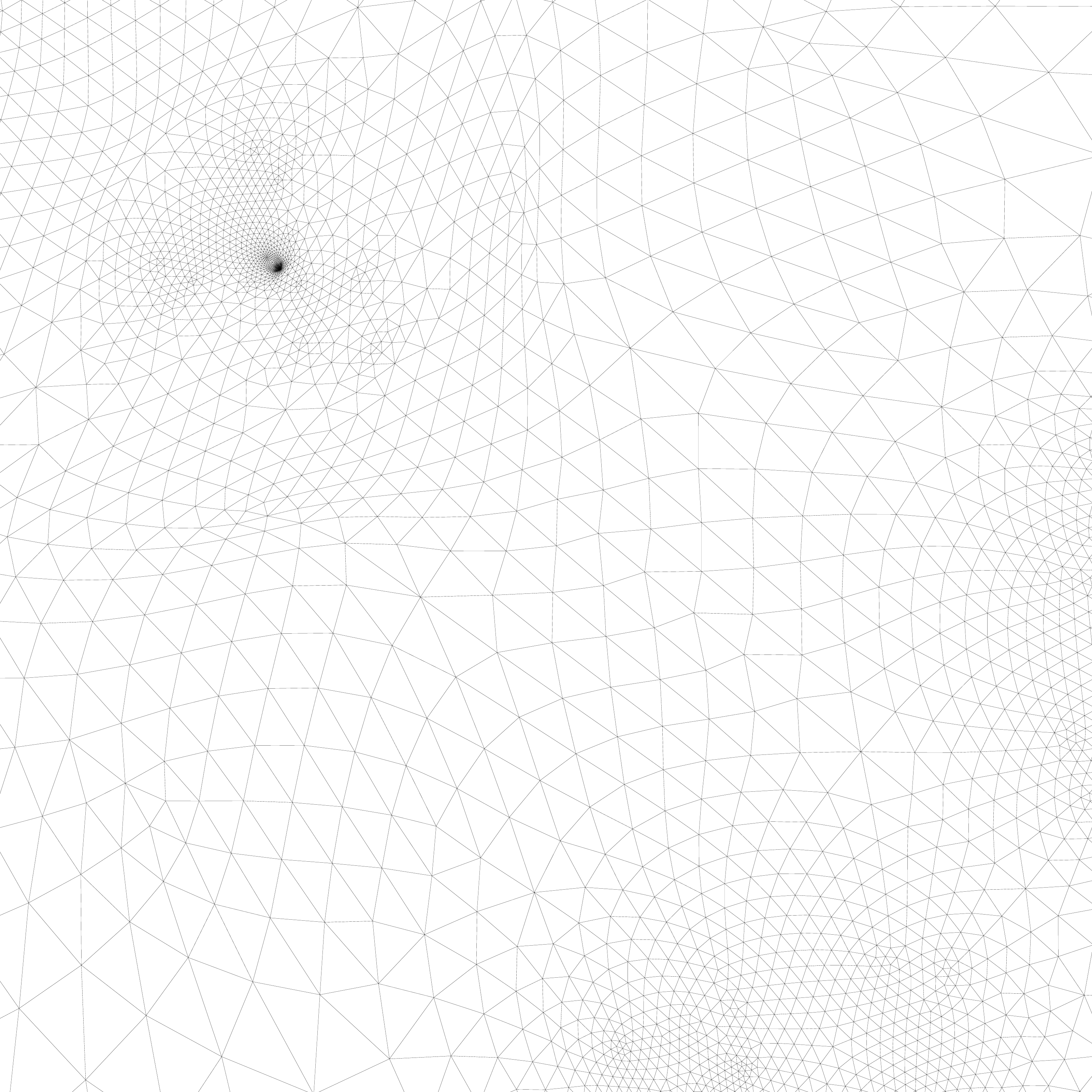}} &
        \colorbox{teal!8}{\includegraphics[width=0.14\linewidth]{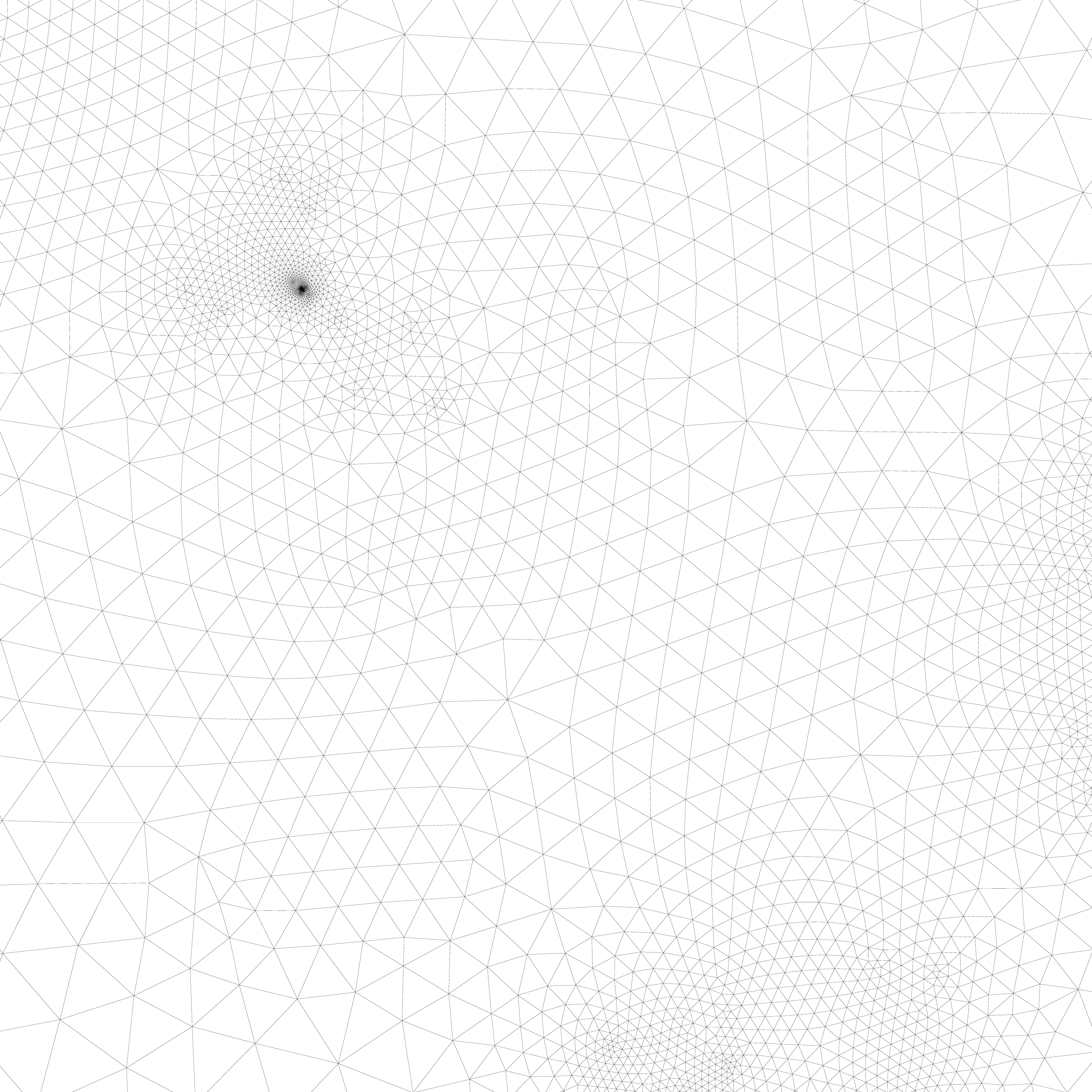}} &
        \colorbox{teal!8}{\includegraphics[width=0.14\linewidth]{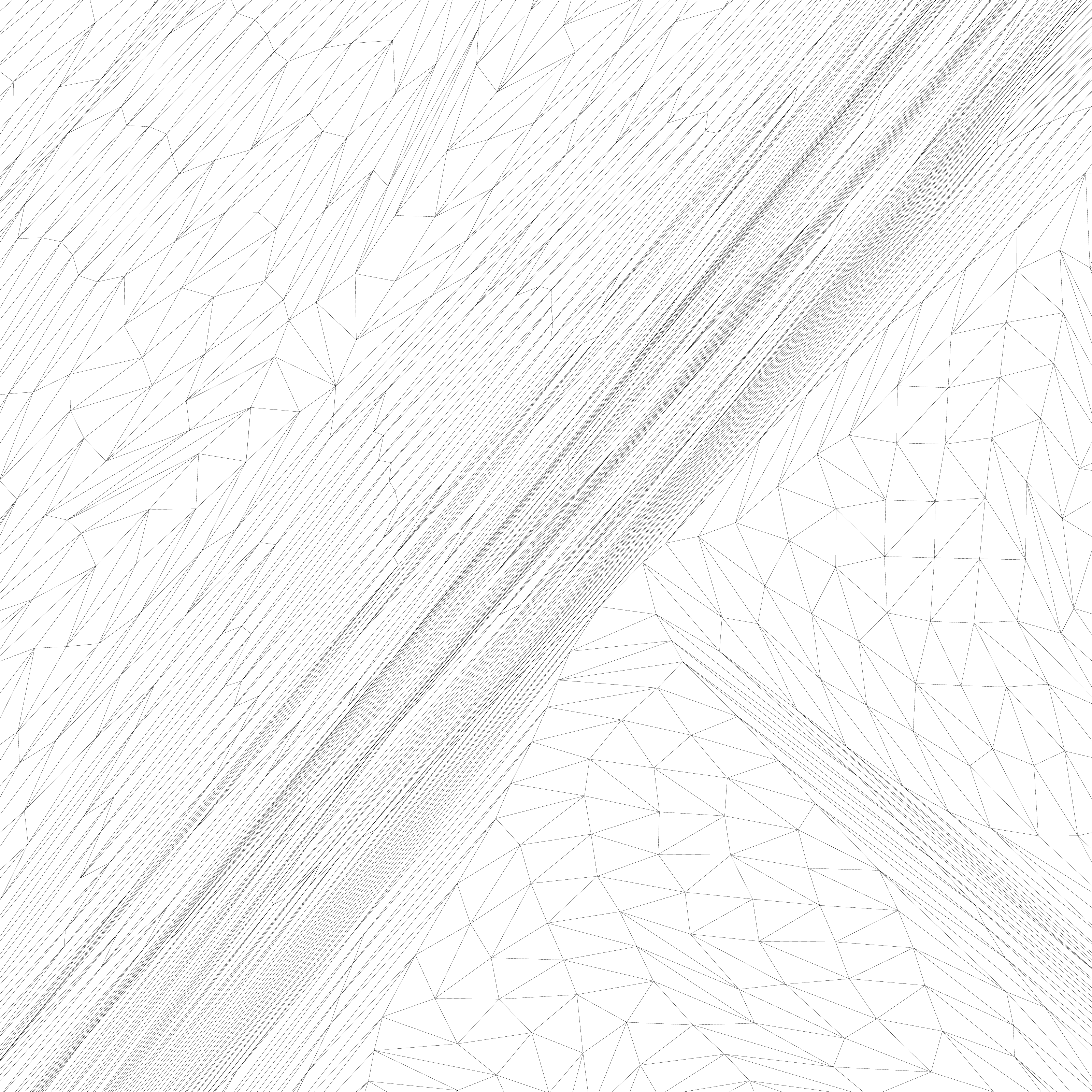}} \\

        \scriptsize 3D Distributions: & \scriptsize 2D Angles Matches 3D & \scriptsize 2D Area Matches 3D & \scriptsize Similar Area \& Angle & \scriptsize Maximize 60$\degree$ & \scriptsize All Areas Equal \\

        \put(-20pt,0.02\linewidth){\rotatebox{90}{Angle Dist.}}
        \put(-10pt,0.035\linewidth){\rotatebox{90}{Count}}
        \includegraphics[width=0.14\linewidth]{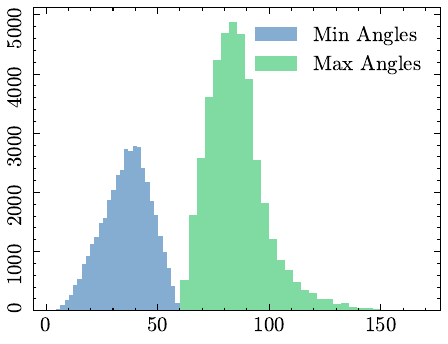} &
        \includegraphics[width=0.14\linewidth]{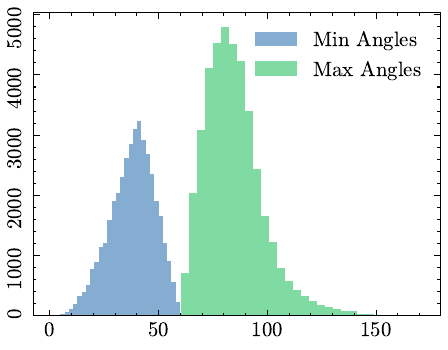} &
        \includegraphics[width=0.14\linewidth]{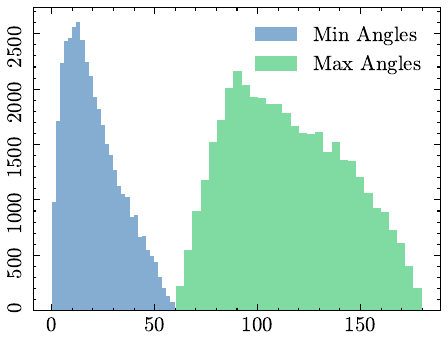} &
        \includegraphics[width=0.14\linewidth]{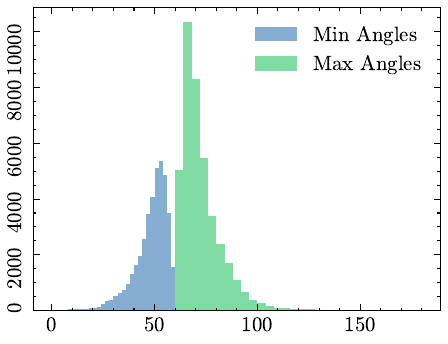} &
        \includegraphics[width=0.14\linewidth]{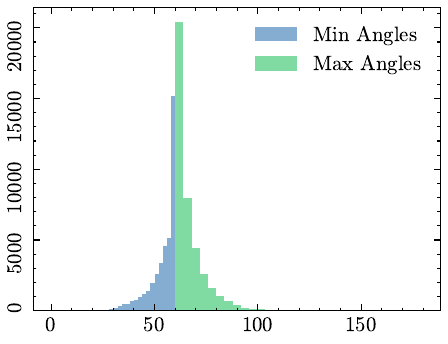} &
        \includegraphics[width=0.14\linewidth]{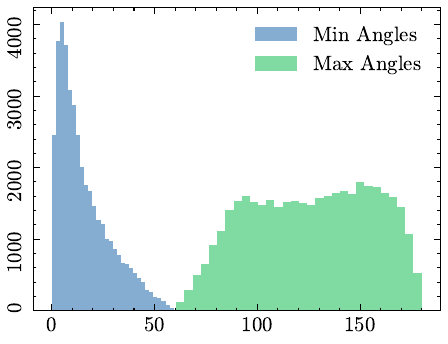} \\
        \multicolumn{6}{c}{\captionsize Angle (degrees), uniform scale} \\

        \put(-20pt,0.02\linewidth){\rotatebox{90}{Area Dist.}}
        \put(-10pt,0.032\linewidth){\rotatebox{90}{Count}}
        \includegraphics[width=0.14\linewidth]{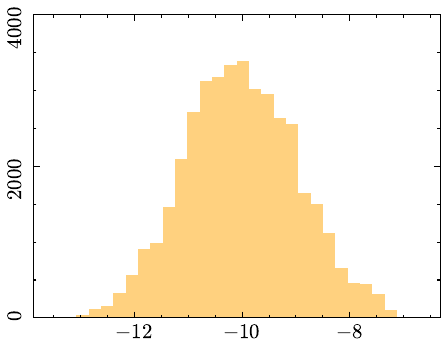} &
        \includegraphics[width=0.14\linewidth]{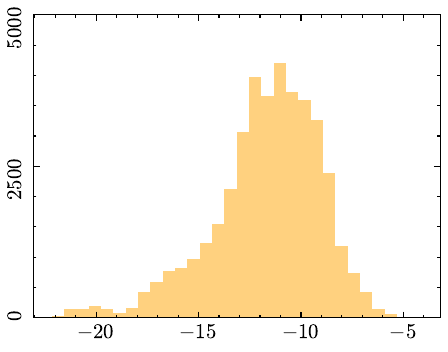} &
        \includegraphics[width=0.14\linewidth]{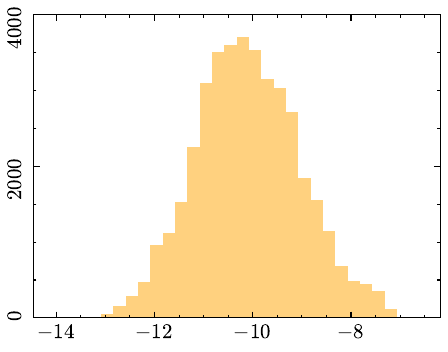} &
        \includegraphics[width=0.14\linewidth]{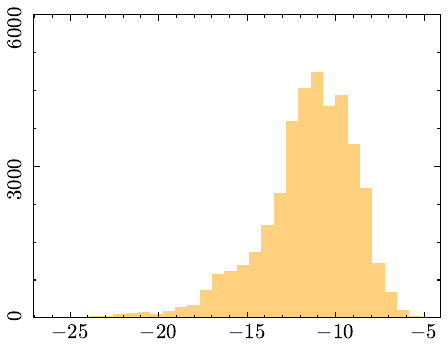} &
        \includegraphics[width=0.14\linewidth]{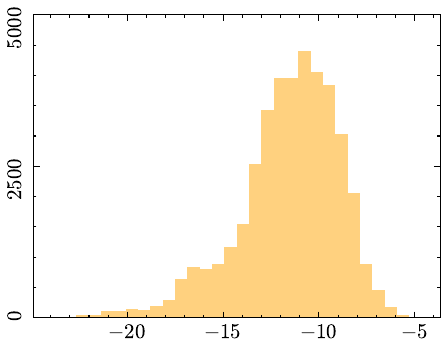} &
        \includegraphics[width=0.14\linewidth]{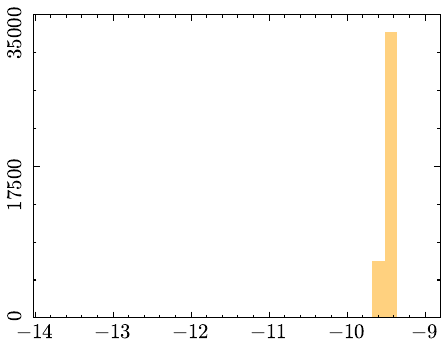} \\
        \multicolumn{6}{c}{\captionsize Log triangle area (distance$^2$), scale varies per plot} \\

    \end{tabular}
    \caption{Comparison of optimized UV Tutte parameterization with different energies. The distribution of angles and areas is shown for the 3D mesh on the left. In each case, there are no inversions, as guaranteed by the alternating optimization approach. The non-zoomed versions are shown in the supplement, as the mesh is quite dense. The angle-preserving energy has a distribution of angles similar to the 3D distribution. The area-preserving energy has a distribution of 2D triangle areas close to the 3D area distribution. The area-weighted Symmetric Dirichlet energy is a balance between area and angle similarity. Equilateral energy leads to angles near $60\degree$, and the equiareal energy leads to most triangles with near equal area. \cczero Ogre (Keenan Crane).}
    \label{fig:tutte-optimization}
    \Description{
    For a single ogre mesh, zooms of different UV parameterizations with optimization with a number of different energies. In the first row, top-left the input mesh is shown. For the rest of the columns, 8x zoom of the UV parameterization is shown. For the second row, there are two distributions shown, one of the max angle in each triangle, and one of the min angle. In the third row, just the distribution of areas is shown.
    }
\end{figure*}

\subsection{UV Optimization}

To start, we demonstrate grid optimization on arbitrary 2D UV meshes with alternating optimization and convex sums. For a single mesh which is parameterized with a convex boundary using a Tutte parameterization~\citep{tutte}, we proceed to optimize a number of different energies, and show inversion-free results for each. This is compared with optimization by directly deforming vertices as $x' = x + \delta$ (both with alternating optimization). The mesh is initialized to a Tutte parameterization with uniform weights and boundary vertices pinned to the unit circle. Then, each vertex is optimized as a convex combination of its neighbors, and inversion is prevented using the alternating optimization approach outlined. Vertices on the boundary can move, but are reprojected onto the unit circle after each step. We then optimize 5 different energies, to demonstrate the broad range of convex combinations and alternating optimizations. Specifically, we optimize the following energies:
\begin{align*}
    \ell_\text{angle-pres.} &= \mathbb{E}(|\angle_\text{3D} - \angle_\text{2D}|) \\
    \ell_\text{area-pres.} &= \text{Var}(\frac{\text{area}_\text{2d}}{\text{area}_\text{3d}}) \\
    \ell_\text{sym-dirich.} &= \mathbb{E}(\text{area}_\text{3d}\cdot(1 +\sigma_1^{-2}\sigma_2^{-2})(\sigma_1^2 + \sigma_2^2)) \\
    \ell_\text{equilateral} &= \mathbb{E}(|\angle_{2D} - 60\degree|) \\
    \ell_\text{equiareal} &= \text{Var}(\text{area}_\text{2D})
\end{align*}
where $\angle_\text{3D}, \angle_\text{2D}$ correspond to the angle in 3D and 2D, and $\sigma_1,\sigma_2$ corresponds to the singular values of the parameterization. The first three energies intend to optimize either angles or area of the parameterization to match the 3D mesh's structure, whereas the last two energies are intended to make the 2D parameterization uniform by making all triangles near equilateral or making all triangles have equal area. The results of optimization on a single mesh are shown in Fig.~\ref{fig:tutte-optimization}. Some of the energies such as the area-preserving or equiareal energy produce extremely stretched triangles, but no inversions are introduced.

Results vary per energy, but by comparing input and output distributions of angles and areas, the results are close to what is expected. For each energy, we can see that the qualities of the input triangles are preserved well. For the conformal angle-preserving energy, the distribution of minimum and maximum angles are similar to the 3D distribution. For the area-preserving metric, the distribution of triangle areas in 2D is close to the distribution of areas in 3D. The area-weighted Symmetric Dirichlet energy is a balance between these two. The equilateral triangle angles are closely centered around $60\degree$, and the equiareal triangle areas are heavily centered around a single area. The optimization approach outlined performs better on most energies, as shown in Tab.~\ref{tab:uv-opt-cmp}. The one exception is the Symmetric Dirichlet energy, which may already provide sufficient regularization during direct deformation that it does not benefit from the convex representation. These results generally show that representing vertices differentially can smooth optimization, regardless of the target energy. For completeness, the distribution of area and angles for direct deformation are shown in the supplement.

\begin{table}
    \centering
    \begin{tabular}{|c|c|c|}
        \hline
        Energy$\downarrow$ & Direct Deform & Convex Sum \\\hline
        $\ell_\text{angle-pres.}$ & 0.126 & 0.0741 \\\hline
        $\ell_\text{area-pres.}$ & 0.0223 & 0.00550 \\\hline
        $\ell_\text{sym-dirich.}$ & $\num{1.58e-4}$ & $\num{1.69e-4}$ \\\hline
        $\ell_\text{equilateral}$ & 0.118 & 0.0676 \\\hline
        $\ell_\text{equiareal}$ & 0.0253 & 0.00174 \\\hline
    \end{tabular}
    \caption{Comparison results for UV optimization of a single on five different energies (both with alternating optimization). For all energies other than Symmetric Dirichlet~\citep{bijective}, vertices represented as a convex sum optimize more smoothly than direct deformation.}
    \label{tab:uv-opt-cmp}
    \vspace{-3em}
\end{table}

\paragraph{Implementation Details for Mesh Vertex Coloring} While the results of optimization with these techniques have good quality, the efficiency depends heavily on the coloring of the UV parameterization. Optimally, when the input is a planar graph, a 4-coloring can be used to color the UV parameterization~\citep{four_color_theorem}, with complexity of finding the coloring quadratic in the number of vertices~\citep{efficiently_four_coloring_planar_graphs}. For simplicity, we use a greedy coloring, which is at most $\max(\text{degree})+1$. For the mesh tested, the greedy coloring is a 6-coloring. Each additional color reduces the efficiency of optimization, requiring $\#\text{iters} \times\#{\text{colors}}$ to optimize each vertex with an equal amount of steps. For grids, this is less of a concern as a 2-coloring is trivial. For planar meshes, the quadratic algorithm may be too costly for large meshes. Furthermore, for meshes embedded in arbitrary dimensions including tetrahedral meshes or hex meshes, an arbitrary number of colors may be required.

\begin{figure*}[ht]
    \centering
    \begin{tabular}{c c}
        Input Image ($768\times512$) & Pixel Grid Deformation ($384\times256$) \\
        \includegraphics[width=0.4\linewidth]{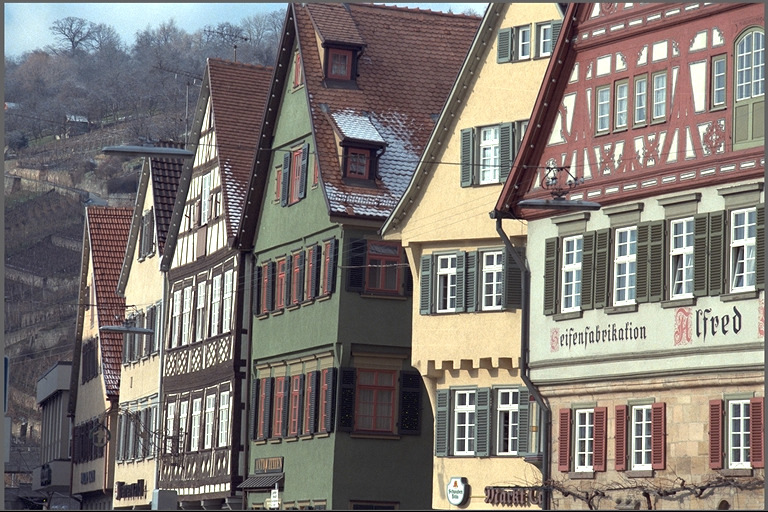} &
        \includegraphics[width=0.4\linewidth]{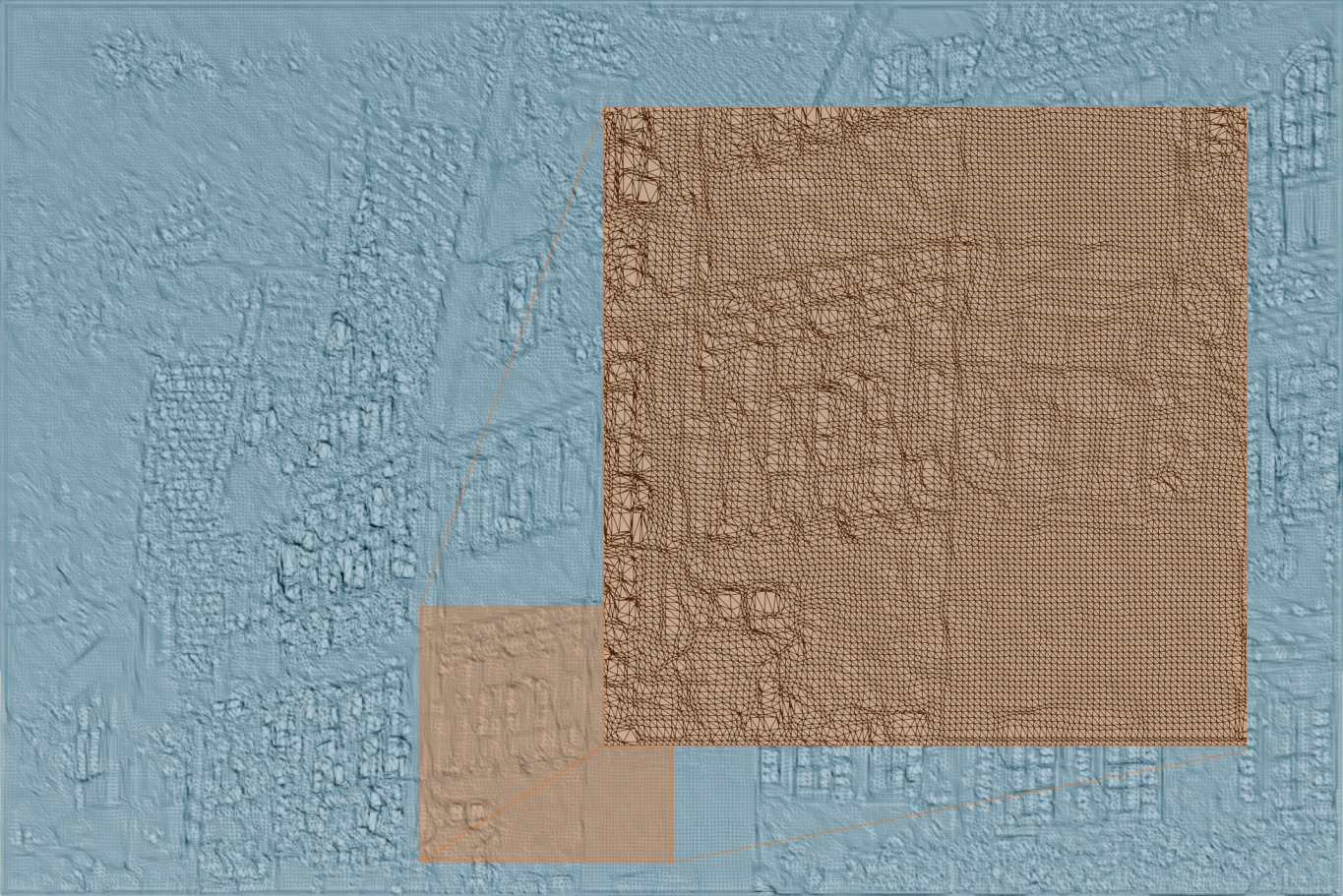} \\
        Optimized Colors ($384\times256$) & Reconstructed Image ($768\times512$) \\
        \includegraphics[width=0.4\linewidth]{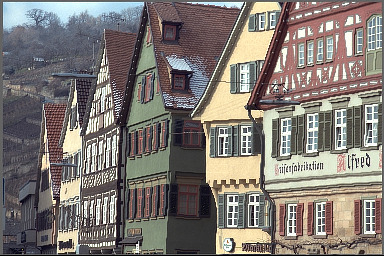} &
        \includegraphics[width=0.4\linewidth]{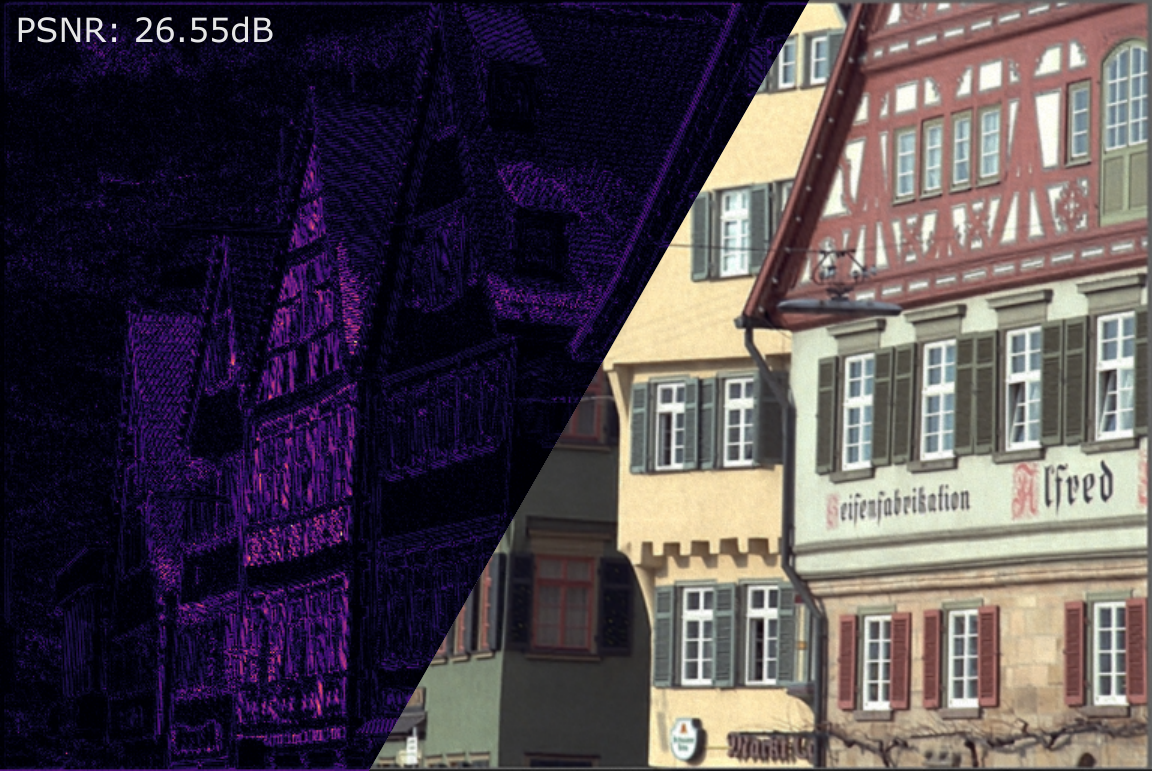} \\
    \end{tabular}
    \caption{Pipeline for image compaction with grid deformation. Given an input high resolution image, a deformed grid with colors per vertex is optimized. Then, a high-resolution image can be reconstructed by resampling each grid cell with its corresponding colors, improving the correspondence with the original compared to direct downsampling. \cczero Kodak \#8 (Alfons Rudolph), triangulated in rendering.}
    \label{fig:image-compaction-example}
    \Description{
    First column, a picture of german style buildings, second a blue image of deformations, third column is colors, which looks quite similar to the input but is more pixelated, final column is reconstructed image with a split showing areas of high error.
    }
\end{figure*}

\subsection{Image Compaction}

We also test grid optimization by compacting images from a high resolution grid to a lower resolution with deformed pixel shapes to align with image features. Pixels usually correspond to evenly spaced, square sample regions in an image, but instead we allow pixels to deform to arbitrary shapes, and evaluate the effectiveness of deformed pixels at capturing higher resolution. An example of this is shown in Fig.~\ref{fig:image-compaction-example}.

\paragraph{Compaction versus Compression}
We focus on \textit{compaction} of an image, by shrinking the amount of parameters used in an image but still capturing high-frequency detail, similar image-deformation in ~\citep{feature_aware_texturing, optimized_scale_and_stretch}. This is in contrast to image \textit{compression}, which uses any technique to reduce the total amount of memory used. While compaction can be used for compression, many other tools are used in compression such as quantization, lossy representations, and removing samples in the frequency-domain. These steps can be orthogonal to compaction, and its possible they can be used concurrently, but we do not explore that in this work.

\paragraph{Method} A deformable image is defined as a rectangular grid of size $H\times W$, with one color defined per vertex $C\in\mathbb{R}^{H\times W\times3}$, as well as a deformed position $\delta\in\mathbb{R}^{H\times W\times2}, \delta\in[-1,1]$, optimized using the approach outlined. Then, at each point in $[-1,1]$, the color is defined as an interpolated value of the enclosing vertices: $c(x,y) = \sum_{i,j\in\square} w_{i,j}(x,y) C[i,j]$. To get the weights $w_{ij}$, we sample barycentric coordinates within each grid cell. Our current implementation samples triangulated grid cells, giving a zero-padded barycentric coordinate in one of the two subdivided triangles. By stochastically sampling values, and using a reconstruction loss $\ell(x,y) = \lVert c_\text{ref}(x,y) - c_\text{curr}(x,y)\rVert_1$ the compacted image is optimized to match the high-resolution image.

\paragraph{Barrier Energy} To prevent grid cells from inverting, we compute the signed area of the 4 possible triangles within each grid cell, and compute the IPC energy~\citep{ipc} with a distance of $\frac{\num{1e-10}}{\sqrt{H\times W}}$. We additionally apply a linear term at $2\times\frac{\num{1e-10}}{\sqrt{H\times W}}$ to push triangles further away from degeneracy.

\paragraph{Evaluation} To evaluate the efficiency of compaction, we test our approach on the 24 images from the Kodak Image dataset~\citep{kodak_dataset}. For each image in the dataset, we deform a grid of half the resolution of the input image to match the ground truth. We compare deformation to bilinear downsampling and upsampling the input. Compared to bilinear downsampling and upsampling without deformation, pixel grid deformation increases quality by 5 dB PSNR on both the mean and median of the dataset. We also compare to JPEG compression and decompression, but note that the approaches are fundamentally different, as JPEG takes a limited number of coefficients in the frequency domain, and the two approaches can be combined. We find that grid deformation at half the input resolution is comparable to JPEG compaction at 50\% quality, with $32.36$ dB PSNR for deformed images and $32.1$ for JPEG at 50\%, and median at $33.14$ versus $32.61$. Full results are shown in Tab.~\ref{tab:image-compact-compare}.

\begin{figure}
    \centering
    \setlength{\tabcolsep}{\arraystretch pt}
    \begin{tabular}{c c}
        \multicolumn{2}{c}{Coarse-to-Fine Blurring Ablation} \\
        Without & With \\
        \includegraphics[width=0.42\linewidth]{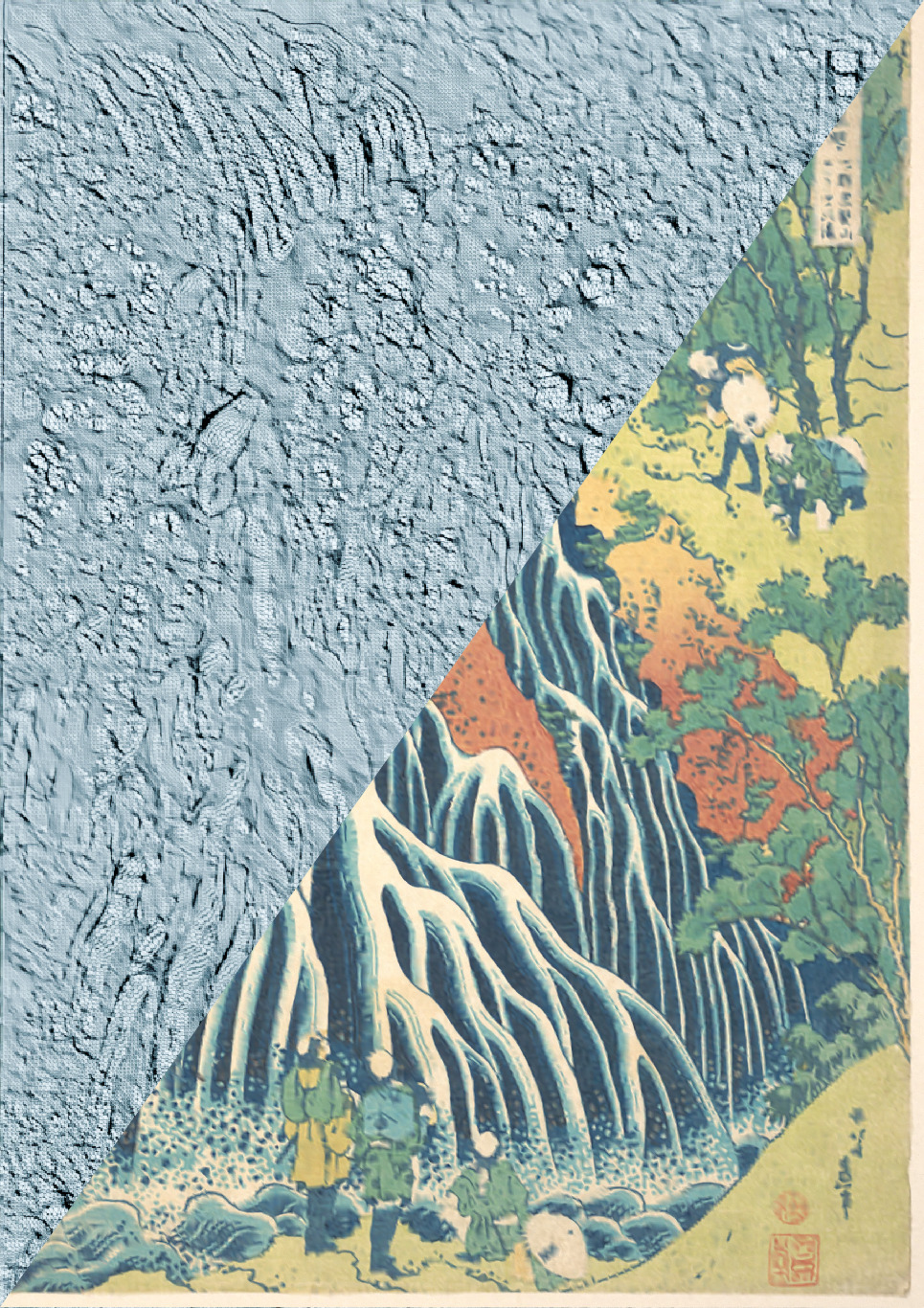} &
        \includegraphics[width=0.42\linewidth]{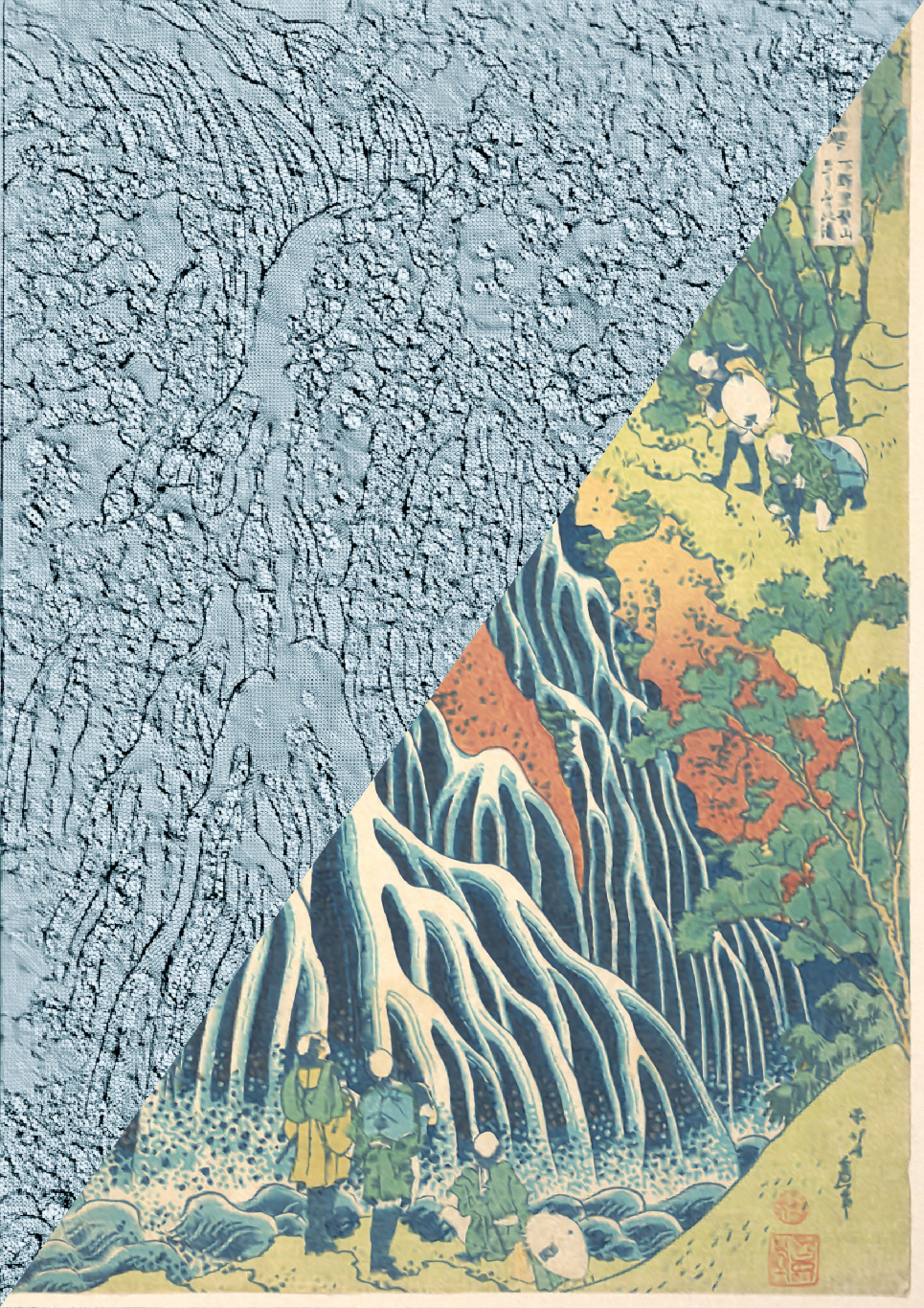} \\
        %$H\times W$: $3844 \times 2716$ &
        $384\times271$ & $384\times 271$ \\
        % PSNR$^\uparrow$ / MS-SSIM$^\uparrow$: &
        23.17 dB / 0.7938 & 24.29 dB / 0.8292 \\
    \end{tabular}
    \caption{By preconditioning the target image with a gaussian blur, it is easier for grid vertices to deform, thus improving the reconstruction quality (original image is $3844 \times 2716$). Please zoom in. \cczero Katsushika Hokusai (Kirifuri Waterfall at Kurokami Mountain, The Metropolitan Art Museum).}
    \label{fig:image-blur-cmp}
    \Description{Japanese Woodblock print of a waterfall, with people standing at the bottom and the top-right. The two right columns are split showing the deformation. The middle column has much stronger deformation than the right column.}
\end{figure}

\paragraph{Coarse-To-Fine Blur Preconditioning}
During optimization, the color within each grid cell is independent of other grid cells, making gradients sparse with respect to vertices of adjacent cells. To more smoothly distribute the influence of samples, we precondition optimization by Gaussian blurring the target image, and slowly decay the amount of blur until the original image is recovered. This allows for distant vertices to be moved towards regions with high loss, as shown in Fig.~\ref{fig:image-blur-cmp}. The target grid is deformed more heavily when using blurring, as shown by the darker regions where the wireframe is more clumped, and leads to improvement of both PSNR and MS-SSIM.

\paragraph{Representation \& Pixel Conversion} After the image has been compacted, it can be represented as a 5-channel image $\mathbb{R}^{H\times W\times (3 + 2)}$, with the first 3 channels corresponding to RGB, and the last 2 corresponding to deformation. For this work, we do not look into compressing this tensor like JPEG, which may be done using Fourier coefficients, and note that quantizing deformation may require a different approach than quantizing colors, since incorrect deformation values may introduce more visible error.

To recover an image at the original resolution with square pixels, we take a naive approach of sampling each pixel of the low-resolution image, and making each high-resolution pixel the average of all low-resolutions samples contained in the pixel. This is not suitable for real-time viewing since it introduces sampling noise and requires many samples. Instead, alternatives such as explicitly computing the intersection between low-resolution deformed pixels and high-resolution uniform pixels may work to get around those two issues. We leave it to future work to explore real-time implementations of image compaction.

\subsection{Differentiable Rendering for Isosurface Extraction}
Another application of grid deformation is for isosurface extraction in differentiable rendering~\citep{nvdiffrec, tetweave, dmtet, flexicubes}. Most recently, \cite{tetweave} use a Delauney tetrahedralization of sampled points near a surface to effectively capture high-frequency detail, which is not possible with uniform grids. In this work though, we show that uniform grids with adaptivity can match the performance of this alternative in efficacy and time, albeit still using far more parameters.

Specifically, we show that deformed grids are capable of capturing sharp mechanical features and spindly/thin organic features, which are difficult to recover in \citep{tetweave} since sampling may miss small structures.

\paragraph{Implementation Details}
While the key ingredient of this work is the fully-adaptive grid, we found many low-level details heavily impact performance of isosurface extraction. To make it easy for follow-up work, we explain a number of small design choices which heavily impact the result.

\paragraph{Isosurface Extraction Algorithm} Even though this work maintains an injective grid, Marching Cubes~\citep{marching_cubes} does not guarantee the output mesh is intersection-free when the input grid cells are non-convex (an example is shown in the supplement). This is true even when using previous semi-adaptive grids~\citep{flexicubes} where deformation is limited to half a grid cell. To prevent self-intersections in isosurface extraction, we split each cube into 5 tetrahedron, and use differentiable marching tetrahedra~\citep{dmtet}. Since each tetrahedron has positive signed volume, the output is guaranteed to be watertight, manifold, and intersection-free.

\paragraph{Barrier Energy} To prevent grid-inversions, we compute the IPC barrier energy~\citep{ipc}. To do so on a regular 3D grid, we tetrahedralize each grid cell, in all possible combinations: when splitting along a fixed axis, it is possible to split a cube into 5 tetrahedrons, and we split along 2 alternative axes, giving 10 possible tetrahedrons. The signed volume of each tetrahedron with vertices $v_0, v_1, v_2, v_3$ is then $\frac{1}{6}(v_1 - v_0) \cdot ((v_2 - v_0) \times (v_3 - v_0))$. To allow for smooth optimization, we also add in a linear term in addition to IPC that pushes each tetrahedron to have $2\times$ the area before the IPC energy applies. Since this step is computed for $10 N^3$ tetrahedra, it has a large performance cost.

\paragraph{Warm-Up \& Cool-Down} Before deforming the grid, we first optimize without deformation for $3250$ iterations Then, for the last $1000$ iterations, we turn off deformation, and only optimize the SDF values, similar to ~\citep{tetweave}.

\paragraph{Removing Floaters} During optimization it is possible that invisible, degenerate elements are in the space inside \textit{and} outside the mesh. To remove elements outside we found it sufficient to slightly push each vertex in the direction of its normal: $v_\text{puffed}' = v + \epsilon\cdot\text{normal}(v), \epsilon\stackrel{\text{default}}{=}\num{1e-3}$. This prevents invisible external floaters which impair AABB refitting. We slowly decay $\epsilon$ to 0 over the first $\frac{2}{15}$ iterations of grid deformation.

For floaters inside the mesh which cannot be seen, we use multiple steps. First, we attempt to prevent their construction by initializing the input as a valid cube SDF. We found that explicit SDF regularization used by prior work is insufficient, and initializing as a valid SDF does not entirely prevent floaters. To further remove floaters, we post-process the output by identifying visible connected components from 48 viewpoints. Then, all triangles which are not in any visible component are discarded.

\paragraph{Image-Level Preconditioning} When modifying the embedding of a grid, it is difficult to use coarse-to-fine optimization approaches since changing grid resolution is not straightforward. Instead, we use two kinds of preconditioning at the \textit{image} level, while getting some benefits of hierarchical optimization. These same techniques can likely be applied to other isosurface and differentiable rendering techniques.

$\bullet$ Gradual Resolution Increase: To speed up optimization, we start optimizing from a low-resolution $256\times256$ image, and linearly increase to the full resolution $2048\times2048$ image, over the first $90\%$ of optimization. Early iterations optimize low-resolution features, and gradually high resolution features are optimized, motivating this gradual upscaling. For efficiency, we change the image resolution at every pixel multiple of 8 to reuse tensor allocations, and flush the GPU cache every 500 iterations.

$\bullet$ Coarse-to-Fine Blurring: One shortcoming of prior work is difficulty in learning thin features, such as ropes, parts of flora and fauna, and small building details. It is difficult to learn these features because they have minimal effect on loss as they only appear in a few pixels. Prior work introduced preconditioning for neural architectures~\citep{stochastic_preconditioning} for learning high-frequency detail, but as our approach does not sample continuous functions, we use image-level preconditioning by Gaussian blurring both the masked reconstruction \textit{and} ground truth~\citep{normalized_and_differential_convolution}. This blurring acts as a preconditioning by making the loss affect a larger region and thus easier to learn. Since uniform Gaussian blurring's cost scales linearly with resolution, blurring is only performed in the first $4750$ iterations for efficiency.

\paragraph{Loss Function} In our implementation, we use only the masked $\ell_1$ depth as a loss function, in contrast to prior work~\citep{tetweave, neural_geometry_fields} which combines both a depth loss and a normal loss. We find that the normal loss often introduces noise, whereas depth alone converges smoothly.
Previous implementations also use a masked loss $\frac{1}{\lvert \text{Image}\rvert}\sum_{x\in\text{Mask}} \lVert x_\text{new} - x_\text{gt}\rVert$, where $x$ is each image's pixel. For each image though, the size of the mask varies, despite normalizing by \textit{all} pixels. Thus, even when a mask is only a few pixels, those pixels will be weighted equivalently to views with many pixels. Instead we normalize by the number of pixels in the mask, $\frac{1}{\lvert \text{Mask}\rvert}$. An ablation of both these changes are shown in the Supplement, showing one example each where this change improves performance slightly.

\paragraph{Evaluation \& Dataset} To test our method against static grids and half-cell semi-adaptive approaches, we collect a dataset of 50 meshes from Sketchfab with features including mechanical parts, spindly features, and high-genus topology with 30 of the meshes shown in the supplement.
We test the described approach against TetWeave~\citep{tetweave}, and to do so we consider two factors for a fair comparison: the number of optimization parameters and the runtime. When considering the number of optimizable parameters, TetWeave uses $P$ points in $\mathbb{R}^3$, each optimized with 4 spherical harmonic coefficients and one SDF coefficient, for a total of $8P$. This work uses a differential grid of resolution $R$, with 6 scalar weights for a total of $6R^3$.

At the same time, the number of optimizable parameters does not reflect the ``working set'', or the memory in use at any given time. For this work, the working set is half of the optimizable parameters, as exactly half of the weights are optimized at one time. For TetWeave, the size of the working set depends on how many edges have a sign change which is generally $O(P)$ but can be up to $O(P^2)$. Thus, while it is clear that TetWeave is more parameter efficient, memory usage during optimization depends on the specific Delaunay Tetrahedralization.

When considering TetWeave with 128K points (1024K), using one weight per vertex at $55^3/56^3$ grid resolution (998K/1054K) is an equivalent number of parameter. When considering an equal number of parameters, it is difficult for a uniform grid to match the quality of TetWeave. Therefore, we take into account another criteria of ``fairness'' by considering runtime for both approaches. Experimentally, we observe 64K points and $64^3$ resolution have a similar runtime, and 128K points and $88^3$ resolution are similar. We compare our approach with $64^3$ resolution and 128K points, where our approach has faster runtime but more parameters, and the supplement also compares approaches with similar runtime. We test our approach both with per-vertex weights and per-dimension weights, and find similar quality, but per-vertex weights are faster. Additional results comparing $88^3$ resolution and $64^3$ resolution against 128K/64K points are shown in the supplement.

\paragraph{Results}

To evaluate quality, we measure the Chamfer and Hausdorff distance of the output. This work and TetWeave both output watertight, manifold meshes without self-intersections, as guaranteed by using deep-marching tetrahedra. While not explicitly laid out in its work, TetWeave prevents self-intersection by fixing the positions of tetrahedra in its final Delaunay tetrahedralization, although it would be straightforward to guarantee local injectivity for TetWeave with barrier energies. This work does not introduce inversions at any point.

\begin{figure}[h]
    \centering
    \small
    \setlength{\tabcolsep}{\arraystretch pt}
    \begin{tabular}{c c}
        Input & TetWeave (64K) \\
        \colorbox{teal!3}{\includegraphics[width=0.34\linewidth]{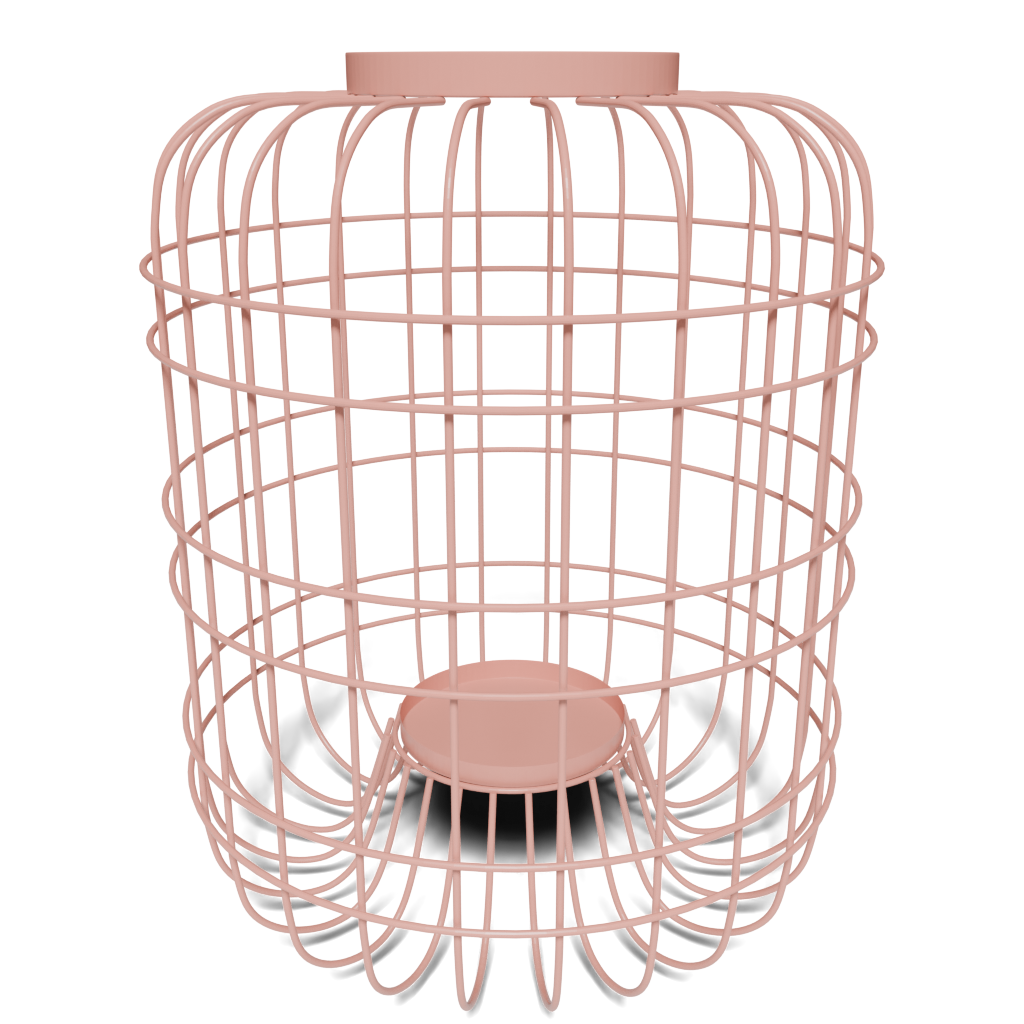}} &
        \colorbox{teal!3}{\includegraphics[width=0.34\linewidth]{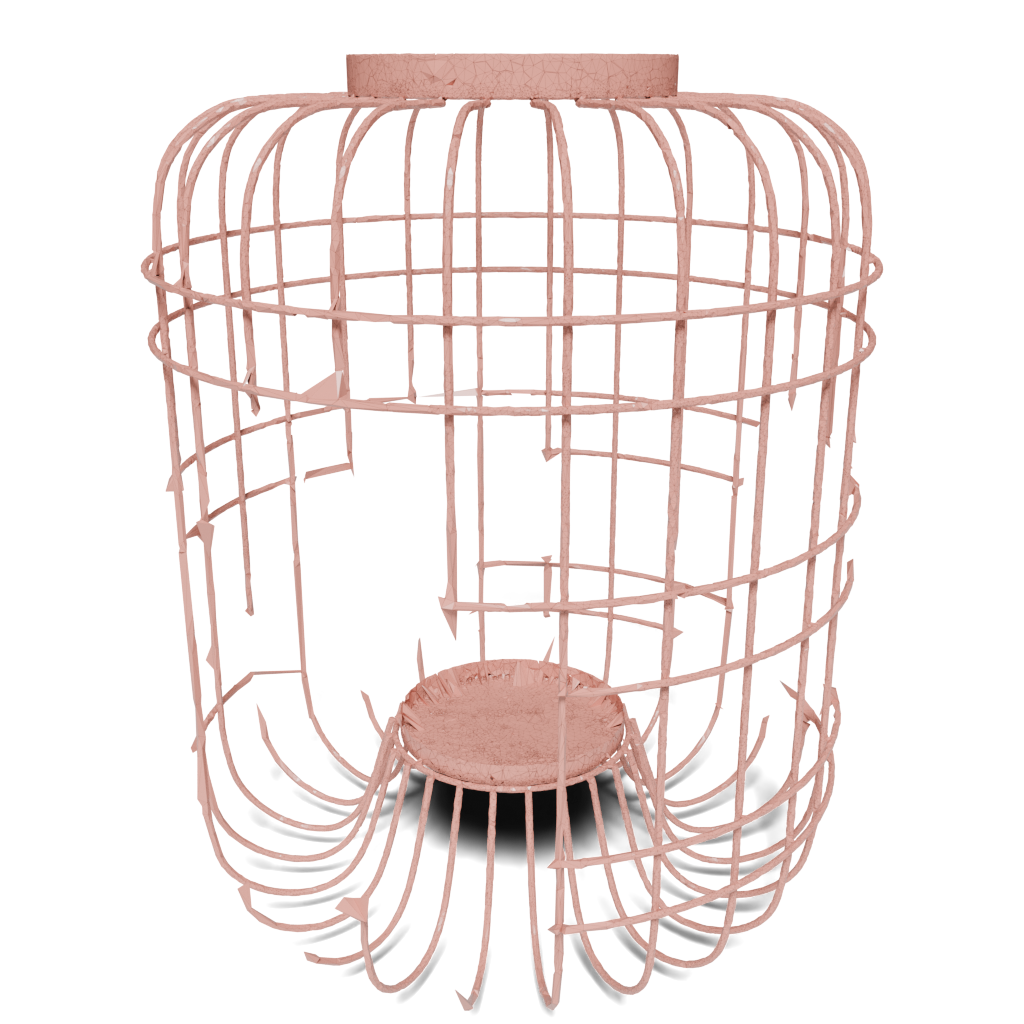}} \\

        Chamfer$^\downarrow$ / Hausdorff$^\downarrow$: & $\num{6.16e-3}$ / $\num{1.25e-1}$ \\
        Parameter Count: & $64\text{K}\times(3 + 1+ 4) = 512\text{K}$ \\
        \colorbox{teal!3}{\includegraphics[width=0.34\linewidth]{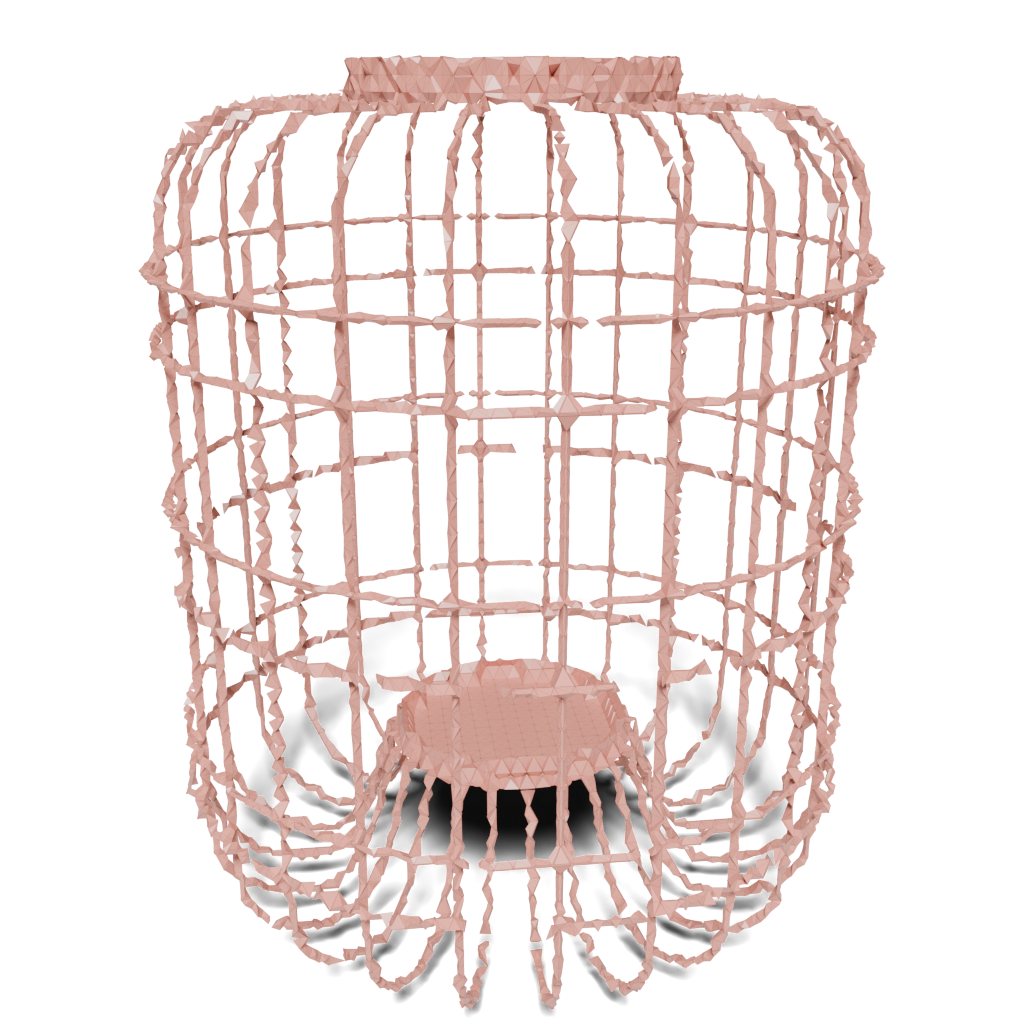}} &
        \colorbox{teal!3}{\includegraphics[width=0.34\linewidth]{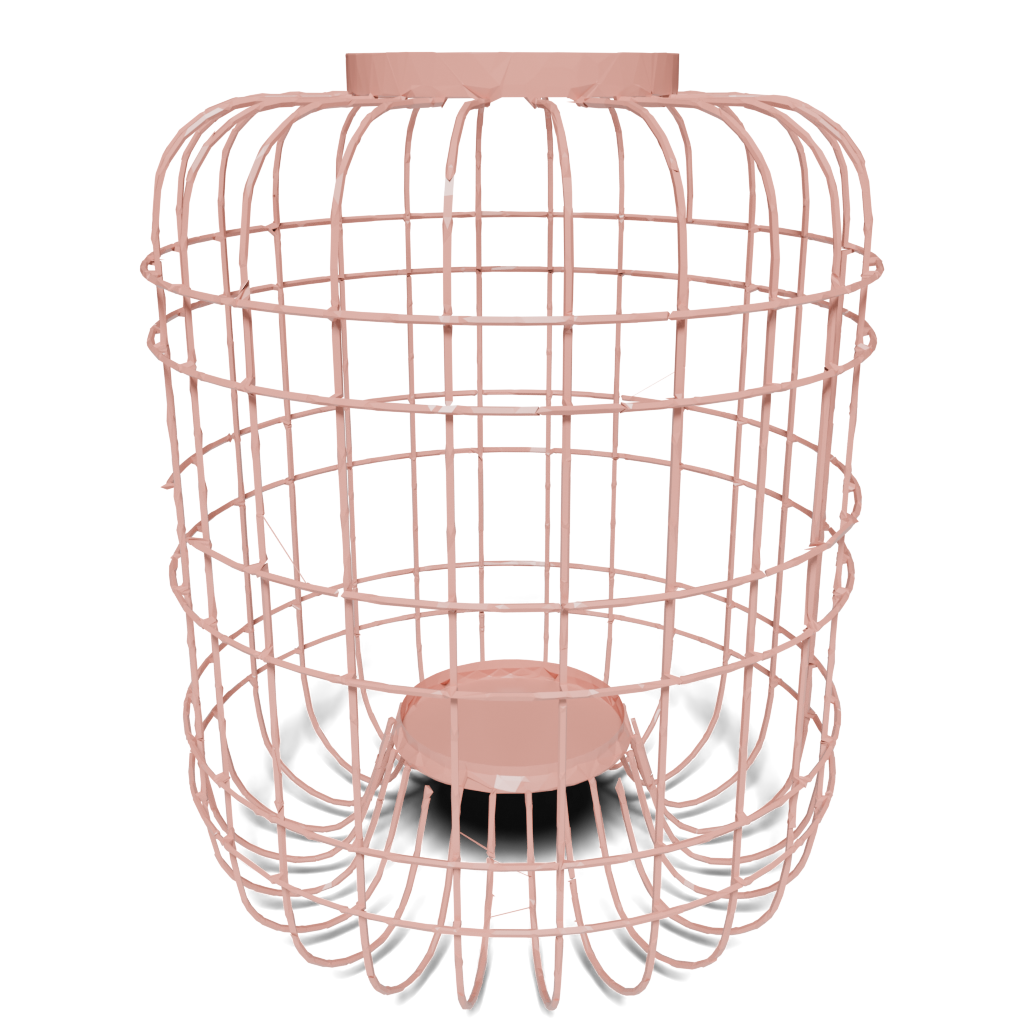}} \\
        Static ($64^3$) & Fully-Adaptive ($40^3$) \\
        $\num{1.89e-3}$ / $\num{3.03e-2}$ & $\num{7.08e-4}$ / $\num{3.27e-2}$ \\
        $64^3 = 262\text{K}$ & $40^3 \times (6 \times 3) \text{ (Weights)} = 1152\text{K}$ \\
    \end{tabular}
    \caption{Adaptive grid deformation can better recover thin structures as compared to static grids and TetWeave~\citep{tetweave}. The underlying isosurface extraction algorithm for all approaches is DMTet~\citep{dmtet}, which prevents the extracted mesh from self-intersecting, as long as tetrahedra have positive volume. \ccby Annie\_3D (Wired Lantern).}
    \label{fig:spindly-objects}
    \Description{Visual comparison of a wired lantern. The static grid and deformed grid both fully reconstruct the lantern, while TetWeave's reconstruction leaves a gap in the middle.
    }
\end{figure}

At 128K points and $64^3$ resolution, this work has a lower mean and median of both Chamfer and Hausdorff distance compared to TetWeave, while taking around $\frac{2}{3}$ of the time. Full results are shown in Tab.~\ref{tab:comp-128K-64}, and results for a thin mesh are shown in Fig.~\ref{fig:spindly-objects}. The median of both approaches is relatively close though, since most objects tested are reconstructed well for both approaches. Despite the aggregate metric being better for our approach, for many objects TetWeave outperforms this work. The difference in aggregate metric is due to TetWeave performing poorly on those meshes with difficult to reconstruct parts, which make up a large part of our dataset. When omitting the three largest outliers though, this work still outperforms TetWeave on average.

\begin{figure}
    \centering
    \setlength{\tabcolsep}{\arraystretch pt}
    \begin{tabular}{c c c}
        Input & Line Search & Alternating Opt. \\
        \colorbox{teal!3}{\includegraphics[width=0.31\linewidth]{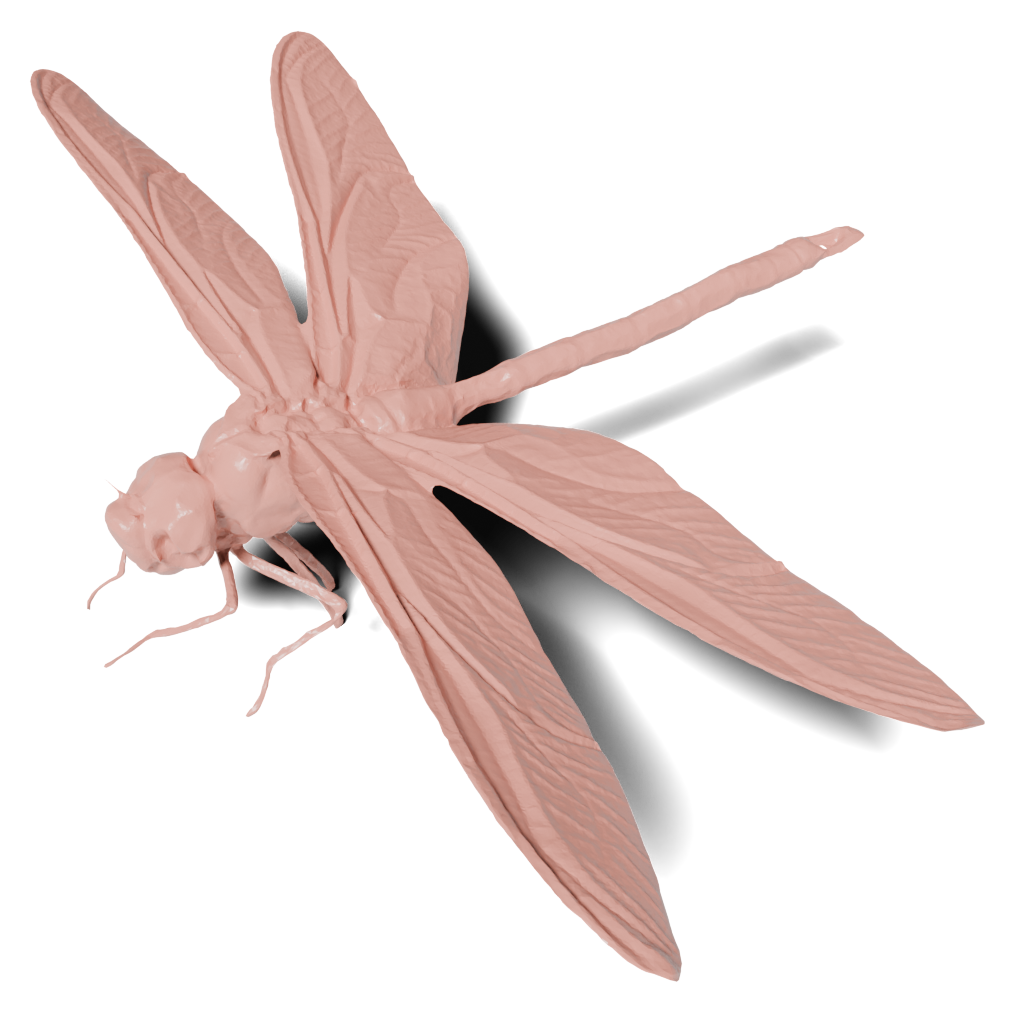}} &
        \colorbox{teal!3}{\includegraphics[width=0.31\linewidth]{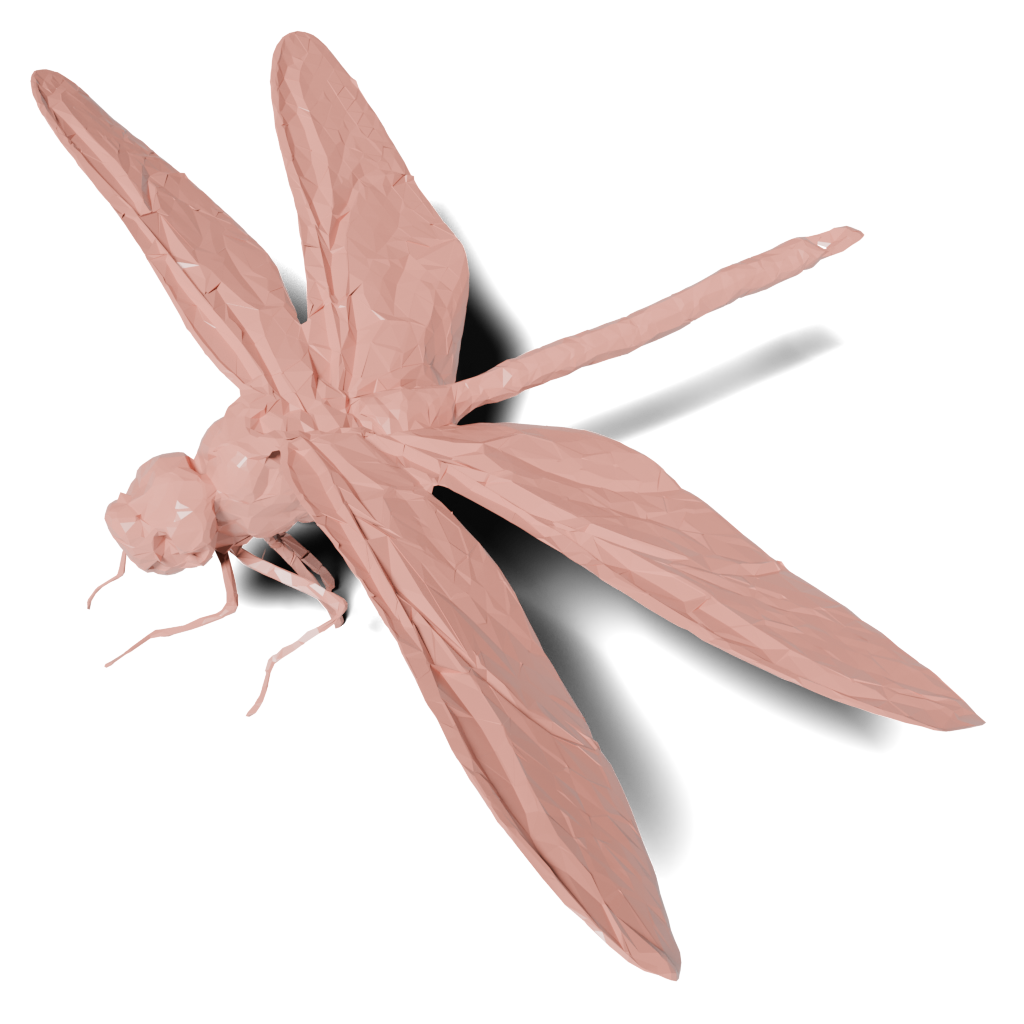}} &
        \colorbox{teal!3}{\includegraphics[width=0.31\linewidth]{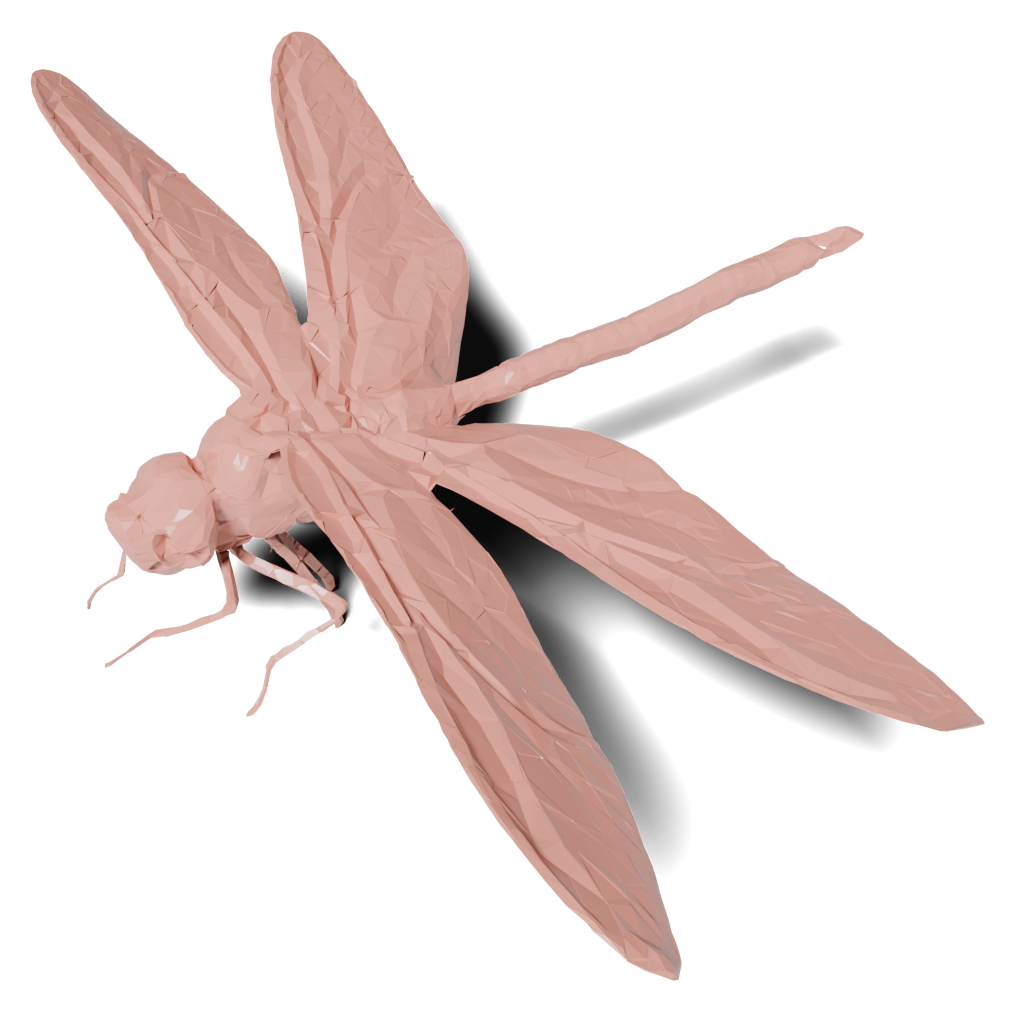}} \\
        Time (seconds) $^\downarrow$: & 1306.8 & 745.4 \\
        \footnotesize Chamfer$^\downarrow$ / Hausdorff$^\downarrow$: & \footnotesize $\num{3.00e-4}$ / $\num{1.45e-2}$ & \footnotesize $\num{3.56e-4}$ / $\num{1.35e-2}$ \\
    \end{tabular}
    \caption{Comparison of line search with backtracking by 0.8 and an IPC threshold of $\num{2e-6}$ versus alternating optimization with resetting vertices. At this threshold both have similar results, but line-search takes longer. When the IPC threshold is raised, both approaches take similar time. \ccbyncsa Lesser Emperor Dragonfly (ffish.asia).}
    \label{fig:cmp-alt-opt-line-search}
    \Description{Reconstruction of a dragonfly with 4 wings, left is the input. The salient detail is the time for this example.}
\end{figure}

\paragraph{Impact of Alternating Optimization}
In differentiable rendering, the use of momentum during optimization of thin elements may lead to inversion, but with a small step size and barrier energy inversion is infrequent, except for meshes with thin structures. To demonstrate the impact of alternating optimization on a single model, we lower the IPC threshold to $\num{2e-6}$ from $\num{2e-5}$, and perform optimization with global line search with backtracking, versus alternating color optimization with reseting of vertices which have inverted. We find that optimization is slowed at the lower IPC for line search due to backtracking, but good performance is maintained for the alternating color optimization, as shown in Fig.~\ref{fig:cmp-alt-opt-line-search}

\begin{figure}
    \centering
    \setlength{\tabcolsep}{\arraystretch pt}
    \begin{tabular}{c c c}
        Input & w/o Blur & w/ Blur \\
        \colorbox{teal!3}{\includegraphics[width=0.3\linewidth]{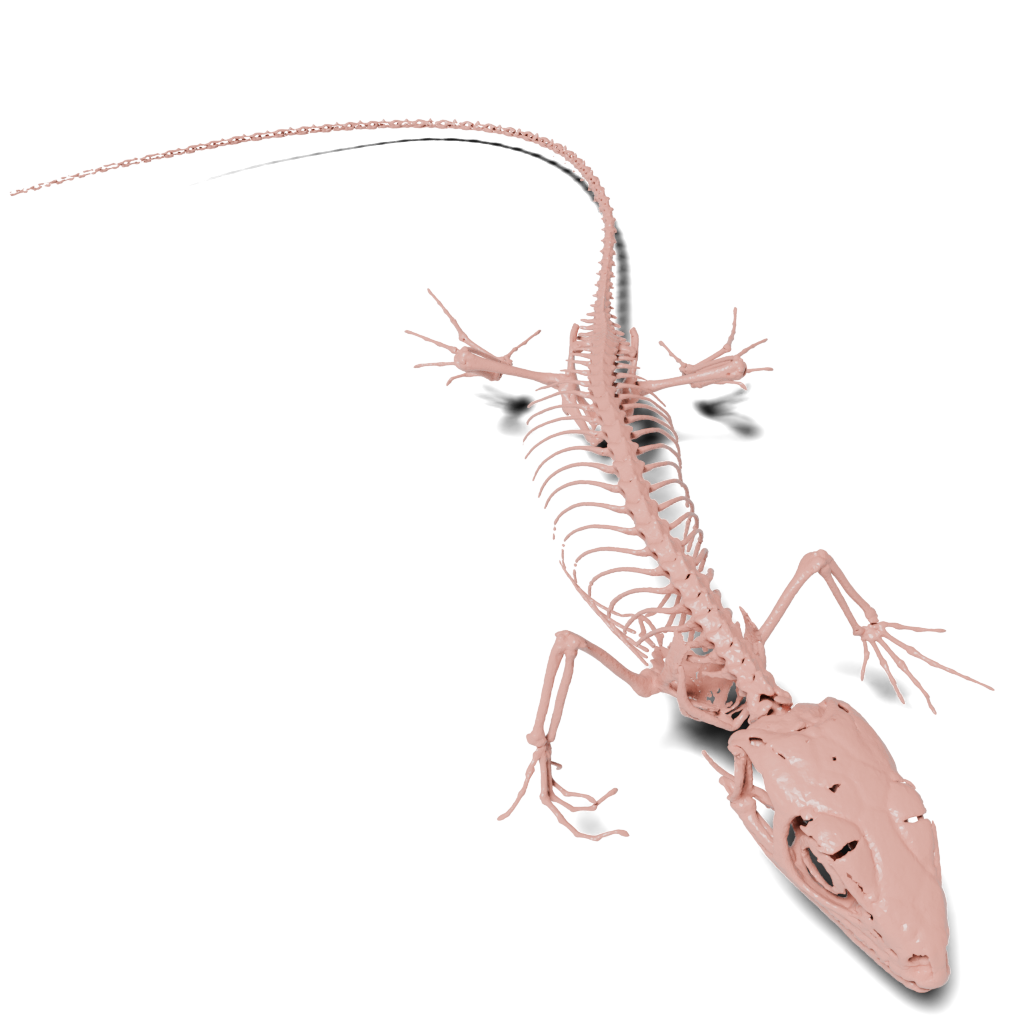}} &
        \colorbox{teal!3}{\includegraphics[width=0.3\linewidth]{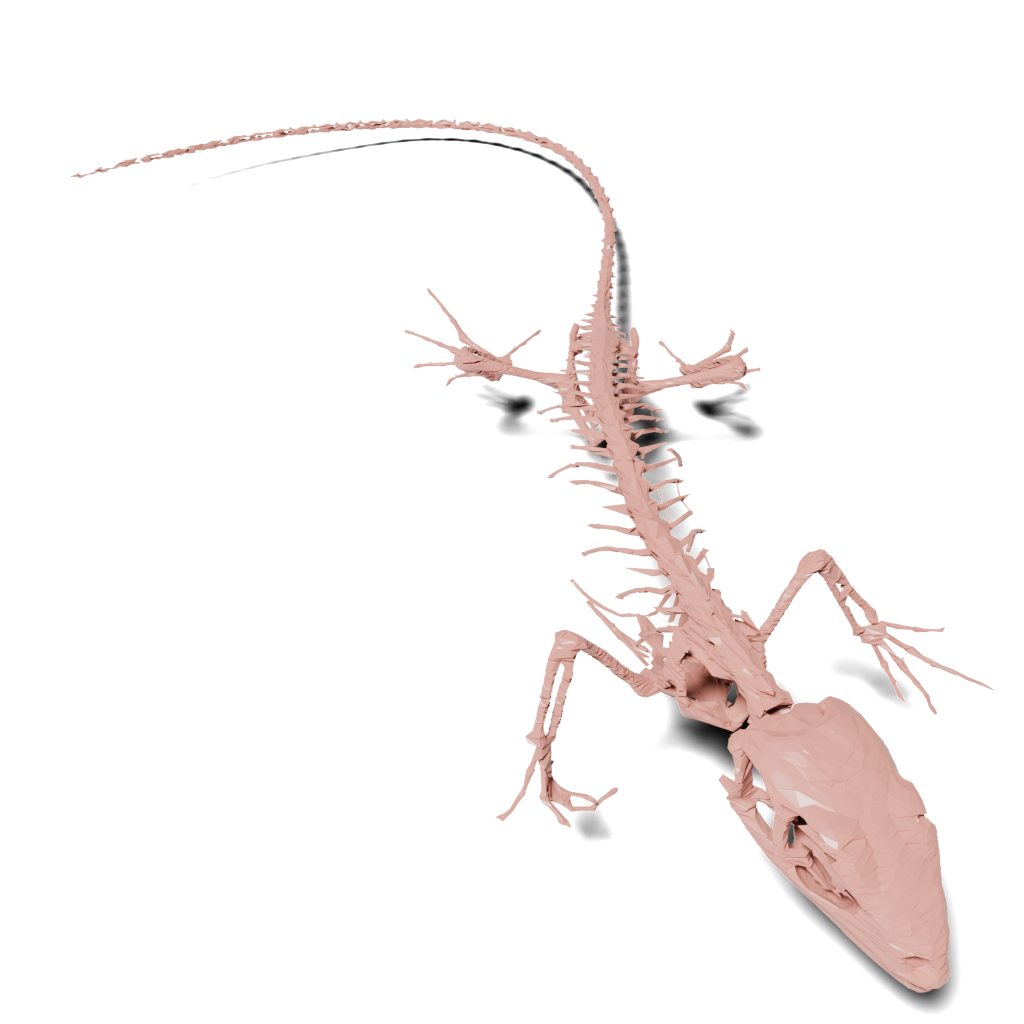}} &
        \colorbox{teal!3}{\includegraphics[width=0.3\linewidth]{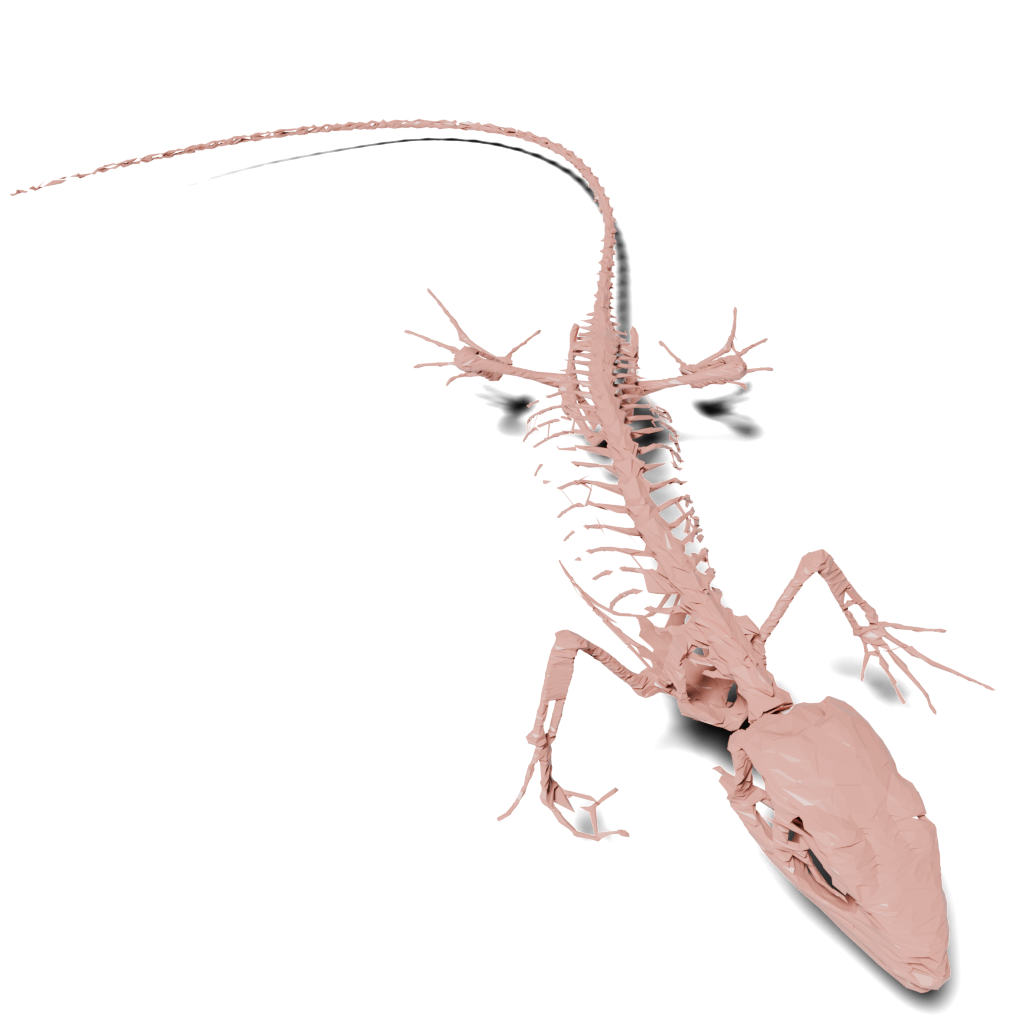}} \\
        \footnotesize Chamfer$^\downarrow$ / Hausdorff$^\downarrow$: & \footnotesize $\num{1.39e-3}$ / $\num{6.39e-2}$ & \footnotesize $\num{9.66e-4}$ / $\num{1.67e-2}$ \\
    \end{tabular}
    \caption{Comparison of differentiable isosurface extraction with and without blurring. Blurring helps recover details of the lizard fingers and bones, but sometimes introduces small artifacts between thin features. \ccbyncsa Japanese Grass Lizard (ffish.asia).}
    \label{fig:blur-precond}
    \Description{Reconstruction of the skeleton of a lizard. When zoomed, the ribcage of the lizard is visible, and a larger portion is reconstructed in the blurred version.}
\end{figure}

\paragraph{Ablating Coarse-To-Fine Image Blur}
To demonstrate the effect of using blurring as a coarse-to-fine-loss, we show an ablation in Fig.~\ref{fig:blur-precond}. For this example, we perform isosurface extraction with and without blurring on the skeleton of a lizard at 80 grid resolution. With blurring enabled, reconstruction improves around small features such as the thin bones.

\begin{figure}[h]
    \centering
    \setlength{\tabcolsep}{-4pt}
    \begin{tabular}{c c c}
        \multicolumn{3}{c}{Grid Resolution Comparison} \\
        Input & 16$^3$ & 32$^3$ \\
        \colorbox{teal!3}{\includegraphics[width=0.35\linewidth]{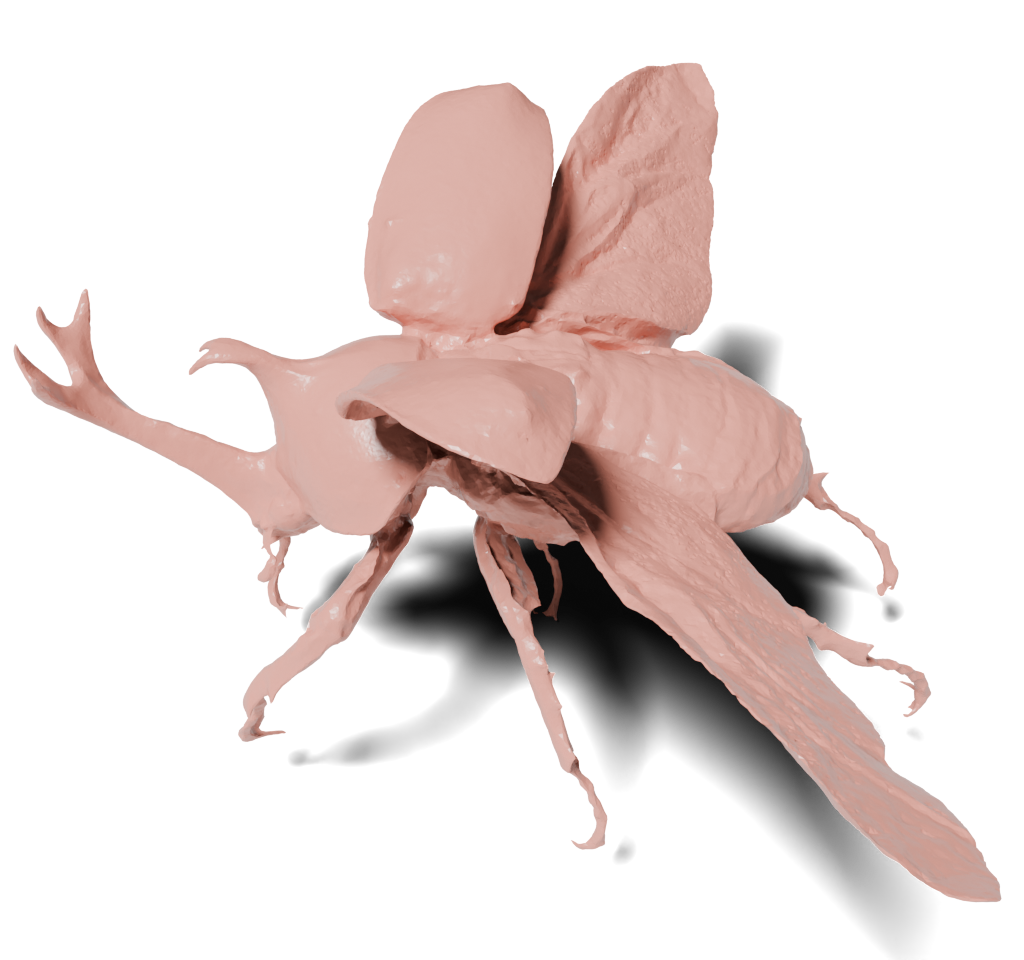}} &
        \colorbox{teal!3}{\includegraphics[width=0.35\linewidth]{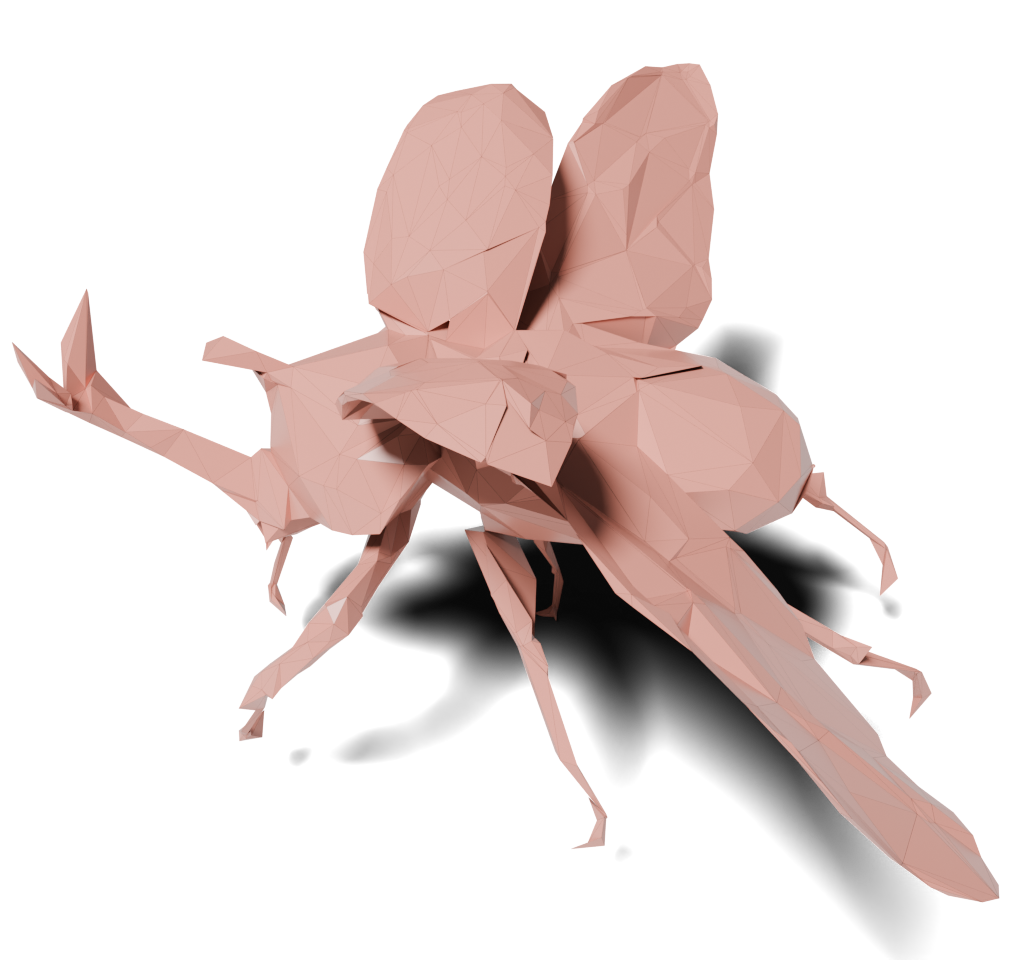}} &
        \colorbox{teal!3}{\includegraphics[width=0.35\linewidth]{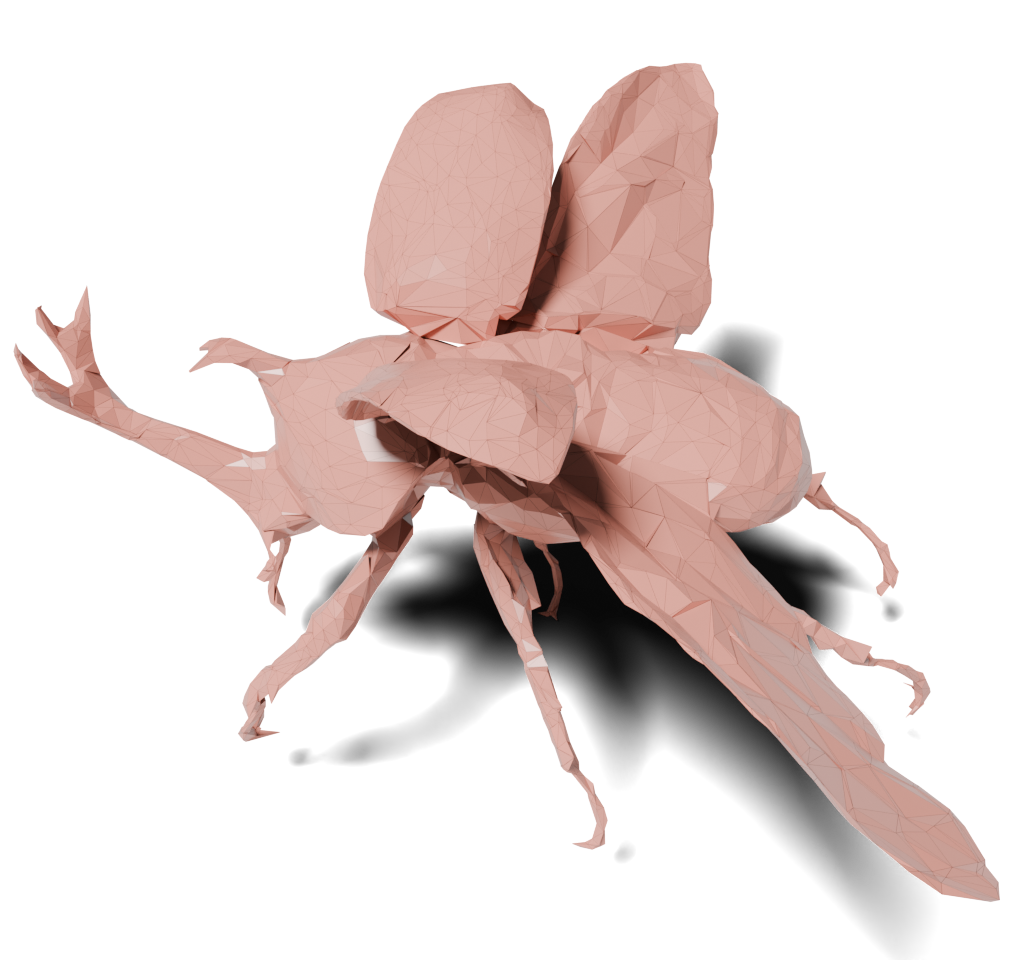}} \\
        \multicolumn{3}{c}{
            \begin{tabular}{c c}
                \colorbox{teal!3}{\includegraphics[width=0.35\linewidth]{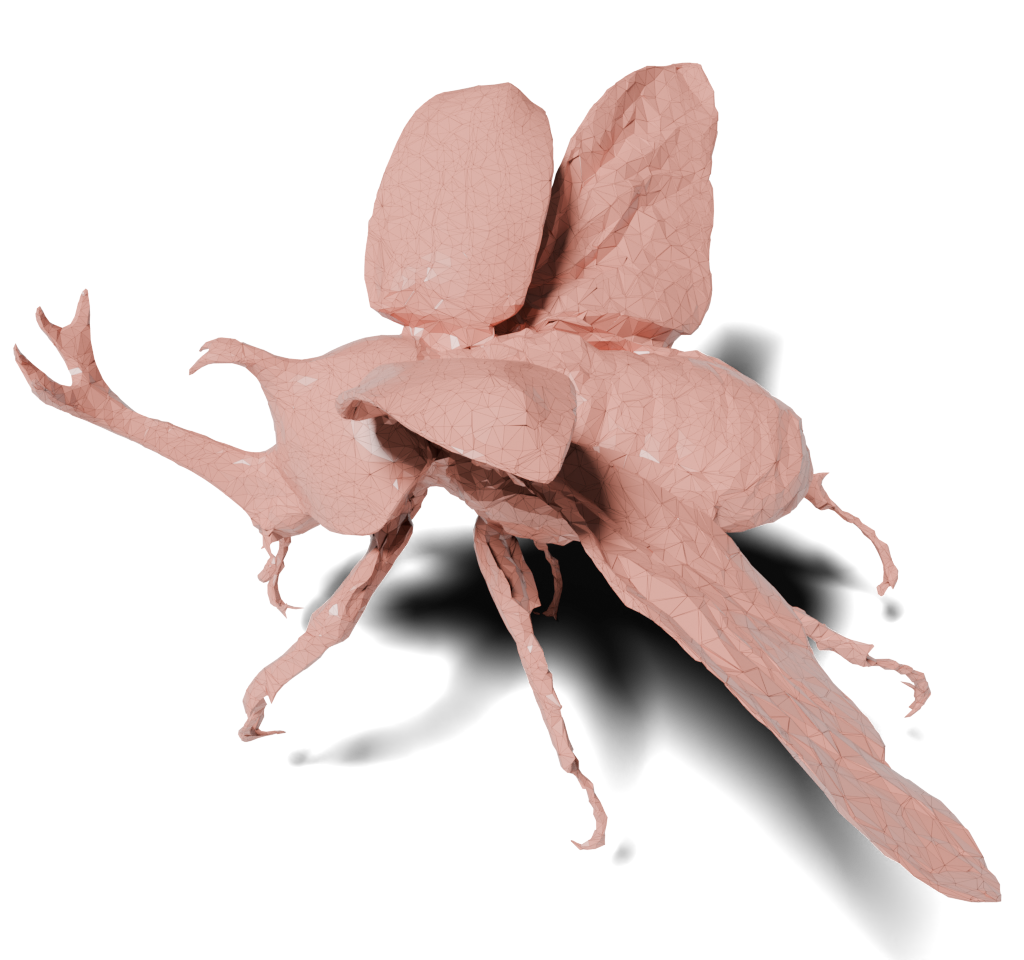}} &
                \colorbox{teal!3}{\includegraphics[width=0.35\linewidth]{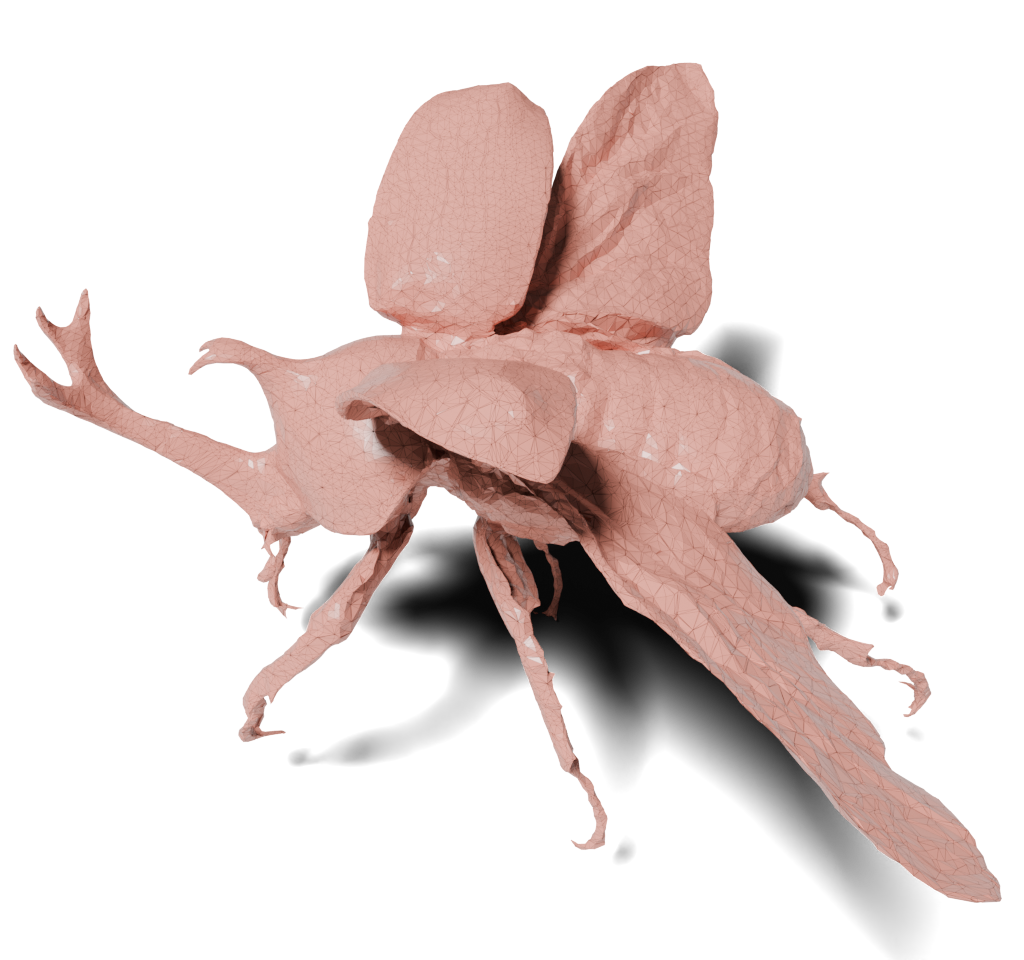}} \\
                 64$^3$ & 96$^3$ \\
            \end{tabular}
        } \\
    \end{tabular}
    \includegraphics[width=0.72\linewidth]{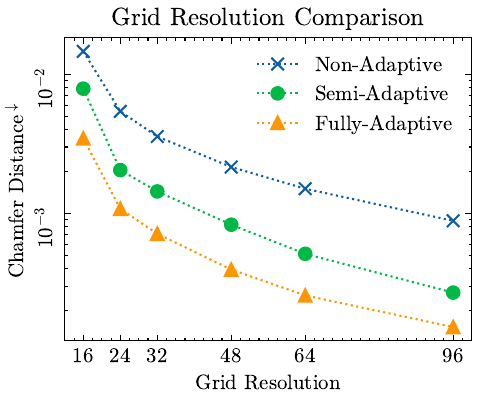}
    \caption{Comparison of different grid resolutions used in isosurface-extraction. Using a fully adaptive grid greatly improves output quality, and lessens the reliance on high grid-resolution for output quality. \cczero ffish.asia (Japanese Rhinoceros Beetle).}
    \label{fig:grid-res-ablation}
    \Description{5 images of a rhinoceros beetle, at increasing levels of detail. The plot has x-axis which is grid resolution, y-axis which is Chamfer distance, and three plots for non-adaptive, semi-adaptive, and fully-adaptive grids. The plots have similar trends, where increasing resolution leads, but the fully-adaptive plot is flatter than the prior approaches.}
\end{figure}

\paragraph{Ablating Effect of Grid Resolution}
We ablate the impact of varying the grid resolution on output quality. Intuitively, we expect grid resolution to have less impact compared to grid resolutions for static grids (non-adaptive) and grids where vertices can deform by half of each cell (semi-adaptive). The results from this experiment are shown  in Fig.~\ref{fig:grid-res-ablation}, where fully adaptive grids increases the efficiency of capturing high frequency detail in iso-surface extraction. Ideally, we hope that the quality of the result is close to linear with the total number of parameters, as each parameter should be used in the output. While we do not observe such a pattern, making the grid fully-adaptive improves the usage of parameters, flattening the influence of grid-resolution on output quality.

\section{Discussion \& Limitations}

%The approach outlined allows for efficient, inversion-free optimization of grids, with arbitrary optimizers. This optimization approach has been shown to be efficient for UV optimization, image-compression, and isosurface-extraction for differentiable rendering. For UV optimization, the approach is able to optimize arbitrary energies efficiently while remaining inversion-free. For image-compression, grid-deformation can improve the PSNR of a compressed image consistently as compared to rescaling without pixel deformation. Finally, for isosurface extraction, using fully deformable grids performs similar to adaptive tetrahedral grids, and is more consistent on difficult thin and mechanical features.

\paragraph{Additional Optimization Steps} Since only a subset of vertices are updated concurrently, it requires $C$ times as many optimizations to explicitly update all vertices for an equal number of iterations, where $C$ is the number of colors. On the other hand, the differential representation implicitly causes vertices to update even when they are not optimized directly.

\paragraph{Sparse Influence} For image compaction and differentiable rendering, vertices are deformed purely using reconstruction loss. This is problematic because vertex gradients only affects elements locally. Furthermore convergence depends mostly on a small number of high-frequency regions, where vertices should be denser, but reconstruction loss alone does not heavily increase density of vertices in those regions. To mitigate this, this work uses image level blurring, which spreads the loss of one pixel to neighboring regions. While this works, it is likely not optimal and it's possible that an alternative loss, possibly an optimal transport loss~\citep{optimal_transport, drot}, can be applied to each grid cell that may allow for faster convergence, but we are unsure what such a loss would look like. We leave it to future work to find such approaches.

\paragraph{Barrier Energies} Another limitation of our approach is that it still requires a barrier energy to prevent inversions. Due to tetrahedral subdivision, this becomes a memory bottleneck when it is computed in parallel since it scales cubically with a high constant factor, hindering the scalability of the approach. It is also unclear how much the barrier energy prevents convergence of the target loss, which was noted as problematic in \citep{rectangular_surface_parameterization}.

\paragraph{Fixed Convex Boundaries} Grids have non-overlapping and locked boundaries. In our experiments, vertices are permitted to slide along boundaries by reprojection to already-known boundary elements. Instead, if boundaries are fully-fixed with an existing injective parameterization, then arbitrary input boundaries can be used. It may be possible to extend this approach to allow for boundary deformation, similar to \citep{simplicial_complex_augmentation_framework}. To do so, additional padding cells can be introduced surrounding the boundary, and the new boundary of these padding cells can be fixed. After optimization, these layers can peeled away, leaving a deformed shape. For meshes (both tetrahedral and 2D meshes embedded in 3D), a similar approach may be possible with erosion and dilation around boundaries, although it may be difficult to prevent dilations or erosions from intersecting.

\paragraph{Slivers in Isosurface Extraction} Due to subdivisions of the input cube into tetrahedra, the output may contain slivers which are nearly degenerate. We did not explore additional mesh triangle optimization, but the approach outlined can be combined with triangle regularization losses.

\paragraph{Deformable Grids versus Adaptive Representations} While the approach in this work shows benefits over the tetrahedral mesh used in TetWeave, they are not necessarily orthogonal. Specifically, an adaptive tetrahedral grid can also optimize vertices differentially, although the implementation is more complicated than for grids. Furthermore, it's likely that missing small features can be fixed by first optimizing a grid, and then switching to an adaptive sampling approach. We leave exploration of these to future work.

\section{Conclusion}

This work describes a straightforward way to make grid optimization fully-adaptive and more efficient using a differential representation, as opposed to prior work which was only semi-adaptive. The techniques used in this work can also be applied to efficient mesh optimization, and we hope they can be adopted into traditional optimization pipelines.

\begin{acks}
This work was partly supported by the National Research Foundation (NRF) grants (RS-2024-00438532, RS-2023-00211658, RS-2024-00357548, RS-2025-00518643), the Institute of Information \& Communications Technology Planning \& Evaluation (IITP) grant (RS-2024-0045788), and ITRC (Information Technology Research Center) grant (IITP-2025-RS-2024-00437866), funded by the Korean government (MSIT).
\end{acks}
\bibliographystyle{./ref/ACM-Reference-Format}
\bibliography{citations}
\appendix

%\clearpage

\paragraph{Image Compaction Results}
The complete results for image compaction against bilinear resizing and JPEG compression are shown in Tab.~\ref{tab:image-compact-compare}.

\paragraph{Differentiable Rendering Results} The full comparison on all models is shown in Tab.~\ref{tab:comp-128K-64} for TetWeave at 128K points and this work at $64^3$. Omitted outliers are shown with \textcolor{red}{$\dagger$}.

\begin{table}[!h]
        \centering
        \setlength{\tabcolsep}{1.5pt}
        \renewcommand{\arraystretch}{0.5}
        \begin{tabular}{|c|c|c||c|c|c|}
        \multicolumn{6}{c}{Image Compaction Comparison} \\\hline
        Input & Bilinear & Deform & JPEG (25) & JPEG (50) & JPEG (75) \\\hline
        Img. 01 & 23.95 & 29.25 & 27.66 & 29.87 & 32.40 \\\hline
        Img. 02 & 30.49 & 35.26 & 30.80 & 32.84 & 34.85 \\\hline
        Img. 03 & 31.17 & 35.91 & 32.19 & 34.56 & 36.86 \\\hline
        Img. 04 & 30.48 & 35.27 & 31.10 & 33.26 & 35.26 \\\hline
        Img. 05 & 23.81 & 29.52 & 27.08 & 29.59 & 32.31 \\\hline
        Img. 06 & 25.52 & 30.13 & 28.83 & 31.16 & 33.79 \\\hline
        Img. 07 & 29.70 & 35.65 & 31.45 & 33.92 & 36.27 \\\hline
        Img. 08 & 21.54 & 26.55 & 26.84 & 29.43 & 32.16 \\\hline
        Img. 09 & 29.36 & 34.79 & 32.20 & 34.53 & 36.70 \\\hline
        Img. 10 & 29.37 & 34.81 & 31.78 & 34.18 & 36.44 \\\hline
        Img. 11 & 26.98 & 31.81 & 29.45 & 31.71 & 34.22 \\\hline
        Img. 12 & 30.11 & 34.98 & 32.16 & 34.60 & 36.81 \\\hline
        Img. 13 & 22.10 & 26.02 & 25.32 & 27.60 & 30.51 \\\hline
        Img. 14 & 26.53 & 31.54 & 28.19 & 30.29 & 32.60 \\\hline
        Img. 15 & 29.08 & 33.65 & 30.96 & 33.07 & 35.26 \\\hline
        Img. 16 & 29.05 & 33.40 & 31.16 & 33.45 & 35.79 \\\hline
        Img. 17 & 29.69 & 34.32 & 31.38 & 33.69 & 35.94 \\\hline
        Img. 18 & 25.84 & 30.00 & 28.05 & 30.29 & 32.72 \\\hline
        Img. 19 & 25.93 & 31.14 & 30.10 & 32.37 & 34.62 \\\hline
        Img. 20 & 28.22 & 33.37 & 31.38 & 33.53 & 35.75 \\\hline
        Img. 21 & 26.26 & 31.06 & 29.29 & 31.47 & 33.90 \\\hline
        Img. 22 & 28.20 & 32.92 & 29.66 & 31.72 & 33.82 \\\hline
        Img. 23 & 31.27 & 36.62 & 32.70 & 35.08 & 37.12 \\\hline
        Img. 24 & 24.69 & 28.67 & 27.69 & 29.98 & 32.44 \\\hline
        \hline
        Mean    & 27.47 & 32.36 & 29.89 & 32.17 & 34.52 \\\hline
        Median  & 28.21 & 33.14 & 30.45 & 32.61 & 34.74 \\\hline
        \end{tabular}
\caption{Comparison of different approaches for image compaction/compression on the Kodak image
dataset. Allowing for deformation of grid cells greatly increases quality compared to standard
bilinear resizing, and performs similarly to JPEG compression with 50\% quality. \cczero Kodak ~\citep{kodak_dataset}.}
\label{tab:image-compact-compare}
\end{table}

\begin{table*}\centering
\renewcommand{\arraystretch}{0.07}
\setlength{\tabcolsep}{0.5pt}
\begin{tabular}{|c| |c|c||c|c| |c|c| |c|c|}\hline
& \multicolumn{2}{c|}{Chamfer} & \multicolumn{2}{c|}{Hausdorff} & \multicolumn{2}{|c|}{Output Faces} &   \multicolumn{2}{|c|}{Time (sec.)} \\\hline
\textbf{Mesh} & TetWeave (128K) & Deformed(64$^3$) & TetWeave (128K) & Deformed(64$^3$) & T & D & T & D \\\hline
Armillary Sphere & \cellcolor{orange!10}$\num{3.64e-04}$ & $\num{5.70e-04}$ & \cellcolor{orange!10}$\num{3.01e-02}$ & $\num{5.28e-02}$ & 642752 & 96020 & 1358.6 & 736.0 \\\hline
Barley Hordeum Vulgare\textcolor{red}{$\dagger$} & $\num{4.75e-01}$ & \cellcolor{orange!10}$\num{3.58e-03}$ & $\num{6.37e-01}$ & \cellcolor{orange!10}$\num{1.08e-01}$ & 1106 & 14712 & 1169.2 & 694.9 \\\hline
Black Pine & $\num{4.92e-03}$ & \cellcolor{orange!10}$\num{7.14e-04}$ & $\num{1.87e-01}$ & \cellcolor{orange!10}$\num{1.58e-02}$ & 272300 & 40376 & 1255.7 & 743.6 \\\hline
Block & $\num{4.32e-05}$ & \cellcolor{orange!10}$\num{2.67e-05}$ & $\num{4.47e-02}$ & \cellcolor{orange!10}$\num{1.35e-03}$ & 290868 & 90560 & 1220.6 & 685.6 \\\hline
Calcite Druse & $\num{1.29e-03}$ & \cellcolor{orange!10}$\num{2.77e-04}$ & $\num{1.09e-01}$ & \cellcolor{orange!10}$\num{1.97e-02}$ & 302772 & 129128 & 1243.0 & 759.7 \\\hline
Candelabrum & \cellcolor{orange!10}$\num{1.39e-04}$ & $\num{3.91e-04}$ & \cellcolor{orange!10}$\num{7.11e-03}$ & $\num{9.78e-03}$ & 352626 & 104652 & 1223.6 & 699.2 \\\hline
Celestial Globe & \cellcolor{orange!10}$\num{6.05e-04}$ & $\num{6.76e-04}$ & $\num{1.44e-01}$ & \cellcolor{orange!10}$\num{1.88e-02}$ & 343914 & 94900 & 1257.1 & 733.6 \\\hline
Chinese Pagoda & $\num{2.62e-03}$ & \cellcolor{orange!10}$\num{1.72e-03}$ & $\num{4.92e-02}$ & \cellcolor{orange!10}$\num{3.54e-02}$ & 434564 & 243020 & 1375.7 & 938.1 \\\hline
Closed Helmet & \cellcolor{orange!10}$\num{1.95e-04}$ & $\num{2.20e-04}$ & $\num{1.03e-02}$ & \cellcolor{orange!10}$\num{9.64e-03}$ & 416566 & 138308 & 1303.1 & 733.0 \\\hline
Common Water Strider & $\num{1.35e-02}$ & \cellcolor{orange!10}$\num{2.82e-04}$ & $\num{2.49e-01}$ & \cellcolor{orange!10}$\num{1.56e-02}$ & 110508 & 11492 & 1199.4 & 613.0 \\\hline
Coral & \cellcolor{orange!10}$\num{1.36e-04}$ & $\num{4.22e-04}$ & $\num{1.41e-01}$ & \cellcolor{orange!10}$\num{1.07e-02}$ & 384552 & 151376 & 1291.4 & 764.4 \\\hline
Corinthian Helmet & $\num{5.45e-05}$ & $\num{6.24e-05}$ & $\num{4.63e-03}$ & \cellcolor{orange!10}$\num{3.81e-03}$ & 427996 & 123544 & 1294.0 & 740.3 \\\hline
Deathwing & \cellcolor{orange!10}$\num{8.48e-04}$ & $\num{1.20e-03}$ & $\num{2.88e-02}$ & \cellcolor{orange!10}$\num{1.67e-02}$ & 427852 & 45968 & 1247.9 & 645.1 \\\hline
Deepwater Carrier Crab & \cellcolor{orange!10}$\num{1.47e-04}$ & $\num{4.59e-04}$ & $\num{6.87e-02}$ & \cellcolor{orange!10}$\num{6.87e-02}$ & 216818 & 35248 & 1280.1 & 745.1 \\\hline
Delta Fractal & \cellcolor{orange!10}$\num{2.55e-03}$ & $\num{3.05e-03}$ & \cellcolor{orange!10}$\num{1.13e-02}$ & $\num{1.38e-02}$ & 636920 & 193576 & 1550.9 & 1041.7 \\\hline
Dragon Sculpture & $\num{1.69e-04}$ & $\num{1.74e-04}$ & $\num{5.61e-02}$ & \cellcolor{orange!10}$\num{1.43e-02}$ & 327240 & 157232 & 1230.3 & 743.8 \\\hline
Garuda Terminal & \cellcolor{orange!10}$\num{1.63e-04}$ & $\num{2.54e-04}$ & $\num{1.71e-02}$ & \cellcolor{orange!10}$\num{1.57e-02}$ & 293748 & 110140 & 1205.9 & 681.9 \\\hline
Gekko Japonicus & \cellcolor{orange!10}$\num{7.28e-04}$ & $\num{1.34e-03}$ & $\num{1.05e-01}$ & \cellcolor{orange!10}$\num{5.03e-02}$ & 184678 & 45000 & 1287.7 & 773.9 \\\hline
George Washington & $\num{3.70e-04}$ & \cellcolor{orange!10}$\num{3.44e-04}$ & $\num{5.69e-02}$ & \cellcolor{orange!10}$\num{1.99e-02}$ & 295528 & 120656 & 1258.6 & 736.3 \\\hline
Golden Bamboo & $\num{1.43e-03}$ & \cellcolor{orange!10}$\num{2.22e-04}$ & $\num{1.21e-01}$ & \cellcolor{orange!10}$\num{1.25e-02}$ & 218850 & 22732 & 1249.3 & 699.1 \\\hline
Halo For Buddha Of The Future & \cellcolor{orange!10}$\num{1.39e-04}$ & $\num{5.70e-04}$ & $\num{9.64e-03}$ & \cellcolor{orange!10}$\num{5.28e-03}$ & 204194 & 40736 & 1198.4 & 635.0 \\\hline
Japanese Grass Lizard\textcolor{red}{$\dagger$} & $\num{1.34e-01}$ & \cellcolor{orange!10}$\num{1.36e-03}$ & $\num{4.99e-01}$ & \cellcolor{orange!10}$\num{2.13e-02}$ & 1618 & 29648 & 1192.6 & 732.3 \\\hline
Japanese Pit Viper & \cellcolor{orange!10}$\num{1.13e-03}$ & $\num{2.02e-03}$ & $\num{1.08e-01}$ & \cellcolor{orange!10}$\num{3.31e-02}$ & 288176 & 71192 & 1602.8 & 1241.9 \\\hline
Kabutomushi Rhino Beetle & \cellcolor{orange!10}$\num{1.15e-04}$ & $\num{2.95e-04}$ & \cellcolor{orange!10}$\num{4.83e-03}$ & $\num{9.00e-03}$ & 465104 & 43492 & 1371.8 & 730.1 \\\hline
Korean Crown & \cellcolor{orange!10}$\num{7.88e-05}$ & $\num{1.98e-04}$ & $\num{1.03e-02}$ & \cellcolor{orange!10}$\num{7.60e-03}$ & 546830 & 116080 & 1358.0 & 723.0 \\\hline
Korean Mantis Angusti Pennis & $\num{9.93e-04}$ & \cellcolor{orange!10}$\num{4.69e-04}$ & $\num{1.80e-01}$ & \cellcolor{orange!10}$\num{1.80e-01}$ & 192260 & 20476 & 1209.1 & 637.0 \\\hline
Kris Dagger & \cellcolor{orange!10}$\num{5.68e-04}$ & $\num{7.16e-04}$ & \cellcolor{orange!10}$\num{2.52e-02}$ & $\num{2.57e-02}$ & 157526 & 52064 & 1232.3 & 636.0 \\\hline
Long Arm Octopus Minor & \cellcolor{orange!10}$\num{1.72e-04}$ & $\num{2.90e-04}$ & $\num{1.30e-02}$ & \cellcolor{orange!10}$\num{1.16e-02}$ & 209200 & 94504 & 1334.5 & 855.3 \\\hline
Mechanical Phoenix & \cellcolor{orange!10}$\num{5.65e-04}$ & $\num{1.11e-03}$ & \cellcolor{orange!10}$\num{9.81e-03}$ & $\num{1.29e-02}$ & 398840 & 42660 & 1308.9 & 745.1 \\\hline
Mechanical Shapes & \cellcolor{orange!10}$\num{1.57e-03}$ & $\num{1.83e-03}$ & \cellcolor{orange!10}$\num{2.96e-02}$ & $\num{3.20e-02}$ & 433788 & 191324 & 1271.0 & 752.6 \\\hline
Mech Head & $\num{4.88e-03}$ & \cellcolor{orange!10}$\num{1.57e-03}$ & $\num{7.96e-02}$ & \cellcolor{orange!10}$\num{5.01e-02}$ & 319388 & 73460 & 1217.7 & 640.2 \\\hline
Nataraja Shiva & \cellcolor{orange!10}$\num{7.95e-05}$ & $\num{2.92e-04}$ & $\num{2.83e-02}$ & \cellcolor{orange!10}$\num{8.64e-03}$ & 287752 & 54012 & 1200.7 & 670.8 \\\hline
Ogre & $\num{4.68e-04}$ & \cellcolor{orange!10}$\num{3.87e-04}$ & $\num{4.96e-02}$ & \cellcolor{orange!10}$\num{1.53e-02}$ & 772914 & 121040 & 1421.2 & 731.7 \\\hline
Oil Lamp & \cellcolor{orange!10}$\num{9.55e-05}$ & $\num{2.08e-04}$ & $\num{3.56e-02}$ & \cellcolor{orange!10}$\num{1.51e-02}$ & 281798 & 69024 & 1199.1 & 683.8 \\\hline
Old Camera & \cellcolor{orange!10}$\num{6.80e-04}$ & $\num{8.11e-04}$ & \cellcolor{orange!10}$\num{2.89e-02}$ & $\num{2.91e-02}$ & 365108 & 107980 & 1247.9 & 734.2 \\\hline
Pocillopora Damicornis & \cellcolor{orange!10}$\num{1.40e-04}$ & $\num{4.34e-04}$ & $\num{7.71e-02}$ & \cellcolor{orange!10}$\num{7.69e-02}$ & 279908 & 74268 & 1185.8 & 678.2 \\\hline
Polarimeter\textcolor{red}{$\dagger$} & $\num{3.26e-01}$ & \cellcolor{orange!10}$\num{8.94e-04}$ & $\num{7.84e-01}$ & \cellcolor{orange!10}$\num{1.79e-02}$ & 181432 & 30404 & 1237.2 & 643.9 \\\hline
Project Spider & \cellcolor{orange!10}$\num{7.99e-04}$ & $\num{1.14e-03}$ & \cellcolor{orange!10}$\num{1.33e-02}$ & $\num{1.46e-02}$ & 286324 & 59916 & 1222.0 & 654.5 \\\hline
Red Spider Lily & $\num{8.44e-04}$ & \cellcolor{orange!10}$\num{6.87e-04}$ & $\num{8.16e-02}$ & \cellcolor{orange!10}$\num{2.94e-02}$ & 306915 & 37492 & 1231.7 & 688.5 \\\hline
Ritual Bell & $\num{1.83e-04}$ & $\num{1.80e-04}$ & \cellcolor{orange!10}$\num{1.22e-02}$ & $\num{1.26e-02}$ & 225158 & 116852 & 1201.6 & 689.1 \\\hline
Rope Ladder & $\num{2.33e-02}$ & \cellcolor{orange!10}$\num{4.19e-04}$ & $\num{1.37e-01}$ & \cellcolor{orange!10}$\num{6.97e-03}$ & 23240 & 167192 & 1110.4 & 690.0 \\\hline
Saxilby Memorial Helmet & \cellcolor{orange!10}$\num{7.97e-05}$ & $\num{1.45e-04}$ & \cellcolor{orange!10}$\num{3.87e-03}$ & $\num{4.72e-03}$ & 366894 & 114912 & 1303.1 & 783.9 \\\hline
Side Chair & \cellcolor{orange!10}$\num{1.68e-04}$ & $\num{3.88e-04}$ & $\num{4.02e-02}$ & \cellcolor{orange!10}$\num{1.98e-02}$ & 275946 & 93912 & 1193.5 & 686.2 \\\hline
Snow Crab & \cellcolor{orange!10}$\num{1.98e-04}$ & $\num{2.70e-04}$ & $\num{3.75e-02}$ & \cellcolor{orange!10}$\num{1.05e-02}$ & 254882 & 43048 & 1266.9 & 735.0 \\\hline
Spider Brake & $\num{7.37e-02}$ & \cellcolor{orange!10}$\num{2.90e-04}$ & $\num{3.35e-01}$ & \cellcolor{orange!10}$\num{1.96e-02}$ & 34170 & 18196 & 1123.4 & 613.3 \\\hline
Starling & \cellcolor{orange!10}$\num{1.84e-04}$ & $\num{4.11e-04}$ & \cellcolor{orange!10}$\num{2.48e-02}$ & $\num{2.54e-02}$ & 256276 & 50476 & 1375.0 & 909.4 \\\hline
Stick Insect & $\num{9.81e-02}$ & \cellcolor{orange!10}$\num{2.65e-04}$ & $\num{3.79e-01}$ & \cellcolor{orange!10}$\num{3.68e-03}$ & 1490 & 17592 & 1110.5 & 602.4 \\\hline
Table And Tea Service & \cellcolor{orange!10}$\num{5.08e-04}$ & $\num{9.52e-04}$ & $\num{2.14e-02}$ & \cellcolor{orange!10}$\num{2.14e-02}$ & 390342 & 115516 & 1252.7 & 734.7 \\\hline
Table For Paints & \cellcolor{orange!10}$\num{5.30e-04}$ & $\num{6.28e-04}$ & \cellcolor{orange!10}$\num{1.20e-02}$ & $\num{1.33e-02}$ & 479130 & 135672 & 1282.0 & 728.6 \\\hline
Table Fountain & \cellcolor{orange!10}$\num{4.16e-04}$ & $\num{6.73e-04}$ & \cellcolor{orange!10}$\num{1.28e-02}$ & $\num{1.50e-02}$ & 431598 & 132416 & 1356.2 & 842.4 \\\hline
Transitional Rapier Dutch & $\num{3.45e-03}$ & \cellcolor{orange!10}$\num{4.83e-04}$ & $\num{1.52e-01}$ & \cellcolor{orange!10}$\num{1.26e-02}$ & 87232 & 44076 & 1197.9 & 664.0 \\\hline
Vesperbild & \cellcolor{orange!10}$\num{1.50e-04}$ & $\num{2.49e-04}$ & \cellcolor{orange!10}$\num{1.28e-02}$ & $\num{1.55e-02}$ & 334898 & 143292 & 1241.5 & 752.6 \\\hline
Water Clover Marsilea Quadrifolia & \cellcolor{orange!10}$\num{6.15e-05}$ & $\num{1.61e-04}$ & \cellcolor{orange!10}$\num{1.89e-03}$ & $\num{5.76e-03}$ & 392786 & 29684 & 1269.2 & 660.2 \\\hline
Water Scorpion & $\num{4.64e-03}$ & \cellcolor{orange!10}$\num{2.69e-04}$ & $\num{1.32e-01}$ & \cellcolor{orange!10}$\num{7.21e-03}$ & 133948 & 15720 & 1192.6 & 616.7 \\\hline
Wired Lantern & $\num{1.05e-03}$ & \cellcolor{orange!10}$\num{4.64e-04}$ & $\num{6.49e-02}$ & \cellcolor{orange!10}$\num{4.50e-03}$ & 338958 & 84992 & 1201.8 & 699.5 \\\hline
Zakopane Style Chair & \cellcolor{orange!10}$\num{6.46e-05}$ & $\num{1.32e-04}$ & $\num{1.96e-02}$ & \cellcolor{orange!10}$\num{8.28e-03}$ & 283494 & 74948 & 1204.9 & 689.6 \\\hline
Zentsuji Gojyunotou & $\num{1.69e-03}$ & \cellcolor{orange!10}$\num{1.66e-03}$ & $\num{5.37e-02}$ & \cellcolor{orange!10}$\num{2.65e-02}$ & 328240 & 266116 & 1434.5 & 1118.1 \\\hline\hline
Mean & \num{2.08e-02} & \textbf{\num{6.89e-04}} & \num{9.83e-02} & \textbf{\num{2.35e-02}} & N/A & N/A & 1264.3 & 735.8 \\\hline
Median & \num{5.30e-04} & \textbf{\num{4.22e-04}} & \num{4.02e-02} & \textbf{\num{1.53e-02}} & N/A & N/A & 1247.9 & 728.6 \\\hline
Mean w/o \textcolor{red}{$\dagger$} & \num{4.67e-03} & \textbf{\num{6.20e-04}} & \num{6.81e-02} & \textbf{\num{2.21e-02}} & N/A & N/A & 1267.8 & 738.3 \\\hline
Median w/o \textcolor{red}{$\dagger$} & \num{4.88e-04} & \textbf{\num{4.15e-04}} & \num{3.65e-02} & \textbf{\num{1.50e-02}} & N/A & N/A & 1248.6 & 729.4 \\\hline
\end{tabular}
\caption{\footnotesize Comparison of TetWeave and Grid Deformation at 128K points and with $64^3$ resolution. At this resolution, optimization takes about 21 minutes for TetWeave, and 12 minutes for the fully-adaptive grid. The Chamfer and Hausdorff distance for the deformable grid is better on average than TetWeave at this resolution, although the majority of models are better for TetWeave, due to models with thin structures which TetWeave performs poorly on. \textcolor{red}{$\dagger$} marked as an outlier.}
\label{tab:comp-128K-64}
\end{table*}

\end{document}